\documentclass[11pt]{article}
\usepackage{epsfig}
\renewcommand{\theequation}{\thesection.\arabic{equation}}

\newcommand{\MSbar}{\overline{\mbox{MS}}}
\newcommand{\Nc}{N_{c}}
\newcommand{\Nf}{N_{\!f}}

\newcommand{\NA}{N_{\!A}}

\newcommand{\Dslash}{D \! \! \! \! /}
\newcommand{\kslash}{k \! \! \! /}
\newcommand{\pslash}{p \! \! \! /}
\newcommand{\xslash}{x \! \! \! /}
\newcommand{\yslash}{y \! \! \! /}
\newcommand{\zslash}{z \! \! \! /}
\newcommand{\partialslash}{\partial \! \! \! /}
\newcommand{\half}{\mbox{\small{$\frac{1}{2}$}}}

\begin{document}

\title{Large $\Nf$ quantum field theory}

\author{J.A. Gracey, \\ Theoretical Physics Division, \\
Department of Mathematical Sciences, \\ University of Liverpool, \\ P.O. Box
147, \\ Liverpool, \\ L69 3BX, \\ United Kingdom.}
\date{}

\maketitle

\vspace{5cm}
\noindent
{\bf Abstract.} We review the development of the large $N$ method, where $N$ 
indicates the number of flavours, used to study perturbative and 
nonperturbative properties of quantum field theories. The relevant historical 
background is summarized as a prelude to the introduction of the large $N$ 
critical point formalism. This is used to compute large $N$ corrections to 
$d$-dimensional critical exponents of the universal quantum field theory 
present at the Wilson-Fisher fixed point. While pedagogical in part the 
application to gauge theories is also covered and the use of the large $N$ 
method to complement explicit high order perturbative computations in gauge 
theories is also highlighted. The usefulness of the technique in relation to 
other methods currently used to study quantum field theories in $d$-dimensions 
is also summarized. 

\vspace{-17.0cm}
\hspace{13.0cm}
{\bf LTH 1189}

\newpage

\section{Introduction}	

The use of perturbation theory was established in the very early years after 
the successful marriage of quantum mechanics and relativity into quantum field 
theory. Indeed one example of the many successes is the high loop order
perturbative computation of the electron anomalous magnetic moment over many 
years using Quantum Electrodynamics (QED) \cite{1,2,3,4,5,6}. Of late the 
experimental value of the fine structure constant has been extracted from
the value of the magnetic moment to an astonishing accuracy \cite{7,8}. In sum
both theory and experiment are in virtual agreement to eleven orders of
precision for the fine structure constant which is a constant of Nature. While 
this is one example others abound such as the recent intense activity into 
computing the $\beta$-function of Quantum Chromodynamics (QCD) to five loop 
accuracy \cite{9,10,11,12}. Again this was built up from successive loop order
computations over nearly half a century \cite{13,14,15,16,17,18,19}. Moreover 
such examples are not restricted to field theories in particle physics as there
have been parallel precise applications in condensed matter theory. 
Mathematically perturbation theory can be summarized by saying that Feynman
graphs contributing to a Green's function or a physical process are ordered
with respect to some counting parameter and then the individual regularized
graphs are evaluated and summed at a particular order. This procedure is then 
repeated at the next order. In perturbation theory the counting parameter is 
the coupling constant. It is assumed to be small with the hope that the series 
is convergent or at best asymptotic in a certain range of the coupling 
constant. While this may appear simplistic or unnecessary to state, the 
ordering criterion is in fact the first organizational step of any perturbative
analysis. However, in certain field theories the conventional perturbative 
expansion is not the only one available.  

If the fields of a theory are elements of a symmetry group, such as a flavour
or colour symmetry, then parameters associated with that group could play the 
role of an ordering parameter. An obvious example of such a scenario arises in 
QCD which describes the strong interactions. It involves $\Nf$ flavours of 
spin-$\half$ quarks which are the matter fields that interact via the force 
quanta which are the gluons. These take values in the nonabelian colour group 
$SU(3)$ but the more general group $SU(\Nc)$ can be considered where $\Nc$ is 
the number of colours. Either parameter, $\Nf$ or $\Nc$, can be used as a way 
of ordering graphs contributing to a Green's function. Then the associated 
small parameter in the respective cases are $1/\Nf$ and $1/\Nc$ when $\Nf$ and 
$\Nc$ are large. Once either ordering is defined one has to follow the earlier 
prescription and use a means to evaluate the full set of (regularized) graphs 
at each order. In general for both these cases and in the coupling constant 
expansion this is not always straightforward to achieve. For instance, in QCD 
at leading order in $1/\Nc$ there are an infinite number of graphs to determine
\cite{20,21}. For the other two cases the number of low loop order graphs is 
small but can increase rapidly the deeper one goes in the expansion. That said 
it is crucial to understand that whatever choice of graph ordering is selected 
{\em no} graphs are ever omitted. One still has an infinite number to determine
to all orders in the expansion parameter of choice. For the problem at hand, 
however, the approximation of the first few orders may suffice to gain 
satisfactory insight. In viewing the exercise of analysing Green's function 
computations in this way we can say at the outset that the formalism for 
coupling constant perturbation theory is very well established and standard 
text book fodder. Equally the $1/\Nc$ expansion of QCD has been well covered in
the literature on nonabelian gauge theories \cite{20,21} and in some sense is 
restricted to that class.

By contrast the remaining ordering parameter, $\Nf$, when $\Nf$ is large is the
topic of this review. There are various reasons for this. Aside from the fact 
that the technique is not widely used it does undergo revival from time to time
in light of new developments in other areas of quantum field theory. This has 
been the case in recent years. For instance, it is now becoming accepted as a 
complementary tool to studying conformal field theories in $d$-dimensions. This
area has also been examined using, for example, the functional or exact
renormalization group developed in Refs. \cite{23,24} and \cite{25} based on 
ideas of Wilson \cite{26} as well as the more recent conformal bootstrap 
programme. This was introduced in its modern guise in
Refs. \cite{27,28,29,30,31} and \cite{32} with an historical summary also 
available \cite{33}. Each of these techniques can be applied to critical 
phenomena, such as the location of phase transitions in field theories and the 
numerical evaluation of the associated critical exponents. Moreover in the case
of the functional renormalization group this can be carried out in continuous 
spacetime dimensions $d$ where $d$ here is not related to the extension of 
spacetime via dimensional regularization. The connection with large $N$ stems 
from the seminal development of the large $N$ critical point formalism in the 
early eighties by Vasil'ev, Pismak and Honkonen for $O(N)$ scalar field 
theories \cite{34,35,36}. The elegance of the method is that rather than 
consider the large $N$ expansion of an $O(N)$ symmetric theory in a fixed 
integer dimension, one focuses on the theory defining the universality class at
the $d$-dimensional Wilson-Fisher fixed point \cite{37,38,39,40}. Thereby the 
method established exponents as functions of $d$ to {\em three} orders in 
$1/N$. This extended $O(1/N^2)$ fixed three dimensional computations in the 
same model by various groups \cite{41,42,43,44,45,46,47,48} where leading order
arbitrary dimension exponents were determined in Ref. \cite{49}. The use of 
$1/N$ as the perturbative parameter, being dimensionless, allows one to 
transcend the restrictions of a dimensional coupling constant expansion near a 
critical dimension. Moreover, the $d$-dimensional information in the critical 
exponent contains data on each of the field theories in the same universality 
class at each of their respective critical dimensions. By the latter we mean 
the spacetime dimension where the theory is renormalizable. In more recent 
years with the extensions of perturbative calculations in $O(N)$ $\phi^4$ 
theory, for example, from three to seven loops 
\cite{50,51,52,53,54,55,56,57,58,59} such data in the exponents of 
Refs. \cite{34,35} and \cite{36} has proved to be invaluable independent 
checks on these high order results. Equally the summation of several $1/N$ 
orders of the exponents have produced values for relatively {\em low} $N$ which
are competitive with functional renormalization group and conformal bootstrap 
studies of the same quantities. In addition since the large $N$ critical point 
technique corresponds to the systematic construction of perturbation theory in 
powers of $1/N$ at the Wilson-Fisher fixed point it is a tool with which to 
directly examine properties of $d$-dimensional conformal field theories. At 
fixed points there is a scale and conformal symmetry.

With these general considerations it seems apt to devote a review to the topic 
of the $1/N$ expansion where by $N$ we will mean the flavour expansion. At 
points where we discuss a gauge theory connection we will use $\Nf$ to
distinguish it from $\Nc$. This is the reason for the appearance of $\Nf$ in 
the article title and we will regard $N$ and $\Nf$ as synonyms throughout but
not with $\Nc$. We will restrict $\Nf$ to gauge theory studies as is the 
convention in the literature. Moreover, as the critical point method of 
Refs. \cite{34,35} and \cite{36} is not as widely established as the other 
techniques we note that main part of the article is aimed at being pedagogical 
with pointers to the relevant background and consequences at appropriate 
junctures. In this way the reader interested in understanding what has been 
achieved or wishing to get a flavour of the technique can benefit from the 
signposts. Though it is worth stressing that the large $N$ formalism is by no 
means complete. For instance, in the context of early large $N$ studies, which 
we will discuss here, this article is not intended to replace classic articles 
by Coleman \cite{60,61} as well as other reviews \cite{62,63}. Indeed there is 
little doubt that what is reviewed here will be overtaken at some time by new 
ideas to extend large $N$ methods. We do consider, however, the different early
theory approaches. One of these, which is that of the explicit bubble summation
for massless fields, and in particular in QED \cite{64,65}, has become 
fashionable again particularly in studying new ideas such as asymptotic safety 
and possible extensions of the Standard Model. A recent review on the latter 
can be found in Ref. \cite{66}. By bubble summation we mean the inclusion of 
closed matter field loops, which contribute a factor of $N$, in force field 
propagators. Equally this summation approach re-invigorated the study of 
renormalons in gauge theories in the nineties which has been reviewed in 
Ref. \cite{67}. Again we briefly discuss various aspects of both these 
applications of the explicit bubble sum approach in the $1/N$ expansion. The 
problems where this large $N$ technique is used are primarily for situations 
which are not directly connected with fixed points or conformal symmetry of the
underlying universal theory. So the construction of 
Refs. \cite{34,35} and \cite{36} is not directly applicable. Instead the 
bubble sum method primarily operates in a theory in a fixed integer rather than
arbitrary dimension and which possibly requires dimensional regularization 
\cite{64,65}.

Therefore in saying this we are revealing some of the richness of the $1/N$
approach which is sometimes referred to as being nonperturbative. This needs
to be qualified since nonperturbative can have different interpretations. One 
is that it just simply means not perturbation theory in the sense of the 
conventional coupling constant expansion. Another connotation of its meaning is 
nonanalytic. By this it is meant that contributions to the quantity being 
computed have an asymptotic expansion in the coupling constant of zero. A 
simple example is the function $e^{-1/g^2}$ where $g$ is a coupling constant. 
This nonanalytic function has zero asymptotic expansion and can never be 
accessed in perturbation theory. This example is particularly apt since it 
relates to one discovery associated with the large $N$ expansion. In early work
on the two dimensional $O(N)$ nonlinear sigma model, summarized in 
Ref. \cite{68}, and $O(N)$ Gross-Neveu model \cite{67} the saddle point 
approximation to the effective potential of a composite field in the large $N$ 
expansion revealed an energetically favourable vacuum solution. In other words 
the usual perturbative vacuum was not an absolute minimum and in fact was 
unstable to fluctuations. Within large $N$ this led to dynamical symmetry 
breaking and mass generation for the matter fields which perturbatively remain 
massless. At leading order in $1/N$ the dependence of the generated mass can be
computed as a function of $g$ and is found to be $e^{-1/g^2}$. Another such 
nonperturbative example accessed in large $N$ is of course the renormalon 
structure which also has similar nonanalytic behaviour in the coupling 
constant. In highlighting several of the topics accessible using large $N$ 
methods here we hope to convey the essence of the background to each as well as
indicate possible avenues to be followed in future. For instance, one early 
hope with the ability to access dynamical symmetry breaking in simple two
dimensional models was that it could give insight into the mechanism for
generating a mass gap and colour confinement in QCD \cite{68}. This clearly has
not been realised but ideas using large $N$, for both $\Nf$ and $\Nc$, have 
both given intriguing suggestions that there is a mechanism which can be 
studied analytically albeit in an approximation. One idea which we will mention
is the adaptation of the effective potential approach in four dimensions for a 
composite of nonabelian gauge fields which retains several key features from 
the toy two dimensional $O(N)$ nonlinear sigma and Gross-Neveu models. 

The article is organized as follows. We recall aspects of the large $N$ bubble
sum formalism and its use in determining that there is dynamical symmetry
breaking in several two dimensional models in section $2$. An extension of the
resummation to extract the perturbative renormalization group functions is the 
main topic of section $3$. Within this there is overlap with studying 
renormalon problems primarily in Green's functions. The subsequent section is 
devoted to the large $N$ critical point formalism where the focus is primarily 
on $O(N)$ scalar field theories. This includes recalling that in this formalism
there is a renormalization procedure which operates at a Wilson-Fisher fixed
point that has parallels with conventional perturbation theory. Aspects of the 
extension of the critical point approach in relation to fermionic theories 
either involving scalar or gauge fields is the focus in section $5$. One 
benefit of the method of Refs. \cite{34} and \cite{35} is the relation the 
$d$-dimensional exponents have with perturbative renormalization group
functions. This is discussed in section $6$ which includes an example in QCD.
Several areas of current activity at large overlap with large $N$ ideas such as
the conformal bootstrap and we discuss these connections in section $7$ before 
concluding with remarks in section $8$. Two appendices detail at length 
techniques used in evaluating massless Feynman integrals with noninteger 
propagator powers. The first of these are general rules while the second gives
of examples of their application.

\section{Background}	

By way of introduction to the area of large $N$ quantum field theory we discuss
the general original approaches outlined in early work in simple theories. For 
the most part these will be the $O(N)$ nonlinear sigma model and the $O(N)$ 
Gross-Neveu model. Several nonexhaustive references to such work include 
Refs. \cite{60,61,62,63,68} and \cite{69}. 

\subsection{Bubble summation}

The basic approach in these low dimensional theories was to exploit the fact 
that the real parameter $N$ of the (flavour) symmetry group could order the 
Feynman graphs contributing to a Green's function differently to the ordering 
defined by the loop expansion of coupling constant perturbation theory. For 
instance in the large $N$ method \cite{69} the generic coupling constant $g$ 
was given a notional $N$ dependence by requiring that the combination $g^2 N$ 
was held fixed in the large $N$ limit which is $N$~$\to$~$\infty$. Therefore 
the leading order or dominant graphs contributing to a Green's function 
corresponded to those where there were chains of bubbles of matter fields which
are illustrated in Fig. \ref{cha1cr}. Consequently one bubble gave a factor 
of $N$ to the value of the graph which with the associated two coupling 
constants meant that a chain of $n$ bubbles had the same notional $N$ 
dependence as any other chain. 

{\begin{figure}[ht]
\begin{center}
\includegraphics[width=9.5cm,height=1.0cm]{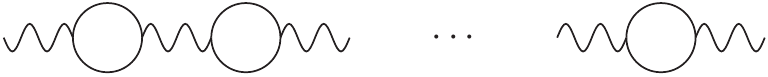}
\end{center}
\caption{Example of a large $N$ bubble chain at leading order.}
\label{cha1cr}
\end{figure}}

To illustrate this we recall the basic formulation of both the $O(N)$ nonlinear
sigma model \cite{68} and $O(N)$ Gross-Neveu model \cite{69}. The former has 
the Lagrangian
\begin{equation}
L^{\mbox{\footnotesize{NLSM}}} ~=~ 
\frac{1}{2} \left( \partial_\mu \phi^i \right)^2 ~+~ \frac{1}{2}
\sigma \left( \phi^i \phi^i ~-~ \frac{1}{g} \right)
\label{lagnlsm}
\end{equation}
where $1$~$\leq$~$i$~$\leq$~$N$, $g$ is the coupling constant and $\sigma$ is a
Lagrange multiplier field which ensures that the scalar fields $\phi^i$ lie on 
the sphere $S^{N-1}$. This is the coordinate free version of the model. 
Eliminating the constraint will introduce a set of coordinates for each
coordinate patch but for the large $N$ approach (\ref{lagnlsm}) is used. In the
case of the Gross-Neveu model we have \cite{69}
\begin{equation}
L^{\mbox{\footnotesize{GN}}} ~=~ 
\frac{1}{2} i \bar{\psi}^i \partialslash \psi^i ~+~ 
\frac{1}{2} \sigma \bar{\psi}^i \psi^i ~-~ \frac{\sigma^2}{2g}
\label{laggn}
\end{equation}
where $\psi^i$ is a fermion, $g$ is the coupling constant but now in this case
$\sigma$ is an auxiliary field. This can be eliminated from (\ref{laggn}) to 
produce a theory with a quartic fermion self-interaction. That formulation of 
(\ref{laggn}) is more appropriate to use if one is carrying out explicit 
perturbative computations in two dimensions which is the theory's critical
dimension. As there is only one vertex in both (\ref{lagnlsm}) and 
(\ref{laggn}) then at leading order in the large $N$ expansion one diagram 
contributing to the $\sigma$ field $2$-point function is given in Fig. 
\ref{cha1cr} where the dots indicate the addition of other bubbles of $\phi^i$ 
or $\psi^i$ fields. Irrespective of whether the fields are massive or not all 
graphs in the sequence are of the same order in large $N$. To summarize, 
summing up the graphs produces a geometric series which is formally illustrated
in Fig. \ref{chasumcr}. Our representation of the process in this figure is 
schematic because for different theories the actual outcome of the summation 
depends on the structure of the underlying Lagrangian. For instance in 
(\ref{laggn}) the quadratic term in $\sigma$ is part of the summation but there
is no corresponding term in (\ref{lagnlsm}). In either the massive or massless 
case the explicit value of the bubble can be determined for the theory of 
interest and the large $N$ form of the $\sigma$ propagator can be found. We
will give an example of this shortly. Also as we will be discussing theories 
which are similar in structure to (\ref{lagnlsm}) and (\ref{laggn}) later from 
this point of view we note that the respective interactions are what we term as
core and in fact each define a separate universality class. Within each a 
sizeable number of field theories are equivalent. In this context we will refer
to the $\sigma$ field and its later equivalents as the force field. The fields 
to which it couples through this common interaction and which lie in the 
flavour group will be termed the matter fields. For our examples these are 
clearly $\phi^i$ and $\psi^i$ of the two respective Lagrangians.

{\begin{figure}[ht]
\begin{center}
\includegraphics[width=12.5cm,height=1.5cm]{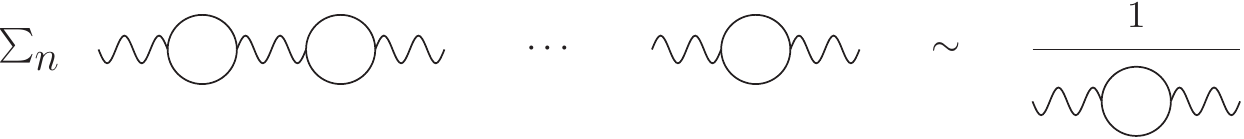}
\end{center}
\caption{Summation of leading order bubble chain.}
\label{chasumcr}
\end{figure}}

\subsection{Dynamical mass generation}

One of the early interests in studying quantum field theories via the large
$N$ expansion rested in the fact that nonperturbative properties can be
studied. At this point it is worth noting that our use of the word 
nonperturbative here means the zero instanton sector where the small coupling 
constant expansion is around the free field solution. In other words we are not
in a sector where there is a perturbative expansion around a classical extended
object or solution such as a breather, kink, monopole or dyon. These latter 
quantities sometimes are included in the overall banner of nonperturbative in 
the nonanalytic sense. So for us nonperturbative now means the summation of 
perturbation theory in the zero instanton and zero extended object sector. 
Given this it transpires that both (\ref{lagnlsm}) and (\ref{laggn}) possess 
rich nonperturbative properties. In the perturbative expansion of the theories 
around zero coupling constant if the matter fields $\phi^i$ and $\psi^i$ are 
originally massless then they remain so. However the perturbative vacuum in 
each case is not stable and there is a more energetically favourable vacuum 
solution which is not perturbatively accessible. In that vacuum there is
dynamical mass generation whereby the matter field of each Lagrangian becomes 
massive. Moreover the apparently nonpropagating field $\sigma$ of both theories
develops a nonfundamental propagator. In effect each becomes a bound state of 
two matter fields as is suggested by the graph of Fig. \ref{chasumcr}. In the 
perturbative formulation of both (\ref{lagnlsm}) and (\ref{laggn}) the $\sigma$
field is not present. In the case of (\ref{lagnlsm}) eliminating the Lagrange 
multiplier constraint by choosing a coordinate system for the underlying 
manifold gives
\begin{equation}
L^{\mbox{\footnotesize{$\sigma$}}} ~=~ 
\frac{1}{2} g_{ab}(\phi) \partial^\mu \phi^a \partial_\mu \phi^b
\label{lagnlsmelim}
\end{equation}
where $1$~$\leq$~$a$~$\leq$~$N-1$ and $g_{ab}(\phi)$ is the metric of the
coordinate system chosen for the patch which includes the free field case. The
coupling constant appears implicitly in the metric. In the case of the
Gross-Neveu model there is a parallel re-expression. Eliminating the $\sigma$
field of (\ref{laggn}), as noted earlier, produces the purely fermionic
Lagrangian \cite{69} 
\begin{equation}
L^{\mbox{\footnotesize{GN}}} ~=~ 
\frac{1}{2} i \bar{\psi}^i \partialslash \psi^i ~+~ 
\frac{g^2}{8} \left( \bar{\psi}^i \psi^i \right)^2 ~.
\label{laggnelim}
\end{equation}
In both cases there is clearly no mass term for each matter field and, 
moreover, {\em none} is generated or apparent in any perturbative expansion. 

{\begin{figure}[ht]
\begin{center}
\includegraphics[width=2.5cm,height=2.5cm]{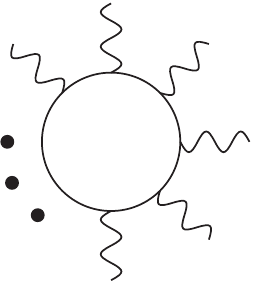}
\end{center}
\caption{$n$-point graphs contributing to the leading order effective
potential.}
\label{efpocr}
\end{figure}}

To access the dynamical mass generation one has to construct the effective
potential for the $\sigma$ field in both cases. For instance, using 
(\ref{laggn}) one sums up the leading order graphs with $n$ $\sigma$ field 
insertions on a closed fermion loop which are illustrated in Fig. 
\ref{efpocr}. This produces the effective potential $V(\langle \sigma \rangle)$
which is
\begin{equation}
V(\langle \sigma \rangle) ~=~ \frac{1}{2} \langle \sigma \rangle^2 ~+~
\frac{N g^2(\bar{\mu}) \langle \sigma \rangle^2}{4\pi} \left[ \ln
\left( \frac{\langle \sigma \rangle^2}{\bar{\mu}^2} \right) ~-~ 3 \right] 
\end{equation}
at leading order for (\ref{laggn}) where $\bar{\mu}$ is the point where the
divergences are subtracted \cite{69}. The first term is clearly the canonical 
quadratic one of the original Lagrangian which would lead to the solution 
$\langle \sigma \rangle$~$=$~$0$ classically for the vacuum expectation value 
of $\sigma$. With quantum corrections included this classical vacuum is not the
only turning point of $V(\sigma)$ as there is a second one at
\begin{equation}
\langle \sigma \rangle ~=~ \bar{\mu} \exp 
\left( 1 ~-~ \frac{\pi}{Ng^2(\bar{\mu})}
\right)
\label{gndynmass}
\end{equation}
which is a global minimum. The one with $\langle \sigma \rangle$~$=$~$0$ 
corresponds to a local maximum meaning that the perturbative vacuum is 
unstable. With the nonzero vacuum expectation value for $\sigma$ the fermion 
$\psi^i$ develops a mass dynamically. From (\ref{gndynmass}) one finds that 
this mass depends on the coupling constant nonanalytically. Such a dependence 
on $g$ means that the dynamical mass has a zero asymptotic expansion near 
$g$~$=$~$0$ and hence not accessible perturbatively. What has transpired is 
that the true vacuum of the theory has been accessed or touched in the large 
$N$ summation of graphs in our nonperturbative meaning. Though it is worth 
stressing that while it has been accessed it has been done so within an 
approximation at {\em leading} order only. Exploring the scenario to higher 
precision requires corrections which are technically very difficult to carry 
out when the nonzero dynamically generated mass is present. Several studies in 
this direction can be found in Refs. \cite{70,71} and \cite{72}. 

However, the existence of the mass gap in this theory given by 
(\ref{gndynmass}), revealed via the leading order large $N$ approach, has been 
refined by the determination of the mass gap of (\ref{lagnlsm}) and 
(\ref{laggn}) {\em exactly} \cite{73,74,75,76} from the explicit form of the 
exact $S$-matrix which was constructed in 
Refs. \cite{77,78,79} and \cite{80}. The large $N$ expansion has been used to
check the exact expression \cite{81,82}. In the former case the $O(1/N)$ 
corrections to the fermion $2$-point function in the Gross-Neveu model were 
computed using (\ref{laggnshift}) in dimensional regularization where tadpole 
graphs have to be included. The latter large $N$ check was carried out by 
computing the $2$-point function for the nonlinear sigma model at $O(1/N^2)$ 
but with lattice perturbation theory. Indeed it is possible to use this large 
$N$ lattice regularization approach to extend analyses to $O(1/N^3)$ for the 
nonlinear sigma model \cite{83}. Equally another check on the exact 
$S$-matrices proposed for (\ref{lagnlsm}) and (\ref{laggn}) centred on the 
assumptions behind the construction. In two Minkowskian spacetime dimensions 
particles can only move along a line which puts a restriction on scattering. 
For theories with an exact $S$-matrix it was assumed that there is no 
$2$~$\to$~$4$ scattering and for three particle scattering the order of the 
constituent $2$~$\to$~$2$ scatterings was immaterial \cite{77,78,79,80}. In the
former case this has been checked explicitly in several cases in the large $N$ 
expansion \cite{77,78,79,80,84} using the K\"{a}ll\'{e}n-Toll cutting rule 
\cite{85}. This rule is specific to massive two dimensional Feynman integrals 
and allows one to express finite one loop integrals as a sum of tree graphs. 
The premise behind its usefulness is the same as that for the exact $S$-matrix 
construction in that particles can move in only one of two directions in two 
dimensions. This allows one to express momenta in a basis determined by the 
light cone and exploit properties of complex analysis \cite{85}. Applying the 
rule in the case of the $2$~$\to$~$4$ scattering the leading order large $N$ 
one loop graph is cut open in such a way that its contribution is equal and 
opposite to the sum of all six tree graphs which ensures that there is no 
production in these theories. By contrast another useful cutting rule was 
provided in Ref. \cite{86} for evaluating a class of massive integrals 
occurring in large $N$ computations. The underlying mathematics is one of the 
Gauss relations of the hypergeometric function. 

Equipped with this knowledge we can return to (\ref{laggn}) and reconsider it.
While the conventional way of calculating perturbatively in the $O(N)$ 
Gross-Neveu model is to use the Lagrangian (\ref{laggnelim}) perturbation
theory can also be carried out in (\ref{laggn}). For instance, this has been 
achieved in Ref. \cite{87} at {\em three} loops. We mention this work as it is 
relevant to the large $N$ approach. While we have noted that the true vacuum 
has $\langle \sigma \rangle$~$\neq$~$0$ one can still perform perturbative 
computations in (\ref{laggn}) with $\langle \sigma \rangle$~$=$~$0$. This is 
achieved by noting that the so-called propagator of $\sigma$ in the classically
unstable vacuum is unity. In other words it is momentum independent. The 
technical exercise in doing such calculations can be viewed in Ref. \cite{87}. 
However, (\ref{laggn}) is not perturbatively renormalizable multiplicatively 
\cite{87,88,89} since the fermion $4$-point function is not finite. Therefore 
an extra interaction needs to be appended to the Lagrangian (\ref{laggn}) for 
perturbative computations which is that of (\ref{laggnelim}) but with a 
different coupling constant. Despite this the three loop $\beta$-function of 
the $O(N)$ Gross-Neveu model was computed for the coupling constant $g$ which 
was in agreement with the independent computation of the $\beta$-function 
carried out at the same time in Ref. \cite{90}. While the original and main 
motivation of Ref. \cite{87} was to construct $V(\sigma)$ in perturbation 
theory to three loops for the present consideration it illustrates that it is 
technically possible to use (\ref{laggn}) for computations in the classical 
vacuum. Earlier loop calculations of the renormalization group functions in 
(\ref{laggn}) were carried out over several years in 
Refs. \cite{91,92} \cite{93}. In noting that (\ref{laggn}) is not formulated 
in a way that is perturbatively renormalizable, we need to clarify that neither
(\ref{laggn}) nor (\ref{laggnelim}) are multiplicatively renormalizable when 
dimensionally regularized \cite{88,89,94,95,96}. This is because in using this 
regularization evanescent $4$-point operators are generated which spoil the 
multiplicative renormalizability and is only a feature of two dimensional 
theories with $4$-fermi interactions. In the case of (\ref{laggnelim}) the 
first order where such an operator appears is three loops \cite{96,97} and its 
effect in the renormalization group functions will not be manifest until 
{\em four} loops \cite{98,99}. However, the formalism developed in 
Refs. \cite{88} and \cite{89} to handle these evanescent operators in two 
dimensional $4$-fermi theories, where they first arise at a lower loop order 
than (\ref{laggn}), can be applied to the $O(N)$ Gross-Neveu case to allow one 
to extract the correct four loop renormalization group functions \cite{98,99}. 
For instance, as the $O(2)$ theory is free the $\beta$-function has to be 
proportional to $(N-2)$. Without knowing about the generation of the evanescent
operators or using a formalism to accommodate their effect, this factor would 
not appear in the $\beta$-function at four loops \cite{99}. 

Given that the effective potential of (\ref{laggn}) determines the true vacuum 
to be $\langle \sigma \rangle$~$\neq$~$0$ we can introduce the new field
$\sigma^\prime$ by
\begin{equation}
\sigma ~=~ m ~+~ \sigma^\prime
\end{equation}
where $m$~$=$~$\langle \sigma \rangle$. Then (\ref{laggn}) becomes 
\begin{equation}
L ~=~ \frac{1}{2} i \bar{\psi}^i \partialslash \psi^i ~+~ 
\frac{1}{2} m \bar{\psi}^i \psi^i ~+~ 
\frac{1}{2} \sigma^\prime \bar{\psi}^i \psi^i ~-~ 
\frac{{\sigma^\prime}^2}{2g} ~-~ \frac{m \sigma^\prime}{g}
\label{laggnshift}
\end{equation}
where we have omitted the constant term. The linear term in $\sigma^\prime$ is
retained as there are tadpole corrections to the $2$-point and other functions 
which play a crucial role in ensuring the quantum theory is consistent for a
nonzero $m$. Further technical details on this can be found in 
Refs. \cite{86} and \cite{100}. As this is the form of the Lagrangian in the 
neighbourhood of the true vacuum we can examine it in the large $N$ formalism. 
With $m$~$\neq$~$0$ the $\sigma^\prime$ field develops a propagator which can 
be deduced from the sum of graphs in Fig. \ref{chasumcr}. In particular at 
leading order in $1/N$ the $\sigma^\prime$ propagator $D_\sigma(p^2)$ is
\begin{equation}
D_\sigma(p^2) ~=~ \frac{2}{[p^2+4m^2] J(p^2)N}
\label{sigprop}
\end{equation}
where
\begin{equation}
J(p^2) ~=~ \frac{1}{2\pi \sqrt{p^2[p^2+4m^2]}} \ln \left[
\frac{\sqrt{p^2+4m^2} + \sqrt{p^2}}{\sqrt{p^2+4m^2} - \sqrt{p^2}} \right]
\end{equation}
in two dimensions. This is clearly a nonfundamental propagator with a pole at
$-4m^2$ from the prefactor. It corresponds to the bound state of two fermions. 
Higher order large $N$ corrections will adjust the bound state mass from this 
leading order value. This can be seen from an alternative point of view. The 
full mass spectrum of the theory can be adduced from the exact $S$-matrix of 
the $O(N)$ Gross-Neveu model which is given in 
Refs. \cite{77,78,79} and \cite{80}. For example it provides the full 
spectrum of particle excitations and their masses including that of dynamically
generated $\sigma$ field. As an aside the {\em exact} $S$-matrix and the bound 
state particle masses are a function of $(N-2)$ for the $O(N)$ theory rather 
than $N$ alone. Indeed it should be the case that if one could compute higher 
order $1/N$ corrections the large $N$ expansion the quantity of interest these 
additional terms should be such as to produce a $1/(N-2)$ expansion. The 
presence of this factor is the dual Coexeter number of the $O(N)$ group and is 
not unrelated to the fact that (\ref{laggnelim}) corresponds to a free field 
theory at $N$~$=$~$2$ as already noted.

One aspect of (\ref{sigprop}) which will be important for later is the form of 
the $\sigma^\prime$ propagator in $d$-dimensions. This can be determined from 
the dimensionally regularized evaluation of the one loop bubble graph on the 
right hand side of the equation in Fig. \ref{chasumcr}. In particular 
\begin{equation}
J(p^2) ~=~ \frac{\Gamma(2-\half d)}{(4\pi)^{\half d}} \left( 
\frac{p^2+4m^2}{4} \right)^{\half d - 2} {}_2F_1 \left( 2 - \frac{d}{2},
\frac{1}{2}; \frac{3}{2}; \frac{p^2}{p^2+4m^2} \right) 
\label{jhyp}
\end{equation}
where ${}_2F_1(a,b;c;z)$ is the hypergeometric function. So in the massless
limit
\begin{equation}
D_\sigma(p^2) ~ \propto ~ \frac{1}{(p^2)^{\half d - 1}} ~. 
\label{gnprophyp}
\end{equation}
While this corresponds to a canonical propagator in four dimensions the 
$4$-fermi interaction is nonrenormalizable in perturbation theory in that
spacetime. However it transpires that this propagating field will play a role
in the critical point dynamics of the Gross-Neveu model at the Wilson-Fisher
fixed point in $d$-dimensions which will be discussed later.

\subsection{Lagrangian reformulation}

As it stands (\ref{laggnshift}) contains a massive fermion with a propagator
which can be deduced by conventional means. While the large $N$ expansion has
revealed that the $\sigma^\prime$ field is dynamical with a propagator albeit
nonfundamental the latter is not present at the outset. In 
Refs. \cite{34,100} and \cite{101} the effective Lagrangian description of 
this scenario in the true vacuum has been addressed. Conventional quantum field
theory at the defining stage writes the action $S$ as a free part $S_0$ and an 
interacting piece $S_I$ with $S$~$=$~$S_0$~$+$~$S_I$. How the split is made is 
entirely an open choice as there is not a unique way of defining $S_0$. For 
instance in (\ref{laggnshift}) the mass term $\half m \bar{\psi}^i \psi^i$ 
could be included in the Lagrangian of $S_0$ or $S_I$. If it is present in the
former one develops the field theory with a massive fermion. By contrast if it 
is present in $S_I$ one has a massless fermion but also a $2$-point vertex 
independent of the coupling constant. In this situation at each loop order one 
has to include an {\em infinite} number of insertions of this $2$-point vertex 
in all the graphs. The consequence of this is to reproduce the massive fermion 
propagator through a geometric series similar to that of the large $N$ bubble 
sum. In other words while the split in the Lagrangian is not unique the results 
obtained are ultimately independent of how it is carried out. An analogous 
process underlies the observation of Refs. \cite{34,100} and \cite{101}. To 
accommodate the fact that the $\sigma^\prime$ field has a nontrivial propagator
in (\ref{laggnshift}) at the outset it is included redundantly in a bilocal way 
\begin{eqnarray}
S_0 &=& \int_x \left[ \frac{1}{2} i \bar{\psi}^i \partialslash \psi^i ~+~ 
\frac{1}{2} m \bar{\psi}^i \psi^i ~-~ \frac{{\sigma^\prime}^2}{2g} \right] ~+~
\frac{N}{2} \int_x \int_y \sigma^\prime(x) {\cal D}(x,y) \sigma^\prime(y)
\nonumber \\
S_I &=& \int_x \left[ \frac{1}{2} \sigma^\prime \bar{\psi}^i \psi^i 
~-~ \frac{m \sigma^\prime}{g} \right] ~-~
\frac{N}{2} \int_x \int_y \sigma^\prime(x) {\cal D}(x,y) \sigma^\prime(y)
\label{laggnsplit}
\end{eqnarray}
where ${\cal D}(x,y)$ is the coordinate space representation of the one loop
bubble of Fig. \ref{chasumcr}. When $m$~$\neq$~$0$ this is a complicated 
function which is related to (\ref{jhyp}) but simplifies substantially when 
$m$~$=$~$0$. This formulation (\ref{laggnsplit}) allows one to carry out the 
formal large $N$ development and renormalization \cite{34,100,101} from a 
Lagrangian standpoint as it corresponds to the (effective) field theory active 
at the true (nonclassical) stable vacuum. While both the Gross-Neveu and 
nonlinear sigma models are two dimensional asymptotically free field theories
with matter fields, which are massless in the classical vacuum, the dynamical 
generation of mass in the true vacuum is one of the reasons they have been 
studied. This is because several of these properties are shared by QCD in four 
dimensions. 

While the $\sigma$ field plays a role similar to the QCD gauge field which is 
massless, low energy lattice gauge theory evidence has accumulated over the 
last decade that indicates that the gluon appears to have an associated nonzero
mass scale. This derives from studies of the gluon propagator in the Landau 
gauge initiated in Ref. \cite{102}. At zero momentum it appears that the 
Landau gauge gluon propagator freezes to a nonzero value. This is not 
indicative of a mass for the gluon in the fundamental sense as the gluon 
propagator does not have a simple or higher order pole at any nonzero momentum.
A simple pole would imply the gluon was visible and contradict the expectation 
that it is a confined quantum. However it does suggest the presence of a mass 
gap for the gluon which has a parallel in the nonlinear sigma and Gross-Neveu 
models. By contrast neither have confinement of the matter or force fields. 
However as colour confinement in QCD is a low energy phenomenon one assumption 
is that it could have a Lagrangian prescription which is not local. This was 
one property uncovered by Gribov in his seminal work \cite{103} on the 
consequences of not being able to globally fix a gauge uniquely in a nonabelian
gauge theory. Another avenue to pursue might be to adapt structures such as 
(\ref{laggnsplit}) to the Yang-Mills or QCD situation. An example of a related 
approach was provided in Refs. \cite{104} and \cite{105} to tackle the 
confinement of quarks. There in a model where the spin-$1$ fields were regarded
as massive the gluon fields were integrated out of the action to produce an 
effective low energy theory involving only quark fields. These were 
subsequently eliminated in favour of bilocal fermionic objects. An underlying 
assumption in that construction was the nonzero mass for the gauge field in the
Lagrangian. While pre-empting the proof of a mass gap for the gluon such 
effective low energy theories could not be countenanced without the presence of
some sort of mass scale, whatever its origin, which clearly is in contradiction
with the gauge action principle. A not unreasonable way to view this would be 
that such masses emerge dynamically or can be accommodated in an effective 
theory situation akin to those seen in the large $N$ expansions of the simple 
two dimensional theories of (\ref{lagnlsm}) and (\ref{laggn}). 

\subsection{Spin-$1$ theories}

To complete this chain of reasoning in the present context one way of accessing 
mass gaps is via large $N$ expansions. For a spin-$1$ nonabelian field the
analogous theory to develop a large $N$ expansion for is the two dimensional 
nonabelian Thirring model (NATM) with Lagrangian \cite{106,107} 
\begin{equation}
L^{\mbox{\footnotesize{NATM}}} ~=~ 
i \bar{\psi}^i \partialslash \psi^i ~+~ 
\frac{g^2}{2} \left( \bar{\psi}^i \gamma^\mu T^A \psi^i \right)^2 ~.
\label{lagnatmelim}
\end{equation}
where $T^a$ are the generators of the Lie group of the colour symmetry. In this
model and the QCD case the range of the flavour index $i$ will be regarded as 
$1$~$\leq$~$i$~$\leq$~$\Nf$ where $\Nf$ is the number of quarks. This is to 
distinguish it from the number of colours of the gauge group as
$1$~$\leq$~$a$~$\leq$~$\NA$ where $\NA$ is the dimension of the adjoint
representation of the Lie group. To effect a Lagrangian formulation which is
analogous to those of (\ref{lagnlsm}) and (\ref{laggn}) a spin-$1$ auxiliary
field $A^a_\mu$ in the adjoint representation of the Lie group is introduced to
give 
\begin{equation}
L^{\mbox{\footnotesize{NATM}}} ~=~ i \bar{\psi}^i \partialslash \psi^i ~+~ 
A^a_\mu \bar{\psi}^i \gamma^\mu T^a \psi^i ~-~ \frac{1}{2g} A^a_\mu A^{a\,\mu}
\label{lagnatm}
\end{equation}
If one were to continue the parallel with (\ref{lagnlsm}) and (\ref{laggn}) in
two dimensions then the $A^a_\mu$ field would become dynamical in the true
vacuum of the theory. This would be apparent via the effective potential of the
composite field analogous to $\sigma$ which is $A^a_\mu$ in (\ref{lagnatm}).
At this stage this reasoning breaks down as a nonzero vacuum expectation value
for the auxiliary field is not possible as the colour symmetry would be 
broken. Moreover this is for a simple two dimensional model of the process
rather than the four dimensional gauge theory. However we will discuss the 
usefulness of (\ref{lagnatm}) in the large $\Nf$ context for extracting 
information on the renormalization group functions in QCD later motivated by
observations provided in Ref. \cite{108}. In addition that article explored 
the consequences of the proposal that the gluon could be regarded as the bound 
state of two quarks. This is not nonsensical since a confined object may not be
a fundamental entity.  

Despite the apparent possibility of extending the dynamical mass generation
properties of (\ref{lagnlsm}) and (\ref{laggn}), exposed via the large $N$ 
expansion, to models with nonabelian composite fields an alternative tack has
been developed in a series of articles \cite{109,110,111,112}. A formalism, now
termed the local composite operator (LCO) technique, was applied to Yang-Mills 
and QCD in Refs. \cite{112} and \cite{113} respectively. In the LCO approach 
an effective potential {\em can} be constructed for the composite field 
$\sigma$ in the case of (\ref{laggn}) in Refs. \cite{109} and \cite{110} and 
$\half A^a_\mu A^{a\,\mu}$ in the gauge theory case in the Landau gauge. An 
early attempt to construct a perturbative potential for the composite operator 
in the Gross-Neveu model was given in Ref. \cite{87}. Although that was a 
three loop computation using (\ref{laggn}) the resulting effective potential 
did not satisfy a homogeneous renormalization group equation. By contrast the 
LCO construction embeds homogeneity within the development of the effective 
potential. Moreover the effective potential can be derived using standard 
perturbative techniques. In the gauge theory case the expectation value of the 
gauge field composite is nonzero and leads to a new more energetically 
favourable vacuum where the gluon becomes massive. Interestingly the anomalous 
dimension of the composite operator in the nonabelian gauge theory is not 
independent in the Landau gauge and satisfies a simple Slavnov-Taylor identity 
being the sum of the gluon and ghost anomalous dimensions. This was first shown
in Ref. \cite{114} and independently observed in a three loop renormalization 
in Ref. \cite{115}. A more in depth all orders proof was subsequently provided
\cite{116}. Amusingly the mass parameter associated with the nonlocal dimension
zero operator introduced in Gribov's model of colour confinement has the 
{\em same} renormalization property as the dimension two gluon mass operator in
the Landau gauge \cite{117,118}. 

\section{Large $N$ renormalization group functions via summation}

While we have concentrated on the application of the large $N$ expansion to
reveal properties of the true vacuum of (\ref{lagnlsm}) and (\ref{laggn}) the
technique can also be used to extract information on the renormalization group 
functions. Early work in this respect was pioneered in 
Refs. \cite{64} and \cite{65} for the case of QED. Later applications have 
been to supersymmetric theories \cite{119,120}, before enjoying a renaissance 
for gauge-Yukawa theories in order to explore ideas about physics beyond the 
Standard Model \cite{121,122,123,124,125}. One idea is to understand the effect
a large number of fermions with a vector rather than Yukawa interaction has in 
a gauge theory and whether asymptotically safe scenarios can be accommodated 
\cite{121,122,123,124,125}.

\subsection{QED}

The basic idea of Refs. \cite{64} and \cite{65} is to determine the 
contribution to the renormalization constants by first evaluating diagrams 
where chains of massless bubbles in the large $N$ approximation are included in
the Green's function such as that shown in Fig. \ref{cha2cr} before carrying 
out the full renormalization at leading order. In other words where a force 
field propagator such as $\sigma$ ordinarily appears in a Feynman diagram the 
propagator is replaced by one with a series of matter field bubbles. The final 
renormalization procedure followed is exactly the same as in perturbation 
theory where coupling constant and wave function counterterms are included 
where appropriate in the chains of graphs. The determination of the leading 
order part of the renormalization group functions is then carried out in the 
$\MSbar$ scheme.

{\begin{figure}[ht]
\begin{center}
\includegraphics[width=9.5cm,height=1.5cm]{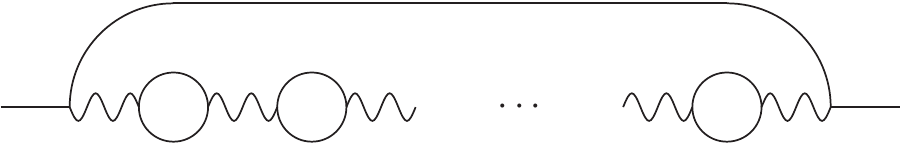}
\end{center}
\caption{Leading order large $N$ bubble chain contributing to the $2$-point 
function.}  
\label{cha2cr}
\end{figure}}

One of the main features of the original approach of 
Refs. \cite{64} and \cite{65} for QED is that these graphs with bubble chains
can be resummed and the overall Feynman diagram evaluated leading to a one 
parameter integral representation of the leading order large $N$ 
renormalization group functions. Here $N$ will be the number of electron 
flavours which we will denote by $\Nf$. Even though there is no colour 
symmetry, leading order QCD data can be deduced in some instances \cite{65}. 
The resummation in effect corresponds to the replacement of the canonical power
of the photon in the chain of Fig. \ref{cha2cr} with the full critical exponent
including the anomalous dimension making it ultimately equivalent to the 
starting point of the critical point large $N$ formalism of 
Refs. \cite{34} and \cite{35} which will be detailed later. However, the 
resummation method has been revisited recently in the context of extending 
gauge theories to include the effect of Gross-Neveu-Yukawa type interactions 
and extensions where the motivation to is to explore the possibility of new 
fixed points in generalizations of the Standard Model. The general area fits 
within exploring the existence and consequence of the asymptotic safety ideas 
introduced by Weinberg \cite{126}. For instance, see 
Refs. \cite{121,122,123,124} and \cite{125} for recent work applying the 
summation method of Refs. \cite{64} and \cite{65} to directly find the large 
$N$ renormalization group functions for such gauge-Yukawa extensions of the 
Standard Model.

{\begin{figure}[ht]
\begin{center}
\includegraphics[width=8.0cm,height=2.0cm]{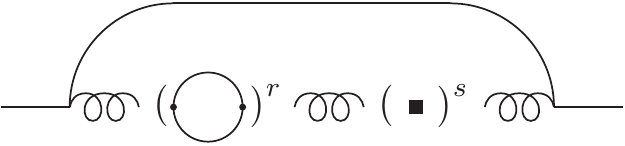}
\end{center}
\caption{Representative graph with bubbles and counterterms contributing to 
electron wave function renormalization.} 
\label{cha3cr}
\end{figure}}

It is best to briefly outline aspects of the procedure of 
Refs. \cite{64} and \cite{65} with an explicit example where the large $N$ 
contribution to a renormalization group function is determined directly by the 
summation method. We consider the contribution to the Landau gauge electron 
wave function renormalization from the sequence of graphs given in Fig. 
\ref{cha2cr}. In practical terms the building block graphs are illustrated in 
Figs. \ref{cha3cr} and \ref{elcr} and are simple to evaluate in QED. For 
instance the graph of Fig. \ref{elcr} gives 
\begin{equation}
\Pi_{\mu\nu}(p) ~=~ \frac{4 (d-2) \Gamma(3-\half d) 
\Gamma^2(\half d - 1)\Nf g^2}{(d-1)(d-4) \Gamma(d-2)} 
\left[ \eta_{\mu\nu} - \frac{p_\mu p_\nu}{p^2} \right] 
\frac{1}{(p^2)^{\half d - 1}}
\label{phocond}
\end{equation}
in general or
\begin{equation}
\Pi_{\mu\nu}(p) ~=~ -~ \frac{4 \Gamma(1+\epsilon) \Gamma^2(2-\epsilon)\Nf g^2}
{(3-2\epsilon)(1-\epsilon) \epsilon \Gamma(2-2\epsilon)} \left[
\eta_{\mu\nu} - \frac{p_\mu p_\nu}{p^2} \right] \frac{1}{(p^2)^{1-\epsilon}}
\label{phocon}
\end{equation}
when $d$~$=$~$4$~$-$~$2\epsilon$. This is included as the amputated photon 
$2$-point bubble in Fig. \ref{cha3cr} prior to completing the integral. The 
evaluation of the graph of Fig. \ref{cha3cr} comprises several parts. The 
momentum independent part of (\ref{phocon}) is factored out of the final 
integration as are the associated counterterms. One aspect of the graph of 
Fig. \ref{cha2cr} worth noting is that we have grouped the bubbles together 
separate from the counterterms. In practice it represents all possible 
different distributions of the bubbles and counterterms from the binomial 
expansion of their sum. As the momentum through each contribution is the same 
within the final Feynman integral then the representation of Fig.
\ref{cha3cr} corresponds to all contributions to the $L$~$=$~$r$~$+$~$s$ loop
order. What remains to complete is the final integration which is of the
form
\begin{equation}
\int_k \frac{\gamma^\mu (\kslash-\pslash) \gamma^\nu}
{(k^2)^{1+r\epsilon}(k-p)^2} \left[ \eta_{\mu\nu} - \frac{k_\mu k_\nu}{k^2}
\right] ~. 
\end{equation}
Consequently there is a complicated contribution to the electron $2$-point
function at $L$ loops given by \cite{64,65}
\begin{equation}
\sum_{i=0}^L \frac{(-1)^i\Gamma(L+1)}{\Gamma(i+1)\Gamma(L+1-i)(6\epsilon)^i}
A^{(L+1-i)\epsilon} B^{L-i} C_{L-i}
\end{equation}
where
\begin{eqnarray}
B &=& \frac{(1-\epsilon)}{2\epsilon(3-2\epsilon)} 
\frac{\Gamma(1+\epsilon)\Gamma^2(1-\epsilon)}{\Gamma(1-2\epsilon)}
\nonumber \\ 
C &=& \frac{(3-2\epsilon)(2-2\epsilon -L(L+1)\epsilon^2)}
{(2-(L+2)\epsilon)(1+L\epsilon)(L+1)\epsilon} \,
\frac{\Gamma(1+L\epsilon+\epsilon)\Gamma(1-L\epsilon-\epsilon)
\Gamma(1-\epsilon)}{\Gamma(1+L\epsilon)\Gamma(2-L\epsilon-2\epsilon)}
\label{bubsum}
\end{eqnarray}
with $A$~$=$~$p^2/(4\pi)$. While this corresponds to the final contribution
of the sequence of graphs to the leading order electron $2$-point functions
the extraction of the part that leads to the renormalization group function
follows the same procedure as in perturbation theory. When the counterterms
are appended one only needs to isolate the simple pole in $\epsilon$. At 
leading order in large $N$ this is possible and the residues of this simple 
pole can be written as a sequence. While the method to achieve this is
involved the solution leads to a one parameter integral representation of the 
electron anomalous dimension. One benefit of this method is its range of 
applicability to several related theories of current interest 
\cite{121,122,123,124,125}. However to go to the next order in $1/N$ by this 
approach would require a significant amount of tedious resummation of graphs 
with not only more than one bubble chain but also $1/N$ corrections to the 
basic bubble of Fig. \ref{elcr}. While this sketch focused on the electron wave
function renormalization it has been extended in QED to the $\beta$-function 
\cite{65} using the same underlying procedure and a one parameter integral 
representation found for the $O(1/N)$ piece. An example of this will be given 
in (\ref{betarep}). The alternative critical point formalism will reproduce
this which we will comment on later. 

{\begin{figure}[ht]
\begin{center}
\includegraphics[width=5.0cm,height=1.6cm]{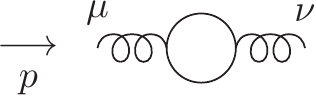}
\end{center}
\caption{Basic one loop bubble correction on photon line denoted by
$\Pi_{\mu\nu}(p)$.}
\label{elcr}
\end{figure}}

\subsection{Renormalon connection}

One related aspect of this summation approach to large $\Nf$ expansions
concerns the presence of renormalons and accessing their effect within a
Green's function. It is widely accepted that the perturbative expansion has 
limitations in its range of applicability. The obvious case is that the 
expansion parameter, which is invariably the coupling constant, is small. 
Ultimately one would want to extend the range of convergence of a perturbative 
series and in an ideal scenario this would mean computing all possible 
contibutions to a Green's function which is clearly unrealistic for the main 
theories describing Nature. An alternative would be to probe beyond the lowest 
orders by summing up classes of graphs contributing to a Green's function 
rather than to gain insight into the coefficients of a renormalization group 
function as the main motivation. In effecting such summations as a function of
the coupling constant the resultant function can have poles for positive
values of the coupling constant. Such poles are an obstruction to the
summation and extracting a value for the Green's function for a value of the
coupling constant beyond the pole closest to the origin will not give reliable
values. Such a pole is usually regarded as a renormalon and indicates that 
there may be a missing contribution to the series \cite{127,128}. In other 
words it indicates the location of a nonperturbative contribution which cannot
be accessed directly via conventional perturbation theory. More recent thinking
\cite{121,122,123,124,125} has suggested that it might be an indication of a 
new type of fixed point applicable to ideas of asymptotic safety and models for 
new physics beyond the Standard Model.  

In practical terms the simplest summation of graphs which can be used to study 
such renormalon poles is the elementary bubble chains of the large $\Nf$ 
expansion such as the graphs in Figs. \ref{cha2cr} and \ref{cha3cr}. While 
the contributions to the bubble chains in a Green's function can be deduced by 
following the resummation process discussed earlier an alternative way is to 
observe that the large $\Nf$ bubbles which are relevant occur in the photon or 
gluon in the case of QCD propagator. Moreover each bubble gives the same 
contribution. Therefore the summation can be efficiently implemented by using 
\begin{eqnarray}
\langle \psi(p) \bar{\psi}(-p) \rangle &=& \frac{i\pslash}{p^2} \nonumber \\
\langle A_\mu(p) A_\nu(-p) \rangle &=& \frac{1}{(p^2)^{1+\delta_r}} 
\left[ \eta_{\mu\nu} ~-~ \frac{p_\mu p_\nu}{p^2} \right]
\end{eqnarray}
rather than the usual propagators where $\delta_r$ would be zero. Here 
$\delta_r$~$=$~$4\Nf T_F g/3$ where $g$ is the coupling constant and $T_F$ is
the Dynkin index of fundamental representation of the QCD colour group. In 
other words $\delta_r$ represents the contribution from each individual bubble 
in the gauge field chain of Fig. \ref{cha2cr}. As an application of this method
we note that the photon propagator in QED has been considered in 
Refs. \cite{129,130} and \cite{131}, for example. Another instance arises in 
deep inelastic scattering where QCD is used to compute the Wilson coefficients 
\cite{132,133}.  For example 
$C^{\mbox{\footnotesize{NS}}}_{\mbox{\footnotesize{long}}}(1,g,L)$ is the 
flavour nonsinglet longitudinal structure function and its coefficients have
been determined at leading order in $1/\Nf$ and expressed compactly in
\cite{131}
\begin{equation}
C^{\mbox{\footnotesize{NS}}}_{\mbox{\footnotesize{long}}}(1,g,L) ~=~
\frac{d^L~}{d\delta_r^L}
\left[ \frac{8 C_F e^{5\delta_r/3}\bar{S}(n,\delta_r)g}
{(2-\delta_r)(1-\delta_r) x^n} \right] ~.
\label{wilcoef}
\end{equation}
Here $L$ corresponds to the number of loops, $x$ is the Bjorken scaling 
variable, $C_F$ is the rank $2$ colour group Casimir and 
$\bar{S}(n,\delta_r)$~$=$~$\Gamma(n+\delta_r)/[\Gamma(n)\Gamma(1+\delta_r)
(n+1+\delta_r)]$. If one effects the $L$th order derivative then the 
coefficients are in exact agreement with the known three loop perturbative 
coefficients \cite{134}. Therefore one can regard (\ref{wilcoef}) as 
representing the resummed set of graphs. What is evident from the expression is
that there are poles when $\delta_r$ is $1$ or $2$ and these correspond to 
renormalons with the first obstruction at unity. 
 
While one can estimate the location of the renormalon from the pole one has to
be aware that in QCD the large $\Nf$ expansion is in effect in the QED sector
of the gauge theory. For the $SU(3)$ colour group when $\Nf$~$\geq$~$17$ then
QCD is no longer asymptotically free. Therefore to circumvent this difficulty
the process of naive nonabelianization was introduced in Ref. \cite{135} and
discussed in Refs. \cite{136} and \cite{137}, for instance. Instead of 
referring to a large $\Nf$ expansion one uses a large $\beta_0$ expansion. Here
$\beta_0$ is the one loop coefficient of the QCD $\beta$-function and given by
\begin{equation}
\beta_0 ~=~ \frac{4}{3} T_F \Nf - \frac{11}{3} C_A 
\end{equation}
where $C_A$ is the rank two colour group Casimir in the adjoint representation.
So when a Green's function has been determined in large $\Nf$ the naive
nonabelianization is implemented by the replacement
\begin{equation}
\Nf ~ \longrightarrow ~ \frac{3}{4T_F} \left[ \beta_0 + \frac{11}{3} C_A
\right]
\end{equation}
and $\beta_0$ is used as the expansion parameter for the Green's function. The
effect of this has been studied in Ref. \cite{135} and an improvement in 
comparison with known two loop estimates has been noted for different 
quantities. As an aside this replacement is not unreminiscent of the use of 
$(N-2)$ rather than $N$ in the large $N$ expansion in the Gross-Neveu and 
nonlinear sigma model in two dimensions noted earlier. In that case, however, 
the $S$-matrix is known exactly and the dependence on $(N-2)$ is evident. In 
higher dimensions there is no guarantee that the $S$-matrix of QCD should be a 
function of $\beta_0$ per se. One final comment on this aspect of large $\Nf$ 
expansions to the properties of Green's functions in general is that the 
examples we have indicated for finding renormalons has been limited to single 
chain cases. While there has been studies of cases where there are two gauge 
propagators corresponding to double chain expansions, the full analysis of such
Green's functions is technically very difficult. This is because one has to 
evaluate the underlying core finite Feynman diagrams where {\em two} of the 
propagators have a power involving $\delta_r$. Moreover since $\delta_r$ 
corresponds to a certain number of inserted bubbles when there are two such 
propagators one has to have two different exponents. This further complicates 
the evaluation of the basic graphs. However progress in this direction in large
$\Nf$ QED has been provided in Ref. \cite{131}. 

\section{Large $N$ critical point formalism}

The large $N$ formalism of Refs. \cite{34} and \cite{35} follows the same 
principles of perturbative quantum field theory in the sense that one has to 
evaluate the core Feynman diagrams contributing to Green's functions. What the 
essential difference though is that in perturbative quantum field theory the 
propagator of a scalar field, for example, has unit power. This is because one 
is perturbing around the Gaussian fixed point of the theory or equivalently in 
the neighbourhood of a free field and one determines the canonical dimensions 
of the fields by ensuring the action is dimensionless in the critical 
dimension. By contrast in the critical point large $N$ formalism \cite{34,35} 
one is close to the critical region of the Wilson-Fisher fixed point 
\cite{37,38,39,40} in $d$-dimensions. At this point the critical theory, being 
away from the critical dimension of the theory, is not free but interacting. 
These two starting points differ as we have discussed earlier in our examples 
of (\ref{lagnlsm}) and (\ref{laggn}). One starting point relates to the theory 
in the conventional perturbative expansion whereas the other corresponds to the
theory in the true stable vacuum. 

\subsection{$O(N)$ scalar theory}

For illustration we focus on the simple case of the $O(N)$ nonlinear $\sigma$ 
model in the formulation of (\ref{lagnlsm}). It has critical dimension $2$ and 
in perturbation theory the free theory involves only the scalar kinetic term
which produces the canonical (massless) propagator $1/p^2$ where $p$ is the 
momentum of the field. This form of the scalar propagator is universal across 
all spacetime dimensions. Equipped with this one computes, for instance, the
renormalization constants of the fields and parameters, such as the mass and
coupling constant, which are the essential quantities for evaluating any 
Green's function. The consequent renormalization group functions can be derived
and these in effect are a measure of the radiative or quantum corrections of 
the underlying theory. For instance, the anomalous dimension of $\phi^i$, 
denoted by $\gamma_\phi(g)$, quantifies the shift of the dimension of the field 
from its canonical or classical dimension. In other words the fields and the
operators built from those fields do not have the classical dimensions when in 
the interacting quantum theory. 

By contrast the large $N$ critical point approach begins from a nonfree 
Wilson-Fisher fixed point, $g_c$, defined by the nontrivial zero of the 
$\beta$-function
\begin{equation}
\beta(g_c) ~=~ 0
\label{betazero}
\end{equation}
and hence the corresponding canonical dimensions of the fields across $d$
dimensions are noninteger. To deduce their canonical values one performs a 
simple dimensional analysis of the Lagrangian in $d$ dimensions with the 
premise that the action remains dimensionless for all $d$. Therefore from 
(\ref{lagnlsm}) the dimension of $\phi^i$ is $(\mu-1)$ since the measure in the
definition of the action introduces the dimension dependence where we will use 
the shorthand that
\begin{equation}
d ~=~ 2 \mu 
\end{equation}
throughout. In addition the interaction of (\ref{lagnlsm}) is noninactive in 
this analysis since we are at the Wilson-Fisher fixed point. It therefore 
defines the canonical dimension of $\sigma$ as $2$ in {\em all} dimensions
since the coupling constant has been scaled into what we call the spectator
term. Therefore the matter field and the core interaction determine the scaling
dimensions of the fields in the universal theory from which one defines the 
propagators used to evaluate Green's functions in the large $N$ formalism 
\cite{34,35}. In addition one can exploit the fact that at criticality the 
fields are massless. Therefore the radiative corrections to a propagator as it 
is iterated order by order within the Green's function can be resummed. The 
consequence of this is that those radiative corrections exponentiate at 
criticality to add an additional term in the exponent of the propagator. It is 
quantified explicitly as the renormalization group anomalous dimension of the 
corresponding field but evaluated at the value of the critical coupling 
constant. In practice this is ordinarily a small number. Within the large $N$ 
construction, however, the corresponding anomalous dimension exponent will be a
function of $d$ \cite{34,35}. Only when evaluated at an integer dimension can 
it be compared to computations by other techniques. The upshot is that the 
propagators of the critical $O(N)$ nonlinear $\sigma$ model are not the usual 
ones associated with perturbation theory but take the leading order form 
\begin{equation}
\langle \phi^i(x) \phi^j(y) \rangle ~\sim~
\frac{A_\phi \delta^{ij}}{((x-y)^2)^{\alpha_\phi}} ~~,~~
\langle \sigma(x) \sigma(y) \rangle ~\sim~
\frac{B_\sigma}{((x-y)^2)^{\beta_\sigma}} 
\label{scalprop}
\end{equation}
in the asymptotic scaling region of the Wilson-Fisher fixed point. It is these
forms which are used in the large $N$ critical point formalism \cite{34,35}. 
They have been formulated in coordinate or $x$-space as the dimensional 
analysis was for the $x$-space version of the Lagrangian. If the momentum space
of (\ref{lagnlsm}) had been used then the propagators would have been
\begin{equation}
\langle \phi^i(p) \phi^j(-p) \rangle ~\sim~
\frac{\tilde{A}_\phi \delta^{ij}}{(p^2)^{\mu-\alpha_\phi}} ~~,~~ 
\langle \sigma(p) \sigma(-p) \rangle ~\sim~
\frac{\tilde{B}_\sigma}{(p^2)^{\mu-\beta_\sigma}} ~. 
\label{momprops}
\end{equation}
These sets of propagators are connected via the Fourier transform which in the
conventions of Refs. \cite{34} and \cite{35} is 
\begin{equation}
\frac{1}{(x^2)^\alpha} ~=~ \frac{a(\alpha)}{2^{2\alpha}} \int_k
\frac{e^{ikx}}{(k^2)^{\mu-\alpha}}
\label{foutra}
\end{equation}
where we define
\begin{equation}
a(\alpha) ~=~ \frac{\Gamma(\mu-\alpha)}{\Gamma(\alpha)}
\end{equation}
as the shorthand for the combination of Euler $\Gamma$-functions which appear
\cite{34,35} and $\int_k$ $=$ $\int d^d k/(2\pi)^d$. In both sets of 
propagators an arbitrary $x$-independent amplitude is present such as $A_\phi$
which can be determined within the large $N$ expansion with the momentum space 
amplitudes related via the Fourier transform. We use the convention that matter
field amplitudes will be denoted by $A_f$ and the corresponding force field 
will be denoted by $B_f$ where $f$ is a generic field. Finally, the exponents 
are defined by
\begin{equation}
\alpha_\phi ~=~ \mu ~-~ 1 ~+~ \half \eta ~~~,~~~
\beta_\sigma ~=~ 2 ~-~ \eta ~-~ \chi_\sigma ~.
\label{expdef}
\end{equation}
In the latter relation the exponent $\chi_\sigma$ represents the anomalous 
dimension of the vertex operator\footnote{In the original work of 
Refs. \cite{34} and \cite{35} the letter $\kappa$ was used for the vertex 
operator anomalous dimension.} $\sigma \phi^i \phi^i$. In general each operator
has an anomalous dimension which corresponds to an independent renormalization 
constant in perturbation theory. Therefore one has to allow for the analogous 
exponent in the critical point large $N$ formalism.

Having defined the asymptotic scaling form of the scalar propagators in both
coordinate and momentum space one can readily deduce analogous expressions for
the related $2$-point functions. These are necessary as one method to deduce
the wave function critical exponent for the matter field $\phi^i$ is to solve
the skeleton Schwinger-Dyson equations for the $2$-point functions of the
theory \cite{34,35}. In momentum space these are given by merely inverting the
scaling forms of (\ref{momprops}) to give 
\begin{equation}
\langle \phi^i(p) \phi^j(-p) \rangle^{-1} ~\sim~
\frac{\delta^{ij}}{(p^2)^{\alpha_\phi-\mu} \tilde{A}_\phi} ~~,~~
\langle \sigma(p) \sigma(-p) \rangle^{-1} ~\sim~
\frac{1}{(p^2)^{\beta_\sigma-\mu}\tilde{B}_\sigma} ~. 
\label{invmomprops}
\end{equation}
The coordinate space counterparts are deduced by applying the inverse Fourier
transform to give 
\begin{equation}
\langle \phi^i(x) \phi^j(0) \rangle^{-1} ~\sim~
\frac{\delta^{ij} p(\alpha_\phi)}{(x^2)^{2\mu-\alpha_\phi} A_\phi} ~~,~~ 
\langle \sigma(x) \sigma(0) \rangle^{-1} ~\sim~
\frac{p(\beta_\sigma)}{(x^2)^{2\mu-\beta_\sigma} B_\sigma} 
\end{equation}
where 
\begin{equation}
p(\alpha) ~=~ \frac{a(\alpha-\mu)}{a(\alpha)} 
\end{equation}
and we have mapped one coordinate to the origin without loss of generality.

While we have introduced the leading order structure of the propagators in the
asymptotic limit to the fixed point in (\ref{scalprop}) corrections to these 
can be included. They have the form 
\begin{eqnarray}
\langle \phi^i(x) \phi^j(y) \rangle &\sim&
\frac{A_\phi \delta^{ij}}{((x-y)^2)^{\alpha_\phi}} 
\left[ 1 ~+~ ((x-y)^2)^\lambda A^\prime_\phi \right] \nonumber \\
\langle \sigma(x) \sigma(y) \rangle &\sim&
\frac{B_\sigma}{((x-y)^2)^{\beta_\sigma}} 
\left[ 1 ~+~ ((x-y)^2)^\lambda B^\prime_\sigma \right] 
\label{inscalcor}
\end{eqnarray}
where $A_\phi^\prime$ and $B_\sigma^\prime$ are the associated coordinate
independent amplitudes. The scaling forms of the $2$-point functions are 
derived in the same way as before producing \cite{34,35} 
\begin{eqnarray}
\langle \phi^i(x) \phi^j(y) \rangle^{-1} &\sim&
\frac{ \delta^{ij}}{A_\phi((x-y)^2)^{2\mu-\alpha_\phi}} 
\left[ 1 ~-~ q(\alpha_\phi) ((x-y)^2)^\lambda A^\prime_\phi \right] 
\nonumber \\
\langle \sigma(x) \sigma(y) \rangle^{-1} &\sim&
\frac{1}{B_\sigma((x-y)^2)^{2\mu-\beta_\sigma}} 
\left[ 1 ~-~ q(\beta_\sigma) ((x-y)^2)^\lambda B^\prime_\sigma \right] 
\end{eqnarray}
where 
\begin{equation}
q(\alpha) ~=~ \frac{a(\alpha-\mu+\lambda)a(\alpha-\lambda)}
{a(\alpha-\mu)a(\alpha)} ~.
\end{equation}
The exponent $\lambda$ is a generic one for the correction term. For instance, 
it could be regarded as the exponent corresponding to the critical slope of the 
$\beta$-function. The evaluation of the $\beta$-function at the fixed point 
clearly cannot correspond to a nontrivial exponent due to (\ref{betazero}). 
However, it could correspond to another exponent such as that relating to the 
specific heat for instance. For scalar theories, for instance, the correction 
term is in effect equivalent to zero momentum insertions on a propagator line 
and therefore such terms identify $3$-point functions. The new amplitudes tag 
the parts of the Feynman graphs which contribute to this. In perturbative field 
theory computations a similar technique is used to extract renormalization 
constants for mass operators where the zero momentum insertion does not give 
rise to infrared divergences. An example of such an application is the three 
loop renormalization of the quark mass operator in the $\MSbar$ scheme in 
Ref. \cite{138}. So the correction terms of (\ref{inscalcor}) should be 
regarded as insertions of $2$-leg operators with possibly derivatives included. 

For instance the $\beta$-function of the nonlinear $\sigma$ model and $\phi^4$ 
theory have both been determined at $O(1/N^2)$ in 
Refs. \cite{34,35} and \cite{139} by this method using the universal theory 
with the same core interaction which is that given in (\ref{lagnlsm}). In the 
former theory the canonical term of the correction exponent $\lambda$ was 
$(\mu-1)$. For the latter it was $(\mu-2)$ as the coupling constant of four 
dimensional $\phi^4$ theory is dimensionless when $\mu$~$=$~$2$. To understand 
the commonality of both theories it is best to recall the $O(N)$ $\phi^4$ 
theory Lagrangian which is the four dimensional partner of the two dimensional 
nonlinear sigma model, (\ref{lagnlsm}). Taking the usual four dimensional 
$O(N)$ $\phi^4$ Lagrangian which is
\begin{equation}
L^{\phi^4} ~=~ \frac{1}{2} \partial_\mu \phi^i \partial^\mu \phi^i ~-~
\frac{1}{8} g^2 \left( \phi^i \phi^i \right)^2
\label{lagphi4}
\end{equation}
then analogous to (\ref{laggnelim}) this can be rewritten as 
\begin{equation}
L^{\phi^4} ~=~ \frac{1}{2} \partial_\mu \phi^i \partial^\mu \phi^i ~+~
\frac{1}{2} \sigma \phi^i \phi^i ~+~ \frac{\sigma^2}{2g^2} 
\label{lagphi4elim}
\end{equation}
which clearly shares the same interaction as (\ref{lagnlsm}). With both 
(\ref{lagnlsm}) and (\ref{lagphi4elim}) the noncommon $\sigma$ dependent 
pieces define the canonical dimensions of the respective coupling constants. In
the former case the dimension is $(\mu-1)$ but $(\mu-2)$ in the latter. It is 
these exponents which are separately used in the correction to scaling parts of
the asymptotic scaling functions of (\ref{inscalcor}) when they are inserted in
the skeleton Schwinger-Dyson equations which will be discussed in detail later. 
The connection of these canonical parts of the exponents with the respective 
$\beta$-functions is that they correspond to the critical slopes of the 
$\beta$-functions. In a dimensionally regularized perturbative computation
the coupling constant has a nonzero dimension in $d$-dimensions. This is 
reflected in the $d$-dependence of the leading order term of the 
$\beta$-function. In mentioning how the exponents of the $\beta$-functions are
computed in relation to the structure of the universal Lagrangian we note that 
these are examples of hyperscaling laws. Another example in (\ref{lagnlsm}) can
be deduced for the dimension of the $\phi^i$ mass dimension. The mass operator 
is $\half \phi^i \phi^i$ which also couples to the $\sigma$ field. Therefore 
the mass anomalous dimension for $\phi^i$ corresponds to the anomalous 
dimension of the $\sigma$ field. By contrast the mass of the $\sigma$ field is 
deduced from its mass operator which is $\half \sigma^2$. In other words when 
the critical dimension is six the anomalous dimension of the mass of $\sigma$, 
which will then be a propagating field, is related to the $\beta$-function of 
$O(N)$ $\phi^4$ theory.  

\subsection{Early connections}

In discussing the critical point large $N$ formalism to this we can now
connect it with other approaches using $1/N$ expansions. The method of 
Refs. \cite{64} and \cite{65} which explicitly summed the chains of bubbles 
in the massless theory to extract the renormalization group functions at 
leading order overlaps with the construction of 
Refs. \cite{34} and \cite{35}. For instance in 
Refs. \cite{64} and \cite{65} the bubble sum produced a propagator with a 
$d$-dependent exponent. Equally in the massive case discussed in the context of
the Gross-Neveu model the auxiliary field becomes dynamical and the 
$d$-dimensional propagator has a $d$-dependent scaling form when the generated 
mass is set to zero as indicated in (\ref{gnprophyp}). Both these consequences 
are accommodated naturally in the asymptotic scaling forms of (\ref{scalprop}).
For instance the $\sigma$ propagators (\ref{gnprophyp}) and (\ref{momprops}) 
are equivalent at leading order. Likewise the overall summed propagator of Fig. 
\ref{cha2cr} leads to an exponent equivalent to that of the $\phi^i$ field of 
(\ref{scalprop}). Moreover the technique of Refs. \cite{64} and \cite{65} 
produces the leading order large $N$ renormalization constants and consequent 
renormalization group functions. These connections are related since they are 
the same quantity but derived in different ways. The advantage of 
(\ref{scalprop}) is that the scaling forms transcend the critical dimensions of
the specific theories and are the propagators in the universal critical theory 
bridging all theories in the same universality class. Moreover while the 
leading order behaviour of the exponent is deduced on dimensional grounds the 
anomalous terms such as $\eta$ and $\chi_\sigma$ in (\ref{expdef}), for 
instance, correspond to quantum corrections. In effect they quantify the 
contributions from graphs of the form of Fig. \ref{cha2cr}. Given this when one
uses (\ref{scalprop}) in critical point large $N$ calculations there is no
dressing of the propagators with bubble chains. These contributions are already
summed in the exponent. Consequently one has to compute massless Feynman graphs
with nonunit powers unlike conventional perturbation theory. Massless field 
theory techniques have been developed for this and we discuss some of these in 
Appendix A with several examples in Appendix B.  

{\begin{figure}[ht]
\begin{center}
\includegraphics[width=9.5cm,height=3.4cm]{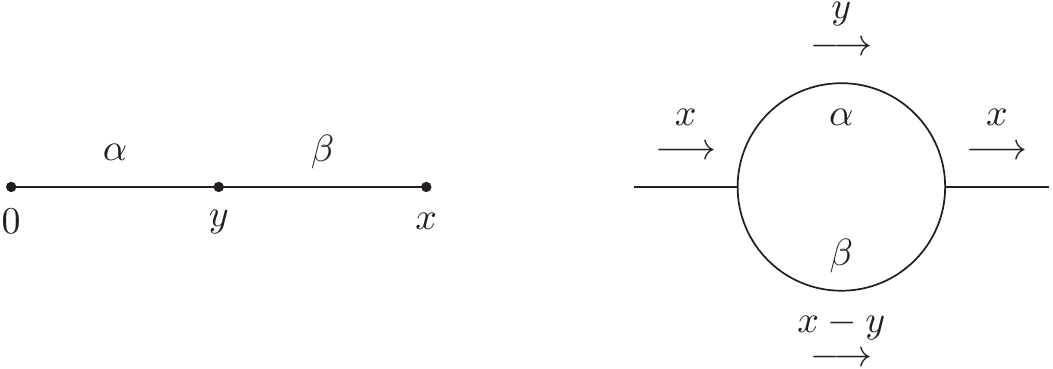}
\end{center}
\caption{Coordinate and momentum space representations of a one loop chain or
bubble integral.}
\label{chaintcr}
\end{figure}}

At this stage it is worthwhile introducing some basic graphical notation 
concerning the evaluation of Feynman graphs for massless theories with nonunit 
propagator powers. We do this in the context of the simple Feynman integrals 
given in Fig. \ref{chaintcr} where $\alpha$ and $\beta$ are arbitrary complex 
parameters here and each graph corresponds to the same integral which is 
\cite{34,35}
\begin{equation}
\int_y \frac{1}{(y^2)^\alpha ((x-y)^2)^\beta} ~=~ 
\nu(\alpha,\beta,2\mu-\alpha-\beta) 
\end{equation}
where 
\begin{equation}
\nu(\alpha,\beta,\gamma) ~=~ \pi^\mu a(\alpha) a(\beta) a(\gamma) ~.
\end{equation}
This integral corresponds to both graphs of Fig. \ref{chaintcr} since they 
are different graphical representations. In the left hand graph the vectors 
comprising the two propagators are placed between points on the plane with 
reference to some origin. The two endpoints $0$ and $x$ are regarded as fixed
and the variable $y$ to be integrated over can be anywhere else on the plane
although we have chosen to place it at the centre of the connecting line. The
more conventional way of graphically representing the integral is given in the
right hand graph of Fig. \ref{chaintcr} where $y$ is the loop momentum and
$x$ the external momentum. It is worth recognising the flexibility which both
representations present. For instance, an integral in momentum space can be
represented in a coordinate space way where the $y$ and $x$ are replaced by
the more conventionally used letters. Irrespective of which is adapted both
will reflect underlying symmetry properties much more clearly than the explicit
integral itself especially in higher loop graphs. Equally some integration 
rules, which we discuss later, are easier to apply in one representation over 
another.  

\subsection{Skeleton Schwinger-Dyson equation}

Equipped with these tools we can now illustrate the core strength of the large
$N$ formalism of Refs. \cite{34} and \cite{35}. For this we use 
(\ref{lagnlsm}) as the basic theory and note that the $2$-point functions of 
the fields are shown in Fig. \ref{sigcr} to $O(1/N^2)$. In the large $N$ 
counting a $\sigma$ field is regarded as one power of $1/N$ and each closed 
$\phi^i$ loop contributes one power of $N$. Therefore the final two graphs of 
each equation in Fig. \ref{sigcr} are the same order in $1/N$. By contrast they
would be different orders in the coupling constant expansion. We have chosen to 
use (\ref{lagnlsm}) rather than (\ref{laggn}) as in the latter case the final 
two graphs of each equation in Fig. \ref{sigcr} would be zero due to the trace 
over an odd number of $\gamma$-matrices and this large $N$ graph ordering would
be less apparent. In either case we have not included any dressings on the 
propagators since the presence of the anomalous dimensions in the critical 
propagators (\ref{scalprop}) and (\ref{momprops}) already represent those 
summations. Using the coordinate space propagators the equations of Fig. 
\ref{sigcr} are formally represented by  
\begin{eqnarray}
0 &=& r(\alpha) ~+~ z Z_\sigma^2 (x^2)^{\chi_\sigma+\Delta} ~+~ 
z^2 \Sigma_1 (x^2)^{2\chi_\sigma+2\Delta} ~+~ 
z^3 N \Sigma_2 (x^2)^{3\chi_\sigma+3\Delta} ~+~ O \left( \frac{1}{N^3} \right) 
\nonumber \\
0 &=& p(\beta_\sigma) ~+~ 
\frac{N}{2} z Z_\sigma^2 (x^2)^{\chi_\sigma+\Delta} ~+~
\frac{N}{2} z^2 \Pi_1 (x^2)^{2\chi_\sigma+2\Delta} ~+~ 
\frac{N^2}{2} z^3 \Pi_2 (x^2)^{3\chi_\sigma+3\Delta} \nonumber \\
&& +~ O \left( \frac{1}{N^2} \right) 
\label{nlsmsde}
\end{eqnarray}
to the first two orders in large $N$ where we have defined the combination of
amplitudes to be $z$~$=$~$A_\phi^2 B_\sigma$ and the external coordinates of
all graphs are $0$ and $x$. The quantities $\Sigma_i$ and $\Pi_i$ are the 
$x$-independent values of the respective graphs of Fig.  \ref{sigcr}. In 
practice their values are divergent not unlike their perturbative counterparts 
and their determination requires a regularization. As the formalism 
\cite{34,35} operates in the universal field theory in $d$-dimensions, where 
$d$ is arbitrary, dimensional regularization cannot be used for this task. 
Instead graphs are regularized by the dimensionless quantity $\Delta$, which is 
regarded as small, and is introduced by shifting the vertex anomalous dimension
via 
\begin{equation}
\chi_f ~ \rightarrow \chi_f ~+~ \Delta
\label{chishift}
\end{equation}
which is the origin of the extra term in each of the powers of the coordinate
$x$ in (\ref{nlsmsde}). 

{\begin{figure}[ht]
\begin{center}
\includegraphics[width=12.5cm,height=4.0cm]{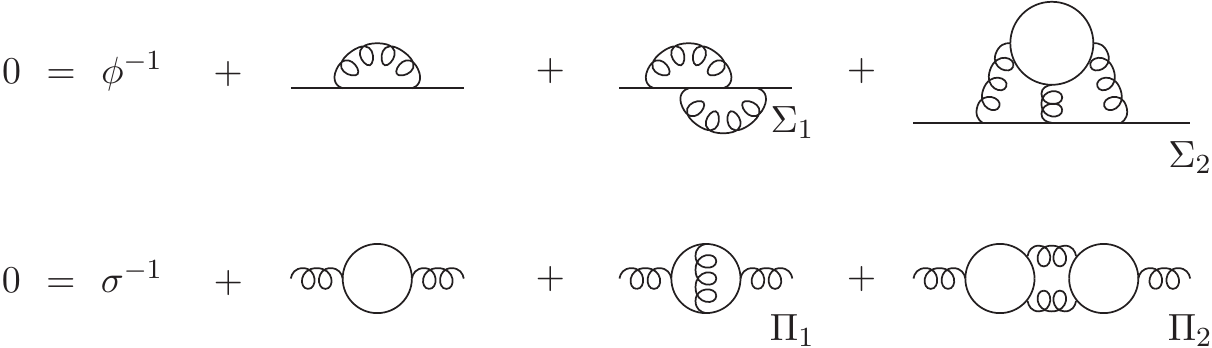}
\end{center}
\caption{Leading order graphs for the $2$-point skeleton Schwinger-Dyson
equations.}
\label{sigcr}
\end{figure}}

In dimensionally regularizing a field theory for perturbative calculations the 
regularizing parameter is introduced by modifying the spacetime dimension
which changes the dimensionality of the coupling constant. This is the 
parameter of the perturbative approximation. Here the situation is similar in 
that one is in effect performing a perturbative expansion in the vertex 
anomalous dimension which is why that parameter is formally modified. Thus we 
can define the general divergence structure of each graph by  
\begin{equation}
\Sigma_i ~=~ \frac{K_i}{\Delta} ~+~ \Sigma_i^\prime ~~~,~~~
\Pi_i ~=~ \frac{K_i}{\Delta} ~+~ \Pi_i^\prime ~.
\end{equation}
There will be $O(\Delta)$ terms but these are not relevant for an $O(1/N^2)$
computation. The equality of the poles is not an error but a reflection of the 
result of the explicit evaluation. More deeply it demonstrates the underlying 
large $N$ renormalizability of the theory which has been discussed in depth in 
the critical point context in Refs. \cite{140,141} and \cite{142}. Similar 
renormalizability arguments apply equally to (\ref{laggn}) and other theories 
which are accessible via the large $N$ critical point formalism. Indeed these 
are not unrelated to the early discussion of the large $N$ renormalizability of
the Gross-Neveu model in dimensions greater than $2$ observed in 
Refs. \cite{63} and \cite{143}. As the equations of Fig. \ref{sigcr} have 
divergences these are removed by the counterterms available from the vertex 
renormalization constant $Z_\sigma$ present in (\ref{nlsmsde}). We have only 
included it in the leading order terms since it plays no role at $O(1/N^2)$. It
would only be required there if one was extending the equations to the next 
order in $1/N$. The structure of $Z_\sigma$ is formally similar to that of 
renormalization constants in conventional perturbative calculations and takes 
the form \cite{141}
\begin{equation}
Z_\sigma ~=~ 1 ~+~ \sum_{l=1}^\infty \sum_{n=1}^l \frac{m_{ln}}{\Delta^n} 
\end{equation}
where $m_{ln}$ are the counterterms and depend on $N$. They can then be  
expanded in large $N$ via \cite{141}
\begin{equation}
m_{ln} ~=~ \sum_{i=1}^\infty \frac{m_{ln,i}}{N^i} ~.
\end{equation}
Aside from the fact that the Schwinger-Dyson equations are divergent and 
require renormalization there is an additional aspect to their structure. As
the ultimate aim is to study them in the critical region in the approach to the
fixed point where there is scaling, we will have to take the $x^2$~$\to$~$0$
limit. While it looks as if this is not a problem from the structure of 
(\ref{nlsmsde}) the divergent graphs mean that there are $\ln(x^2)$ terms
when the exponent is expanded for small $\Delta$. For (\ref{lagnlsm}) these
are excluded by defining the vertex anomalous dimension by \cite{34,35} 
\begin{equation}
\chi_{\sigma,1} ~=~ -~ z_1 K_1 ~-~ 2 z_1^2 K_2 ~. 
\label{chidefnlsm}
\end{equation}
We use the notation that exponents and amplitudes such as $z$ have the $1/N$
expansions
\begin{equation}
\chi_\sigma ~=~ \sum_{i=1}^\infty \frac{\chi_{\sigma,i}}{N^i} ~~~,~~~
z ~=~ \sum_{i=1}^\infty \frac{z_i}{N^i} ~.
\end{equation}
Since (\ref{chidefnlsm}) involves the simple poles of the $O(1/N^2)$ graphs if
the residue was different for the graphs in the separate equations conflicting
values would have emerged from both equations. That this does not happen is
again reflective of the underlying renormalizability. Consequently finite 
equations emerge to represent Fig. \ref{sigcr} and the $x^2$~$\to$~$0$ limit 
can be smoothly taken after the regularization is lifted via the 
$\Delta$~$\to$~$0$ limit. This finally produces \cite{34,35} 
\begin{eqnarray}
0 &=& r(\alpha) ~+~ z ~+~ z^2 \Sigma_1^\prime ~+~ z^3 N \Sigma_2^\prime ~+~ 
O \left( \frac{1}{N^3} \right) \nonumber \\
0 &=& p(\beta_\sigma) ~+~ \frac{N}{2} z ~+~ \frac{N}{2} z^2 \Pi_1^\prime ~+~ 
\frac{N^2}{2} z^3 \Pi_2^\prime ~+~ O \left( \frac{1}{N^2} \right) 
\label{nlsmsdefin}
\end{eqnarray}
which are algebraic equations dependent on the amplitude variable $z$ and 
embedded in the scaling functions the other unknown at this stage which is 
$\eta$. To complete the evaluation requires the explicit values of the four 
graphs of Fig. \ref{sigcr}. Various calculational techniques used to find these
for (\ref{lagnlsm}) and other theories have been summarized in Appendix A. 
Eliminating $z$ in (\ref{nlsmsdefin}) order by order in powers of $1/N$ 
produces a $d$-dimensional expression for the $\phi^i$ field exponent dimension
at $O(1/N^2)$ and given in Refs. \cite{34} and \cite{35}. For instance at 
leading order
\begin{equation}
\eta_1^{\mbox{\footnotesize{NLSM}}} ~=~ -~ 
\frac{2(\mu-1)(\mu-2)\Gamma(2\mu-1)}{\mu\Gamma^3(\mu)\Gamma(2-\mu)} ~.
\end{equation}
One of the reasons for including this is that this leading order combination of
$\Gamma$-functions is common for all $d$-dimensional models accessible via the 
large $N$ method of Refs. \cite{34} and \cite{35}. It is therefore related to
the structure found in the large $N$ explicit bubble summation \cite{64,65}. 
For instance in (\ref{bubsum}) the combination of $\Gamma$-functions in the 
product of $B$ and $C$ with $L$-independent arguments are related to 
$\Gamma(2-\mu) \Gamma^3(\mu-1)/\Gamma(2\mu-1)$ when $\epsilon$ is mapped back 
to the dimension $d$. In extracting this we have used the identity 
$\Gamma(z+1)$~$=$~$z\Gamma(z)$ to highlight the commonality of the structure. 
In addition the $d$-dimensional $O(1/N^2)$ exponents agree with the earlier 
evaluations of Refs. \cite{41,42,43,44,45,46,47} and \cite{48} in strictly 
three dimensions. From the basic exponents $\eta$ and $\nu$ then the full suite
of exponents can be deduced from hyperscaling relations \cite{46}, for 
instance.

\subsection{Large $N$ renormalization}

We have discussed the derivation of the underlying critical point 
representation of the $2$-point functions at length as the same procedure 
applies to the higher point functions. Indeed it has been established 
\cite{141} that one can perform a perturbative expansion in the critical point 
large $N$ expansion formalism \cite{34,35} which is completely analogous to 
conventional perturbation theory in the coupling constant. For instance, if one
computes the scaling dimension of the vertex Green's function of 
(\ref{lagnlsm}) or (\ref{laggn}) using the large $N$ formalism then it 
corresponds to the vertex critical exponent $\chi_\sigma$. While we indicated 
that the leading order value could be deduced from the $O(1/N^2)$ $2$-point 
function evaluating the vertex function at $O(1/N)$ produces the same result. 
This is completely parallel to conventional coupling constant perturbation 
theory. One can deduce the one loop coupling constant counterterm in the two 
loop $2$-point function computation by noting that there cannot be any 
$\ln(p^2)$ terms with a divergence in $\epsilon$ in a renormalizable theory, 
where $p$ is the momentum flowing through the $2$-point function. If one 
examines such terms closely at say two loops then it is apparent that it 
involves the one loop coupling constant counterterm. Renormalizability ensures 
that no such $\ln(p^2)$ terms remain when the counterterms are fixed from the 
vertex renormalization at the lower loop order. If such terms remained they 
would correspond to nonlocal contributions in the action. The main difference 
between large $N$ and conventional perturbation theory is that the latter is 
carried out in the underlying $d$-dimensional universal theory defining the 
Wilson-Fisher fixed point using a parameter, $1/N$, which is dimensionless 
across all dimensions. The formalism not only applies to $n$-point Green's 
functions but also to the renormalization of composite operators. Several early
papers outline the procedure to follow in the context of large $N$ accessible 
scalar field theories \cite{141,142,144,145}. 

One novel aspect of operator renormalization in the large $N$ expansion occurs 
when there is operator mixing and the comparison to the operator anomalous 
dimensions computed in coupling constant perturbation theory. This can be 
illustrated simply in the Gross-Neveu model. In (\ref{laggn}) the fields 
$\psi^i$ and $\sigma$ have the respective canonical dimensions $\half$ and $1$ 
in strictly two dimensions. Therefore the operators $(\bar{\psi}^i \psi^i)^2$,
$\sigma \bar{\psi}^i \psi^i$ and $\sigma^2$ have the same canonical dimensions 
and will mix under renormalization in perturbation theory. To accommodate this 
one has to determine a mixing matrix of renormalization constants, $Z_{pq}$, 
where
\begin{equation}
{\cal O}_{p\,\mbox{\footnotesize{o}}} ~=~ Z_{pq} {\cal O}_q
\label{opmix}
\end{equation}
and ${}_{\mbox{\footnotesize{o}}}$ indicates the operator ${\cal O}_p$ is bare.
From (\ref{opmix}) one can compute the mixing matrix of anomalous dimensions
$\gamma_{pq}(g)$. In perturbation theory this is usually sufficient for the 
problem of immediate interest. However for a system of $n$ operators, for
instance, what is not ordinarily considered is that it defines $n$ key 
operators which are the eigenoperators formed from the original operators and 
have eigen-anomalous dimensions. In other words in the eigen-basis one can 
transform to a diagonal mixing matrix. While there appears to be no use for the
eigen-operators, the Gross-Neveu example we have mentioned illustrates the 
connection with the large $N$ critical point formalism. In $d$-dimensions at 
the Wilson-Fisher fixed point $\psi^i$ and $\sigma$ have {\em different} 
canonical dimensions which are $\half(d-1)$ and $1$ respectively. Therefore the
three operators have different canonical dimensions and cannot mix under 
renormalization in the large $N$ critical point formalism. This appears to 
contradict the perturbative scenario already outlined. However this is not the 
case. In $d$-dimensions the operators correspond to the {\em eigen}-operators 
of the perturbative mixing matrix. Indeed this is evident in examples where 
large $N$ critical exponents are computed explicitly using the formalism of 
Refs. \cite{34} and \cite{35} in $d$-dimensions and compared with the 
perturbative eigen-anomalous dimensions at the Wilson-Fisher fixed point in the
sense that they are in total agreement. Another example is discussed in the 
next section. Finally we note that it is only in the critical dimension of the
underlying quantum field theory that the canonical dimensions of the operators 
are the same.

\section{Gauge-Yukawa theories}

While our discussion has been in respect of scalar spin-$0$ force fields the 
same approach can be established for higher spin force fields. For instance, 
theories with an $O(N)$ or $SU(N)$ symmetry with fermions as matter fields and 
spin-$1$ gauge fields can be treated at the corresponding large $N$ accessible 
Wilson-Fisher fixed point. 

\subsection{Formalism}

The critical point propagators for such cases are consistent with the structure
apparent in perturbation theory aside from the adjustment of the exponent from
unity. For instance, the coordinate space propagators for fermions and a gauge
field are \cite{146,147}
\begin{eqnarray}
\langle \psi^i(x) \bar{\psi}^j(y) \rangle &\sim&
\frac{(\xslash-\yslash) A_\psi \delta^{ij}}{((x-y)^2)^{\alpha_\psi}} 
\nonumber \\
\langle A_\mu(x) A_\nu(y) \rangle &\sim& 
\frac{B_\gamma}{((x-y)^2)^{\beta_\gamma}} \left[ \eta_{\mu\nu} + 
\frac{2( 1 - b )\beta_\gamma}{(2\mu-2\beta_\gamma-1+b)}
\frac{(x-y)_\mu (x-y)_\nu}{(x-y)^2} \right]
\label{ferphoprop}
\end{eqnarray}
where the matter field carries indices reflecting the $O(N)$ symmetry for 
instance. In a supersymmetric theory the spin-$\half$ field could correspond to
a gluino in which it would be associated with the force supermultiplet and
hence have no $O(N)$ indices. For the gauge field we have not included indices
such as those associated with the colour group in QCD for instance as that 
group is not central to the large $N$ expansion. In that case the expansion 
corresponds to the number of quark flavours, $\Nf$, rather than the number of 
colours, $\Nc$. So the symmetry group associated with the expansion would be 
$O(\Nf)$ or $SU(\Nf)$.  

The structure of the critical point gauge field propagator is not the familiar 
one associated with the perturbative case. This is partly because there is a 
nonunit propagator power which means nontrivial exponent dependence is present
after the Fourier transform of the critical momentum space propagator
\begin{equation}
\langle A_\mu(p) A_\nu(-p) \rangle ~\sim~ 
\frac{\tilde{B}_\gamma}{(p^2)^{\mu-\beta_\gamma}} 
\left[ \eta_{\mu\nu} - ( 1 - b ) \frac{p_\mu p_\nu}{p^2} \right] 
\label{gauprop}
\end{equation}
which is the reference point for deriving the $x$-space gauge field scaling
propagator. Throughout we denote the gauge parameter by $b$ to avoid confusion
with $\alpha$ which is regarded as the matter field dimension here. For the 
fermion the momentum space propagator is
\begin{equation}
\langle \psi(p) \bar{\psi}(-p) \rangle ~\sim~
\frac{\tilde{A}_\psi\pslash}{(p^2)^{\mu-\alpha_\psi+1}} 
\end{equation}
where $\tilde{A}_\psi$ and $\tilde{B}_\gamma$ are related to the corresponding
coordinate space amplitudes via the Fourier transform (\ref{foutra}) and its
derivatives with respect to $x$. Here the exponents are defined by 
\begin{equation}
\alpha_\psi ~=~ \mu ~+~ \half \eta ~~~,~~~
\beta_\gamma ~=~ 1 ~-~ \eta ~-~ \chi_\gamma 
\label{expqed}
\end{equation}
and we note that throughout our discussion will be with reference to a gauge 
theory with fermion interactions rather than scalar interactions. While models 
such as $CP(N)$ sigma models and scalar QCD are of interest the large $N$ 
formalism discussed here can be readily adapted to these cases. The asymptotic 
scaling form of the inverse of the propagators can be deduced for the fermion 
in the same way as for the scalar by inverting the momentum space 
representation and then mapping to coordinate space with a Fourier inverse. 
This produces \cite{146} 
\begin{equation}
\langle \psi^i(x) \bar{\psi}^j(y) \rangle^{-1} ~\sim~
\frac{r(\alpha_\psi-1) (\xslash-\yslash) \delta^{ij}}
{A_\psi((x-y)^2)^{2\mu-\alpha_\psi+1}}
\end{equation}
where
\begin{equation}
r(\alpha) ~=~ \frac{\alpha a(\alpha-\mu)}{(\mu-\alpha)a(\alpha)} ~. 
\end{equation}
For the gauge field the inverse scaling form of the propagator is found by
first inverting the momentum space propagator on the transverse subspace
\cite{147} as that is the only contribution to the self-consistency equations 
at criticality which are physically meaningful. So 
\begin{equation}
\langle A_\mu(p) A_\nu(-p) \rangle^{-1} ~\sim~ 
\frac{\tilde{B}_\gamma}{(p^2)^{\beta_\gamma-\mu}} 
\left[ \eta_{\mu\nu} - \frac{p_\mu p_\nu}{p^2} \right] 
\end{equation}
and mapping this to coordinate space produces 
\begin{eqnarray}
\langle A_\mu(x) A_\nu(y) \rangle^{-1} &\sim&
\frac{t(\beta_\gamma)}{B_\gamma((x-y)^2)^{2\mu-\beta_\gamma}} \nonumber \\
&& \times
\left[ \eta_{\mu\nu} + \frac{2(2\mu-\beta_\gamma)}{(2\beta_\gamma-2\mu-1)}
\frac{(x-y)_\mu (x-y)_\nu}{(x-y)^2} \right] 
\end{eqnarray}
where
\begin{equation}
t(\alpha) ~=~ 
\frac{[4(\mu-\alpha)^2-1]a(\alpha-\mu)}{4(\mu-\alpha)a(\alpha)} ~.
\end{equation}
An indication of the correctness of this formalism, for example, is that it was
used to compute the critical exponents of the QCD $\beta$-function and quark 
mass dimension at $O(1/\Nf)$ and $O(1/\Nf^2)$ respectively \cite{148,149,150}. 
Both agree with the recent {\em five} loop four dimensional perturbative 
results of Refs. \cite{9,10,11,12,151,152,153} and \cite{154}. This will be 
illustrated later as an example of how to connect the information in the 
$d$-dimensional critical exponents with known perturbative results. If 
corrections to scaling are included then
\begin{eqnarray}
\langle \psi^i(x) \bar{\psi}^j(0) \rangle &\sim&
\frac{\xslash A_\psi \delta^{ij}}{(x^2)^{\alpha_\psi}} 
\left[ 1 + A_\psi^\prime(x^2)^\lambda \right] \nonumber \\
\langle \psi^i(x) \bar{\psi}^j(0) \rangle^{-1} &\sim&
\frac{r(\alpha_\psi-1) \xslash \delta^{ij}}
{A_\psi(x^2)^{2\mu-\alpha_\psi+1}}
\left[ 1 - A_\psi^\prime s(\alpha_\psi-1)(x^2)^\lambda \right]
\end{eqnarray}
for fermions and
\begin{eqnarray}
\langle A_\mu(x) A_\nu(0) \rangle &\sim& 
\frac{B_\gamma}{(x^2)^{\beta_\gamma}} \left[ \eta_{\mu\nu} + 
\frac{2( 1 - b )\beta_\gamma}{(2\mu-2\beta_\gamma-1+b)}
\frac{x_\mu x_\nu}{x^2} \right. \nonumber \\
&& \left. ~~~~~~~~~+ 
\left[ \eta_{\mu\nu} + 
\frac{2( 1 - b )(\beta_\gamma-\lambda)}{(2\mu-2\beta_\gamma+2\lambda-1+b)}
\frac{x_\mu x_\nu}{x^2} \right] B_\gamma^\prime (x^2)^\lambda \right] 
\nonumber \\
\langle A_\mu(x) A_\nu(0) \rangle^{-1} &\sim&
\frac{t(\beta_\gamma)}{B_\gamma(x^2)^{2\mu-\beta_\gamma}} \nonumber \\
&& 
\left[ \eta_{\mu\nu} + \frac{2(2\mu-\beta_\gamma)}{(2\beta_\gamma-2\mu-1)}
\frac{x_\mu x_\nu}{x^2} \right. \nonumber \\
&& \left. ~ -
\left[ \eta_{\mu\nu} + \frac{2(2\mu-\beta_\gamma-\lambda)}
{(2\beta_\gamma+2\lambda-2\mu-1)}
\frac{x_\mu x_\nu}{x^2} \right] 
u(\beta_\gamma,b) B_\gamma^\prime (x^2)^\lambda 
\right]
\label{gaugecor}
\end{eqnarray}
for a gauge field \cite{155} where
\begin{eqnarray}
s(\alpha) &=& \frac{\alpha(\alpha-\mu)q(\alpha)}{(\alpha-\mu+\lambda)
(\alpha-\lambda)} \nonumber \\
u(\alpha,b) &=& \frac{(\mu-\alpha+\lambda)(2\mu-2\alpha-1+b)
(2\alpha+2\lambda-2\mu-1)q(\alpha)}
{(2\mu-2\alpha+2\lambda-1+b)(\mu-\alpha-\lambda)(2\alpha-2\mu-1)}
\end{eqnarray}
are the associated functions. It is worth stressing that this approach can only
be used when there are no $n$-leg terms with $n$~$>$~$2$ in the operator whose
dimension corresponds to the critical $\beta$-function slope.

One aspect of the massless asymptotic scaling forms of the coordinate space 
gauge field propagator deserves general comment. This concerns the relation to 
a conformal transformation which we take to be
\begin{equation}
x_\mu ~ \rightarrow ~ x^\prime_\mu ~=~ \frac{x_\mu}{x^2} ~.
\label{confmap}
\end{equation}
For a useful early background to conformal field theories in $d$-dimensions see
Ref. \cite{156}. This produces the transformation 
\begin{equation}
\Lambda_{\mu\nu}(x) ~=~ \frac{\partial x^\prime_\mu}{\partial x_\nu} ~=~
\eta_{\mu\nu} ~-~ 2 \frac{x_\mu x_\nu}{x^2} 
\label{conftendef}
\end{equation}
which is sometimes referred to as the conformal tensor. It has the interesting
property that
\begin{equation}
\Lambda_\mu^{~\,\sigma}(x) \Lambda_{\sigma\nu}(x) ~=~ \eta_{\mu\nu} ~.
\label{tenproj}
\end{equation} 
In terms of the parameter $b$ of the tensor defined in (\ref{gauprop}) there is
only one value of $b$ for which such tensors formally satisfy (\ref{tenproj})
which is $b$~$=$~$-$~$1$. While we have a scaling form for the gauge field 
propagator in coordinate space we can determine the relation between the 
scaling dimension of the gauge field and the gauge parameter. Setting the 
second coefficient of the gauge field propagator in the second equation of 
(\ref{ferphoprop}) to $(-2)$ we find 
\begin{equation}
b ~=~ \frac{(2\mu-1-\beta)}{(\beta-1)}
\label{confcond1}
\end{equation}
or
\begin{equation}
\beta ~=~ \frac{(2\mu-1+b)}{(b+1)} ~.
\label{confcond2}
\end{equation}
When these conditions are met for the gauge field then we will refer to that
gauge choice as the conformal gauge. As an aside we note that while the
conformal tensor of (\ref{conftendef}) appears to bear a resemblance to the
structure of the Fried-Yennie gauge introduced in Ref. \cite{157} we note that
they are not the same. The latter corresponds to the choice of gauge parameter
$b$~$=$~$3$ in the {\em momentum} space version of the gauge field propagator
rather than the coordinate space one. 

\subsection{Nonabelian gauge theories}

At this stage we have taken as our starting point the asymptotic scaling forms
for spin-$\half$ and spin-$1$ propagators at the Wilson-Fisher fixed point
without specifying the underlying theory. Morever we have alluded to the
connection of the large $\Nf$ analysis with QCD. More concretely in the larger
view of the universal theory at the Wilson-Fisher fixed point involving
fermionic matter and a nonabelian gauge field the theory corresponds to QCD
when the critical dimension is four. For the gauge theory case here the lower 
dimensional partner of QCD is the nonabelian Thirring model (\ref{lagnatm}). In
order to compare we note that 
\begin{equation}
L^{\mbox{\footnotesize{QCD}}} ~=~ -~ \frac{1}{4} G_{\mu\nu}^a
G^{a \, \mu\nu} ~-~ \bar{c}^a \left( \partial^\mu D_\mu c \right)^a ~+~ 
i \bar{\psi}^i \Dslash \psi^i
\label{lagqcd}
\end{equation}
is the QCD Lagrangian prior to fixing a gauge where $G^a_{\mu\nu}$ is the field
strength of the gauge field $A^a_\mu$ and $c^a$ is the Faddeev-Popov ghost.
The key feature from the large $\Nf$ point of view is the common force-matter
interaction via the group valued covariant derivative $D_\mu$. The equivalence
of (\ref{lagnatm}) to (\ref{lagqcd}) in the large $\Nf$ expansion has been 
discussed in Ref. \cite{108}. One feature of (\ref{lagnatm}) is that using 
solely its interaction, which drives the universal theory at the Wilson-Fisher 
fixed point, ensures that the contribution from the triple and quartic gluon 
vertices of the theory with critical dimension $4$ correctly emerge in that
dimensions. This is a remarkable observation in a gauge theory \cite{108}. More
importantly this critical relation between the two theories has a computational
benefit. If one wishes to compute information about QCD using the critical 
point large $\Nf$ method the core Lagrangian to use is (\ref{lagnatm}) which 
defines the field canonical dimensions and common interaction. This reduces the
number of Feynman graphs to consider. For example, for (\ref{lagnatm}) using 
the asymptotic scaling forms of the propagators the quark and gluon $2$-point 
functions can be evaluated in powers of $1/\Nf$ in complete parallel to 
(\ref{nlsmsde}). Indeed the graphs are formally the same as Fig. \ref{sigcr} 
and both the Landau gauge quark anomalous dimension and quark mass anomalous 
dimension are available at $O(1/\Nf^2)$ in Ref. \cite{150} as a function of 
$d$.

Moreover it transpires that the connection of two dimensional theories with
four dimensional counterparts is not limited to nonabelian gauge theories where
the lower dimensional theory generates the underlying Feynman ruloes of the 
higher dimensional one. A careful examination of the core theories we have 
concentrated on here, (\ref{lagnlsm}) and (\ref{laggn}), have the same 
property. For instance, in the case of (\ref{laggn}) the higher dimensional 
theory to which it is critically related to is that discussed in
Ref. \cite{158} and is now termed the Gross-Neveu-Yukawa (GNY) theory. It has 
the Lagrangian
\begin{equation}
L^{\mbox{\footnotesize{GNY}}} ~=~ i \bar{\psi}^i \partialslash \psi^i ~+~
\frac{1}{2} \partial_\mu \sigma \partial^\mu \sigma ~+~
\frac{1}{2} g_1 \sigma \bar{\psi}^i \psi^i ~+~ \frac{1}{24} g_2^2 \sigma^4 ~.
\label{laggny}
\end{equation}
where there are two coupling constants $g_1$ and $g_2$ and the $\sigma$ field
which was auxiliary in (\ref{laggn}) propagates. As is apparent (\ref{laggny})
has the matter-force interaction, $\sigma \bar{\psi}^i \psi^i$, in common
with (\ref{laggn}). By contrast the quartic $\sigma$ self-energy interaction
cannot be present in two dimensions on dimensional grounds required by
renormalizability but has to be present in four dimensions for the same reason.
In addition (\ref{laggn}) and (\ref{laggny}) provide another example of the 
contrast of operator mixing in perturbation theory and large $N$. For instance 
on dimensional grounds in two dimensions the fermion mass operator 
$\bar{\psi}^i \psi^i$ mixes with $\sigma$ in perturbation theory. By contrast 
in the large $N$ critical point formalism $\bar{\psi}^i \psi^i$ does not mix.

In relating both Lagrangians (\ref{laggn}) and (\ref{laggny}) in this way and 
their connection effected through the $d$-dimensional Wilson-Fisher fixed point
we are describing an example of what is now termed ultraviolet completion 
\cite{159,160}. Another example is the ultraviolet completion of both 
(\ref{lagnlsm}) and (\ref{lagphi4elim}) which was discussed in 
Refs. \cite{161} and \cite{162} which is $O(N)$ $\phi^3$ theory in six 
dimensions. The Lagrangian in that instance is 
\begin{equation}
L^{\phi^3} ~=~ \frac{1}{2} \partial_\mu \phi^i \partial^\mu \phi^i ~+~
\frac{1}{2} \partial_\mu \sigma \partial^\mu \sigma ~+~
\frac{g_1}{2} \sigma \phi^i \phi^i ~+~ \frac{g_2}{2} \sigma^3
\label{lagphi3}
\end{equation}
which clearly shares a common interaction. As a consequence it has been shown
to be in the same universality class as (\ref{lagnlsm}) and (\ref{lagphi4elim}) in Refs. \cite{161,162} and \cite{163}. This has been achieved by computing
the renormalization group functions of (\ref{lagphi3}) to high loop order
\cite{162,163} and showing that the critical exponents derived from them at 
the Wilson-Fisher fixed point are in precise agreement with the $d$-dimensional
critical exponents evaluated to various orders in large $N$ in the underlying
universal field theory. Interestingly both (\ref{laggny}) and (\ref{lagphi3})
share a relatively new property of multicoupling field theories \cite{164,165}.
This is the feature of emergent symmetries. If one examines the location of the
fixed points on the $(g_1,g_2)$ coupling plane then for specific values of $N$,
which is usually low, there is a fixed point where the critical couplings are 
equal. At such a point the theory is symmetric under a larger symmetry. In the 
case of $N$~$=$~$1$ for (\ref{laggny}) there is an emergent 
{\em supersymmetry}, for instance. An in depth study can be found in 
Ref. \cite{160}, for example. Such an emergent phenomenon may be a feature of 
the Standard Model itself indicating the existence of a higher symmetry whether 
supersymmetry or not.  

In terms of the critical point large $N$ formalism of 
Refs. \cite{34,35} and \cite{36} it has been extended to the 
Gross-Neveu-Yukawa universality class at $O(1/N^2)$ by several groups. Leading 
order exponents were deduced by conventional methods. See for example
Refs. \cite{63,143} and \cite{166}. However the formalism discussed
previously coupled with conformal integration techniques summarized in the
Appendices meant that the fermion exponent $\eta$ was computed first at
$O(1/N^2)$ in Ref. \cite{146}. Subsequently there were independent evaluations
of the exponent $\nu$ and fermion mass dimension at $O(1/N^2)$ before $\eta$ 
was extended to $O(1/N^3)$ in Refs. \cite{167,168,169,170,171} and \cite{172}
using the large $N$ conformal bootstrap formalism \cite{36} which we will 
summarize later. As these were in $d$-dimensions the data contained within the 
exponents proved crucial to the explicit evaluation of the perturbative 
renormalization group functions of (\ref{laggn}). These are now known to four 
loops in Refs. \cite{90,91,92,93,98} and \cite{99}. Included in this was the 
exponent relating to the $\beta$-function of (\ref{laggn}) which is $\nu$. 
However for the four dimensional theory in the same universality class, 
(\ref{laggny}), the $\beta$-function exponents were only computed more 
recently. The leading order exponent $\omega$ was given in Ref. \cite{173} 
while the $O(1/N^2)$ extension was provided in Ref. \cite{174}. The latter 
gave the information on the $\beta$-functions of both coupling constants of 
(\ref{laggny}). Again the remarkable connection through the critical 
renormalization group equation, which we will detail later, allows one to 
connect the exponents to the perturbative results and ensure that there is full
agreement in the region of overlap of the perturbative coefficients at 
$O(1/N^2)$. 

\section{Critical exponent connection with renormalization group functions}

Having presented the formalism for evaluating critical exponents in
$d$-dimensions we devote this section to making the indicated connection with 
the renormalization group functions of the theories in the same universality 
class. This is achieved by considering the $\epsilon$ expansion of the 
exponents around the critical dimension of a specific theory. 

\subsection{$\beta$-function}

The starting point is the $\beta$-function. To illustrate this we define a 
generic $\beta$-function that will have a large $N$ accessible fixed point 
which we generally regard as being of the Wilson-Fisher type. In order to 
appreciate the subtlety of where $N$ appears in the $\beta$-function we define 
the coefficients with two labels. One relates to the order of the coupling 
constant and the other to the power of $N$. Therefore we define
\begin{equation}
\beta(g) ~=~ \sum_{n=1}^\infty \sum_{r=0}^{n-1} b_{n-1 \, r} N^r g^n
\label{betagen}
\end{equation}
or more concretely
\begin{eqnarray}
\beta(g) &=& b_{00} g ~+~ \left( b_{11} N + b_{10} \right) g^2 ~+~ 
\left( b_{22} N^2 + b_{21} N + b_{20} \right) g^3 \nonumber \\ 
&& +~ \left( b_{33} N^3 + b_{32} N^2 + b_{31} N + b_{30} \right) g^4 
\nonumber \\
&& +~ \left( b_{44} N^4 + b_{43} N^3 + b_{42} N^2 + b_{41} N 
+ b_{40} \right) g^5 ~+~ \ldots ~.
\label{betapertgen}
\end{eqnarray}
This is the canonical expression of the $\beta$-function in the sense that it
represents the perturbative coupling constant or loop expansion. Each
coefficient of the coupling constant is a polynomial of the parameter $N$ 
where in effect the coefficient of $N$ corresponds to the contribution of 
Feynman graphs at that loop order with various numbers of closed matter 
bubbles. However one can formally rewrite or reorder the terms of the 
$\beta$-function without altering its interpretation. Instead we can regroup
the terms in the following form 
\begin{eqnarray}
\beta(g) &=& \left( b_{00} g ~+~ b_{11} N g^2 ~+~ b_{22} N^2 g^3 ~+~ 
b_{33} N^3 g^4 ~+~ b_{44} N^4 g^5 ~+~ \ldots \right) \nonumber \\
&& +~ \left( b_{10} g^2 ~+~ b_{21} N g^3 ~+~ b_{32} N^2 g^4 ~+~ 
b_{43} N^3 g^5 ~+~ \ldots \right) \nonumber \\
&& +~ \left( b_{20} g^3 ~+~ b_{31} N g^4 ~+~ b_{42} N^2 g^5 ~+~ 
b_{53} N^3 g^6 ~+~ \ldots \right) \nonumber \\
&& +~ \left( b_{30} g^4 ~+~ b_{41} N g^5 ~+~ b_{52} N^2 g^6 ~+~ 
b_{63} N^3 g^7 ~+~ \ldots \right) \nonumber \\
&& +~ \ldots ~. 
\label{betalargengen}
\end{eqnarray}
This version represents the large $N$ ordering of the coefficients of
$\beta(g)$. As will become evident later the large $N$ counting is such that
the coupling constant $g$ is regarded as $1/N$. This means that each of the 
terms in the first line of (\ref{betalargengen}) are $O(1/N)$ whereas in the
next line they are $O(1/N^2)$. In each of the subsequent lines the power of
$1/N$ is one less. While this is not the standard way of looking at a 
renormalization group function it is completely mathematically equivalent to
(\ref{betapertgen}).

To this point we have summarized the general structure of a $\beta$-function
in a theory with a dimensionless parameter $N$ which usually derives from an
$O(N)$ or $SU(N)$ symmetry. The usual procedure to determine the coefficients
$b_{nr}$ is to compute the relevant Green's functions in the underlying
renormalizable quantum field theory order by order in perturbation theory.
Overlooking for the moment the huge technical issues required to achieve this,
this procedure will in principle determine these coefficients with respect to
a particular renormalization scheme. This is usually the $\MSbar$ scheme in
which the five loop QCD $\beta$-function is now available \cite{9,10,11,12} 
which extended the earlier lower loop results of 
Refs. \cite{13,14,15,16,17,18} and \cite{19}. However, the loop expansion is 
not the only tool of deducing the coefficients $b_{nr}$. An alternative is 
available from the large $N$ ordering and is effected through properties of the
critical renormalization group equation as well as the universality properties 
of quantum field theories as already discussed. Since the parameter $N$ derives
from a group it is dimensionless when one dimensionally regularizes and hence 
it can be regarded as an ordering parameter which transcends the spacetime 
dimension. Therefore in the large $N$ expansion one can analyse quantum field 
theories away from their critical dimension. As a side remark we note that 
another technique which can examine quantum field theories in a similar way in 
$d$-dimensions is the exact or functional renormalization group approach which 
is based on early ideas by Wilson \cite{24,25,26}. In this context one ought to
regard large $N$ results as complementary.

To effect the large $N$ expansion in the critical point approach \cite{34,35} 
one needs to locate a fixed point in $d$-dimensions at leading order in $1/N$. 
This is given by solving $\beta(g)$~$=$~$0$ as a power series in $1/N$ where 
the coefficients are functions of $d$. Though we will appeal to the critical 
dimension of the theory, $d_c$, and set $d$~$=$~$d_c$~$-$~$2\epsilon$. Due to 
renormalizability this means that $b_{00}$ is proportional to $\epsilon$. At 
this point we stress that we are computing in $d$-dimensions and $\epsilon$ 
here is {\em not} the regularizing parameter of dimensional regularization. In 
the large $N$ approach \cite{34,35} a theory is analytically and not 
dimensionally regularized. Once $b_{00}$ is fixed the leading order large $N$ 
term to the critical coupling $g_c$ where $\beta(g_c)$~$=$~$0$ is found from 
setting the first term of (\ref{betalargengen}) to zero. This is clearly 
impossible and defeats the purpose of the large $N$ approach which is to 
systematically determine the coefficients $b_{nr}$ by a different method to 
perturbation theory. In other words to find $g_c$ at leading order one already 
needs to know all the coefficients $b_{nn}$ for $n$~$\geq$~$1$. In QCD the 
$\beta$-function has this form with respect to the colour group $SU(\Nc)$ where
the $\Nc$ dependence at two loops is quadratic and cubic at three loops and so 
on. In other words one would have to sum up an infinite number of graphs at 
leading order in a $1/\Nc$ expansion to find the leading order critical 
coupling in large $\Nc$. Another example of a theory which has the same feature
is $O(N)$ $\phi^6$ theory which has a critical dimension of three. However in 
that case the large $N$ expansion has been examined by several authors and 
meaningful information on the critical point structure of the field theory in 
$d$-dimensions has been extracted. See, for example, 
Refs. \cite{175,176} and \cite{177}.

By contrast if one looks at the $\Nf$ dependence in the QCD $\beta$-function 
then the two loop term is {\em linear} in $\Nf$ and quadratic at three loops. 
So at leading order in a $1/\Nf$ expansion there are only two nonzero terms and
this allows one to obtain the starting point for the critical point method 
\cite{34,35}. In light of this from now on we will restrict our discussion to 
quantum field theories with an underlying $O(N)$ symmetry which have 
$\beta$-functions of the form
\begin{eqnarray}
\beta(g) &=& b_{00} g ~+~ b_{11} N g^2 \nonumber \\
&& +~ \left( b_{10} g^2 ~+~ b_{21} N g^3 ~+~ b_{32} N^2 g^4 ~+~ 
b_{43} N^3 g^5 ~+~ \ldots \right) \nonumber \\
&& +~ \left( b_{20} g^3 ~+~ b_{31} N g^4 ~+~ b_{42} N^2 g^5 ~+~ 
b_{53} N^3 g^6 ~+~ \ldots \right) \nonumber \\
&& +~ \left( b_{30} g^4 ~+~ b_{41} N g^5 ~+~ b_{52} N^2 g^6 ~+~ 
b_{63} N^3 g^7 ~+~ \ldots \right) \nonumber \\
&& +~ \ldots ~.
\label{betalargen}
\end{eqnarray}
In other words $b_{nn}$~$=$~$0$ for $n$~$\geq$~$2$. 

Equivalently we can rewrite (\ref{betalargen}) in the generic form 
\begin{equation}
\beta(g) ~=~ -~ \epsilon g ~+~ ( b_{11} N + b_{10} ) g^2 ~+~
N g^2 \sum_{i=1}^\infty \frac{1}{N^i} \beta_i ( b_{11} N g )
\label{betalargenfn}
\end{equation}
which defines the functions $b_i(b_{11}gN)$ and these can be related to the
coefficients in the $1/N$ expansion of the corresponding critical exponent
which is $\beta^\prime(g_c)$. This quantity is sometimes referred to as the 
correction to scaling exponent. For instance with
\begin{equation}
\omega ~=~ -~ \frac{1}{2} \beta^\prime(g_c) 
\end{equation}
one can deduce that $\beta_1(b_{11}gN)$ satisfies the simple differential
equation
\begin{equation}
\omega_1(\epsilon) ~=~ -~ \frac{\epsilon^2}{2b_{11}} 
\beta_1^\prime(\epsilon) ~.
\end{equation}
Similar first order differential equations can be determined for the higher 
order terms in (\ref{betalargenfn}) which means that the $\beta$-function can 
be represented as a one parameter integral where the integrand involves
$\omega_n(\epsilon)$ in a nonlinear way. For example, we have 
\begin{equation}
\beta(g) ~=~ -~ \epsilon g ~+~ ( b_{11} N + b_{10} ) g^2 ~-~
2 b_{11} N g^2 \int_0^{b_{11}Ng} \, d\xi \, \frac{\omega_1(\xi)}{\xi^2} ~+~ 
O \left( \frac{1}{N^2} \right) ~.
\label{betarep}
\end{equation}
In principle this means that for instance the radius of convergence of the 
perturbative series can be explored and the location of the first renormalon 
found or the existence of fixed points beyond the Gaussian one. As noted 
earlier such fixed points may impact upon asymptotic safety ideas for physics 
beyond the Standard Model, \cite{121,122,123,124,125,178,179}. For example the 
latter has been explored recently at leading order in $1/N$ in four dimensional
gauge theories coupled to Yukawa interactions \cite{121,122,123,124,125}. 

In relation to the perturbative expansion of the $\beta$-function $\omega$ 
contains information not only on its slope but encodes the expansion of the 
critical coupling $g_c$. For instance $g_c$ can be written in terms of the
functions $\beta_i(\epsilon)$ but in order to demonstrate the explicit 
connection of the information contained within the exponents and the 
coefficients in the corresponding renormalization group functions we can
deduce 
\begin{eqnarray}
g_c &=&
\frac{\epsilon}{b_{11}} 
- \left[ \frac{b_{10}}{b_{11}^2} \epsilon
+ \frac{b_{21}}{b_{11}^3} \epsilon^2
+ \frac{b_{32}}{b_{11}^4} \epsilon^3
+ \frac{b_{43}}{b_{11}^5} \epsilon^4
+ \frac{b_{54}}{b_{11}^6} \epsilon^5
+ \frac{b_{65}}{b_{11}^7} \epsilon^6
+ \frac{b_{76}}{b_{11}^8} \epsilon^7
\right] \frac{1}{N} \nonumber \\
&& + \left[ 
\frac{b_{10}^2}{b_{11}^3} \epsilon
+ \left[ 3 \frac{b_{10} b_{21}}{b_{11}^4} - \frac{b_{20}}{b_{11}^3} \right]
\epsilon^2
+ \left[ 2 \frac{b_{21}^2}{b_{11}^5} + 4 \frac{b_{10} b_{32}}{b_{11}^5}
- \frac{b_{31}}{b_{11}^4} \right] \epsilon^3
\right. \nonumber \\
&& \left. ~~~
+ \left[ 5 \frac{b_{21} b_{32}}{b_{11}^6} + 5 \frac{b_{10} b_{43}}{b_{11}^6}
- \frac{b_{42}}{b_{11}^5} \right] \epsilon^4
+ \left[ 3 \frac{b_{32}^2}{b_{11}^7} + 6 \frac{b_{21} b_{43}}{b_{11}^7}
+ 6 \frac{b_{10} b_{54}}{b_{11}^7} - \frac{b_{53}}{b_{11}^6} \right] \epsilon^5
\right. \nonumber \\
&& \left. ~~~
+ \left[ 7 \frac{b_{32} b_{43}}{b_{11}^8} + 7 \frac{b_{21} b_{54}}{b_{11}^8}
+ 7 \frac{b_{10} b_{65}}{b_{11}^8} - \frac{b_{64}}{b_{11}^7} \right] \epsilon^6
\right. \nonumber \\
&& \left. ~~~
+ \left[ 4 \frac{b_{43}^2}{b_{11}^9} + 8 \frac{b_{32} b_{54}}{b_{11}^9} 
+ 8 \frac{b_{21} b_{65}}{b_{11}^9} + 8 \frac{b_{10} b_{76}}{b_{11}^9} 
- \frac{b_{75}}{b_{11}^8} \right] \epsilon^7 \right] 
\frac{1}{N^2} \nonumber \\
&& +~ O \left( \epsilon^8 ; \frac{1}{N^3} \right) 
\label{critcn}
\end{eqnarray}
from (\ref{betalargen}) at several orders in $1/N$ where the order symbol
indicates the truncation point of the two independent series.

\subsection{Other renormalization group functions and scheme issues}

While our focus has been on
the $\beta$-function the other renormalization group functions have a 
comparable expansion in $1/N$. If we define the perturbative structure of a 
generic renormalization group function in a similar way to (\ref{betagen}) by
\begin{equation}
\gamma(g) ~=~ \sum_{n=1}^\infty \sum_{r=0}^{n-1} a_{n \, r} N^r g^n
\label{rgegen}
\end{equation}
then the corresponding expression to (\ref{critcn}) is
\begin{eqnarray}
\gamma(g_c) &=&
\left[ \frac{a_{10}}{b_{11}} \epsilon + \frac{a_{21}}{b_{11}^2} \epsilon^2
+ \frac{a_{32}}{b_{11}^3} \epsilon^3 + \frac{a_{43}}{b_{11}^4} \epsilon^4
+ \frac{a_{54}}{b_{11}^5} \epsilon^5 + \frac{a_{65}}{b_{11}^6} \epsilon^6
+ \frac{a_{76}}{b_{11}^7} \epsilon^7 \right] \frac{1}{N} \nonumber \\
&& 
+ \left[ 
-~ \frac{b_{10} a_{10}}{b_{11}^2} \epsilon
+ \left[
- \frac{b_{21} a_{10}}{b_{11}^3}
- 2 \frac{b_{10} a_{21}}{b_{11}^3} 
+ \frac{a_{20}}{b_{11}^2}
\right] \epsilon^2
\right. \nonumber \\
&& \left. ~~~
+ \left[ 
- \frac{b_{32} a_{10}}{b_{11}^4} 
- 2 \frac{b_{21} a_{21}}{b_{11}^4}
- 3 \frac{b_{10} a_{32}}{b_{11}^4}
+ \frac{a_{31}}{b_{11}^3}
\right] \epsilon^3
\right. \nonumber \\
&& \left. ~~~
+ \left[ 
- \frac{b_{43} a_{10}}{b_{11}^5}
- 2 \frac{b_{32} a_{21}}{b_{11}^5}
- 3 \frac{b_{21} a_{32}}{b_{11}^5}
- 4 \frac{b_{10} a_{43}}{b_{11}^5}
+ \frac{a_{42}}{b_{11}^4}
\right] \epsilon^4
\right. \nonumber \\
&& \left. ~~~
+ \left[ 
- \frac{b_{54} a_{10}}{b_{11}^6}
- 2 \frac{b_{43} a_{21}}{b_{11}^6}
- 3 \frac{b_{32} a_{32}}{b_{11}^6}
- 4 \frac{b_{21} a_{43}}{b_{11}^6}
- 5 \frac{b_{10} a_{54}}{b_{11}^6}
+ \frac{a_{53}}{b_{11}^5}
\right] \epsilon^5
\right. \nonumber \\
&& \left. ~~~
+ \left[ 
- \frac{b_{65} a_{10}}{b_{11}^7}
- 2 \frac{b_{54} a_{21}}{b_{11}^7}
- 3 \frac{b_{43} a_{32}}{b_{11}^7}
- 4 \frac{b_{32} a_{43}}{b_{11}^7}
- 5 \frac{b_{21} a_{54}}{b_{11}^7}
\right. \right. \nonumber \\
&& \left. \left. ~~~~~~~
- 6 \frac{b_{10} a_{65}}{b_{11}^7}
+ \frac{a_{64}}{b_{11}^6}
\right] \epsilon^6
\right. \nonumber \\
&& \left. ~~~
+ \left[ 
- \frac{b_{76} a_{10}}{b_{11}^8}
- 2 \frac{b_{65} a_{21}}{b_{11}^8}
- 3 \frac{b_{54} a_{32}}{b_{11}^8}
- 4 \frac{b_{43} a_{43}}{b_{11}^8}
- 5 \frac{b_{32} a_{54}}{b_{11}^8}
- 6 \frac{b_{21} a_{65}}{b_{11}^8}
\right. \right. \nonumber \\
&& \left. \left. ~~~~~~~
- 7 \frac{b_{10} a_{76}}{b_{11}^8}
+ \frac{a_{75}}{b_{11}^7}
\right] \epsilon^7
\right] \frac{1}{N^2} ~+~ O \left( \epsilon^8 ; \frac{1}{N^3} \right) ~.
\label{rgeexpN}
\end{eqnarray}
This illustrates the necessity of knowing the critical point location at each
order in the large $N$ expansion in order to map the $\epsilon$ expansion of a
large $N$ critical exponent to the perturbative coefficients.

\subsection{Renormalization scheme issues}

We devote this section to brief remarks concerning the consequences of 
different renormalization schemes. In considering (\ref{betagen}) and 
(\ref{betapertgen}) at the outset we have made an implicit assumption on the
general structure. This is that the formal expression is for $\beta$-functions
in the $\MSbar$ scheme \cite{180}. While this is the most common scheme used 
for perturbative quantum field theory and moreover the one in which one can 
readily compute to high loop order, it is not a kinematic scheme. In other 
words the subtraction of the infinities can be carried out at a wide range of 
momentum configurations and so it is not tied to one subtraction point. By 
contrast kinematic schemes have a twofold feature. First they are defined at 
some kinematic point. For example, the momentum subtraction (MOM) schemes 
discussed by Celmaster and Gonsalves for QCD in 
Refs. \cite{181} and \cite{182} are centred on the completely symmetric point
where all the squared external momenta of a vertex function are equal. A
similar scheme was discussed in the context of six dimensional $\phi^3$ theory 
in Ref. \cite{183}. Second for MOM at this subtraction point not only are the 
divergences with respect to the regularizing parameter removed by the 
renormalization constant but also the finite part. Although this is one example
there are others such as the onshell scheme \cite{184}. Irrespective of which 
scheme is used the renormalization group functions are scheme dependent after 
some (low) loop order. This would suggest that the critical exponents derived 
from renormalization group functions in different schemes are not the same. In 
other words the coefficients $a_{mn}$ and $b_{mn}$ have different values in 
separate schemes. This cannot be the case, however, since the critical 
exponents are renormalization group invariants or equivalently physical 
observables. The resolution of this in the context of the derivations here is 
in the assumption behind the forms of (\ref{betagen}) and (\ref{betapertgen}). 
In the $\MSbar$ scheme at the subtraction point only the divergences are 
removed aside from the trivial finite part \cite{180} of 
$\ln(4\pi e^{-\gamma})$ to differentiate it from the original MS scheme. Here 
$\gamma$ is the Euler-Mascheroni constant. In a scheme where a finite part is 
removed into the renormalization constant the generic $\beta$-function and 
renormalization group functions (\ref{betapertgen}) and (\ref{rgegen}) are 
strictly only valid in the critical dimension of the theory. In the ultimate 
step to derive these in dimensional regularization the regulator is removed. 
Before this is carried out the $d$-dimensional renormalization group functions 
have $\epsilon$-dependent coefficients at each order \cite{183}. This property
can be accommodated in (\ref{betapertgen}) and (\ref{rgegen}) by making the 
formal shifts
\begin{equation}
b_{n0} ~ \rightarrow ~ b_{n0} ~+~ b^\prime_{n} \epsilon ~~~,~~~
a_{n0} ~ \rightarrow ~ a_{n0} ~+~ a^\prime_{n} \epsilon
\end{equation}
respectively where the prime denotes the finite part. These extra pieces then 
play a crucial role in ensuring the renormalization group invariance of the 
critical exponents. This is not straightforward to see at a formal level since
$a_n^\prime$ and $b_n^\prime$ are {\em not} unrelated to $a_{n0}$ and $b_{n0}$ 
due to the beauty of the underlying renormalization group equation. It can be 
seen, however, in explicit examples where the renormalization group functions 
are known at the orders where scheme dependence is first manifest. One such 
simple case is in Ref. \cite{183} for $\phi^3$ theory while another has been 
discussed in the MOM schemes of QCD at three loops \cite{185}. 

\subsection{Example in QCD}

Returning to the problem of connecting large $N$ critical exponents with the 
perturbative expansion of the renormalization group functions, we can 
illustrate the coefficient mapping formalism of the previous section with an 
example from QCD. We consider the critical exponent corresponding to the quark 
mass anomalous dimension, $\eta_m$. The leading order term was deduced in 
Ref. \cite{65} from the QED result while the $O(1/\Nf^2)$ correction was 
determined in Ref. \cite{150}. The two terms are given by
\begin{equation}
\eta_{m\,1} ~=~ -~ \frac{2 C_F \eta_1}{(\mu-2)} \nonumber \\
\end{equation}
and
\begin{eqnarray}
\eta_{m\,2} &=& -~ \left[ \frac{2 (\mu-1)^2 (\mu-3)}{\mu(\mu-2)}
+ \frac{(2\mu^2-4\mu+1)}{(\mu-2)}
\right. \nonumber \\
&& \left. ~~~~
+~ 3 \mu (\mu-1) \left[ \hat{\Theta}(\mu) - \frac{1}{(\mu-1)^2} \right]
\right] \frac{C_F^2 \eta_1^2}{(\mu-2)^2 (2\mu-1)} \nonumber \\
&& -~ \left[ \frac{[12\mu^4-72\mu^3+126\mu^2-75\mu+11]}
{(2\mu-1)(2 \mu-3)(\mu-2)}
- \mu (\mu-1) \left[ \hat{\Psi}^2(\mu) + \hat{\Phi}(\mu) \right] 
\right. \nonumber \\
&& \left. ~~~~
+~ \frac{[8\mu^5-92\mu^4+270\mu^3-301\mu^2+124\mu-12]}
{2(2\mu-1)(2\mu-3)(\mu-2)} \hat{\Psi}(\mu)
\right. \nonumber \\
&& \left. ~~~~
-~ \frac{\mu^2(2\mu-3)^2}{4(\mu-2)(\mu-1)} \right] 
\frac{C_F C_A \eta_1}{(\mu-2)^2(2\mu-1)} 
\end{eqnarray}
in $d$-dimensions where here 
\begin{equation}
\eta_1 ~=~ -~ \frac{(2\mu-1)(2-\mu)\Gamma(2\mu)}
{4\mu\Gamma(2-\mu)\Gamma^3(\mu)} ~.
\end{equation}
Also the functions
\begin{eqnarray}
\hat{\Theta}(\mu) &=& \psi^\prime(\mu-1) ~-~ \psi^\prime(1) \nonumber \\
\hat{\Psi}(\mu) &=& \psi(2\mu-3) ~+~ \psi(3-\mu) ~-~ \psi(1) ~-~ \psi(\mu-1)
\nonumber \\
\hat{\Phi}(\mu) &=& \psi^\prime(2\mu-3) ~-~ \psi^\prime(3-\mu) ~+~ 
\psi^\prime(1) ~-~ \psi^\prime(\mu-1) 
\end{eqnarray}
involve the Euler polygamma function $\psi(z)$. The $\epsilon$ expansion of 
these functions near $d$~$=$~$4$~$-$~$2\epsilon$ involve the Riemann zeta
function $\zeta_n$ with $n$~$\geq$~$3$. Also to disentangle the information 
encoded in each of $\eta_{m,i}$ the critical slope of the QCD $\beta$-function 
is also required at $O(1/\Nf)$ as is evident from (\ref{rgeexpN}). As this is 
encoded in the exponent $\omega$ we note \cite{149}
\begin{eqnarray}
\omega &=& (\mu-2) ~+~
\left[ (2\mu-3)(\mu-3) C_F
- \frac{[4\mu^4-18\mu^3+44\mu^2-45\mu+14]}{4(2\mu-1)(\mu-1)} C_A \right] 
\frac{\eta_1}{\Nf} \nonumber \\
&& +~ O \left( \frac{1}{\Nf^2} \right) ~.
\end{eqnarray}
In both exponents $\omega$ and $\eta_m$ the computations were carried out for
an arbitrary gauge parameter $b$ which cancelled in the final expression. The 
$O(1/\Nf)$ QED contributions were computed originally in Ref. \cite{65}. It is
straightforward to determine the $\epsilon$ expansion of both sets of 
exponents. To assist the reader with the conventions associated with the
coefficient matching we recall the leading terms of the $\MSbar$ QCD 
$\beta$-function are \cite{13,14,15,16,17}
\begin{eqnarray}
\beta(g) &=& -~ \epsilon
+ \left[ \frac{4}{3} T_F \Nf - \frac{11}{3} C_A \right] g^2
+ \left[ 4 C_F T_F \Nf + \frac{20}{3} C_A T_F \Nf - \frac{34}{3} C_A^2
\right] g^3 \nonumber \\
&& +~ \left[ 2830 C_A^2 T_F \Nf - 2857 C_A^3 + 1230 C_A C_F T_F \Nf 
- 316 C_A T_F^2 \Nf^2 \right. \nonumber \\ 
&& \left. ~~~~-~ 108 C_F^2 T_F \Nf - 264 C_F T_F^2 \Nf^2 \right] 
\frac{g^4}{54} ~+~ O(g^5) ~.
\label{betaqcd}
\end{eqnarray}
For the quark mass anomalous dimension we note \cite{138,186,187} 
\begin{eqnarray}
\gamma_m(g) &=& -~ 3 C_F g
+ \left[ \frac{10}{3} C_F T_F \Nf - \frac{3}{2} C_F^2 - \frac{97}{6} C_A C_F
\right] g^2 \nonumber \\
&& +~ \left[ 3483 C_A C_F^2 - 11413 C_A^2 C_F 
+ [ 5184 \zeta_3 + 2224 ] C_A C_F T_F \Nf - 6966 C_F^3
\right. \nonumber \\
&& \left. ~~~~+~ [ 4968 - 5184 \zeta_3 ] C_F^2 T_F \Nf 
+ 560 C_F T_F^2 \Nf^2 \right] \frac{g^3}{108} ~+~ O(g^4)
\end{eqnarray}
as the reference point for comparing conventions with other work. Like 
(\ref{betaqcd}) the {\em five} loop contribution is known \cite{12,151} 
building on the four loop expression given in 
Refs. \cite{188} and \cite{189}.

If we regard the coefficients of $\gamma_m(g)$ as corresponding to the set 
$\{ a_{nr} \}$ of (\ref{rgegen}) then to determine values beyond those known at
five loops we first need
\begin{eqnarray}
b_{54} &=&
\left[
\frac{856}{243}
+ \frac{128}{27} \zeta_3
\right] C_F
+ \left[
\frac{640}{81} \zeta_3
- \frac{916}{243}
\right] C_A
\nonumber \\
b_{65} &=&
\left[
\frac{1024}{135} \zeta_4
- \frac{4832}{1215}
- \frac{11264}{1215} \zeta_3
\right] C_F
+ \left[
\frac{16064}{3645}
+ \frac{1024}{81} \zeta_4
- \frac{40448}{3645} \zeta_3
\right] C_A
\nonumber \\
b_{76} &=&
\left[
\frac{4288}{729}
+ \frac{4096}{243} \zeta_5
- \frac{11264}{729} \zeta_4
- \frac{78848}{6561} \zeta_3
\right] C_F
\nonumber \\
&&
+ \left[
\frac{20480}{729} \zeta_5
- \frac{9440}{2187}
- \frac{40448}{2187} \zeta_4
- \frac{27136}{6561} \zeta_3
\right] C_A
\end{eqnarray}
which are found from the $\epsilon$-expansion of $\omega$ and we note that the 
$N$ of (\ref{rgegen}) now corresponds to $N$~$=$~$T_F \Nf$. The five loop
coefficient is also provided to assist with comparing conventions. Once these 
six and higher loop leading order large $N$ $\beta$-function coefficients are 
known we find 
\begin{eqnarray}
a_{54} &=&
\left[
\frac{256}{9} \zeta_4
- \frac{1040}{81}
- \frac{1280}{81} \zeta_3
\right] C_F
\nonumber \\
a_{53} &=&
\left[
320 \zeta_4
- \frac{8966}{81}
- \frac{512}{3} \zeta_5
- \frac{352}{3} \zeta_3
\right] C_F^2
\nonumber \\
&&
+ \left[
\frac{67133}{243}
+ \frac{4096}{9} \zeta_5
- 1024 \zeta_4
+ \frac{4576}{27} \zeta_3
\right] C_A C_F
\nonumber \\
a_{65} &=&
\left[
\frac{14432}{729}
- \frac{2048}{27} \zeta_5
+ \frac{2560}{81} \zeta_4
+ \frac{17920}{729} \zeta_3
\right] C_F
\nonumber \\
a_{64} &=&
\left[
\frac{5120}{9} \zeta_5
- \frac{1953064}{3645}
- \frac{2560}{9} \zeta_6
+ \frac{2688}{5} \zeta_4
+ \frac{135424}{1215} \zeta_3
- \frac{4096}{9} \zeta_3^2
\right] C_F^2
\nonumber \\
&&
+ \left[
\frac{33280}{27} \zeta_6
- \frac{3202964}{3645}
- \frac{100352}{81} \zeta_5
+ \frac{54272}{81} \zeta_4
- \frac{292352}{3645} \zeta_3
+ \frac{16384}{27} \zeta_3^2
\right] C_A C_F
\nonumber \\
a_{76} &=&
\left[
\frac{65728}{2187}
- \frac{40960}{243} \zeta_6
+ \frac{20480}{243} \zeta_5
+ \frac{35840}{729} \zeta_4
+ \frac{84992}{2187} \zeta_3
- \frac{8192}{243} \zeta_3^2
\right] C_F
\nonumber \\
a_{75} &=&
\left[
\frac{1773976}{6561}
- \frac{4096}{9} \zeta_7
+ \frac{25600}{27} \zeta_6
- \frac{84992}{81} \zeta_5
+ \frac{75520}{243} \zeta_4
+ \frac{315008}{729} \zeta_3
\right. \nonumber \\
&& \left. ~
- \frac{16384}{9} \zeta_3 \zeta_4
+ \frac{40960}{27} \zeta_3^2
\right] C_F^2
\nonumber \\
&&
+ \left[
\frac{736892}{6561}
+ \frac{327680}{81} \zeta_7
- \frac{773120}{243} \zeta_6
- \frac{1767424}{729} \zeta_5
+ \frac{71680}{243} \zeta_4
+ \frac{2269312}{2187} \zeta_3
\right. \nonumber \\
&& \left. ~~~
+ \frac{65536}{27} \zeta_3 \zeta_4
- \frac{604160}{243} \zeta_3^2
\right] C_A C_F 
\end{eqnarray}
The five loop coefficients are in agreement with the recent corresponding five 
loop $\MSbar$ quark mass anomalous dimension coefficients in 
Refs. \cite{188} and \cite{189}. This example is intended to illustrate the 
power of the large $N$ critical exponents in matching to known perturbative 
results at high loop order. In addition it also provides nontrivial information
on the perturbative structure of renormalization group functions ahead of a
computation beyond a currently available loop order. Similar connections with 
recent high loop perturbative results include $O(N)$ $\phi^4$ theory at six and
seven loops \cite{58,59} $SU(N)$ Gross-Neveu model at four loops \cite{99} and 
$O(N)$ $\phi^3$ theory at three and four loops \cite{162,163}. 

\section{Relation to other areas}

The focus to this point has primarily been with various aspects of the large 
$N$ technique such as its historical development and then the application to
critical phenomena. In the latter case it has been possible to progress the 
$1/N$ expansion to several orders in the critical point approach of 
Refs. \cite{34} and \cite{35} which may not have been as straightforward with
early approaches. Having reviewed that it now seems appropriate to place the 
contribution of the technique in relation to other current methods in quantum 
field theory as well as indicating how it can complement results from those. We
will concentrate on several areas instead of trying to be exhaustive and 
discuss these from a general point of view rather than in depth. 

\subsection{Large $N$ conformal bootstrap}

One of the more recent developments has been the conformal bootstrap method
\cite{27,28,29,30,31,32}  which is the coupling of ideas from conformal field 
theory with numerical methods to extract extremely accurate values for critical
exponents in a variety of models. For instance, the most accurate estimates for
the Ising model were provided in recent years in 
Refs. \cite{28} and \cite{29}. The general idea is that at a fixed point in 
$d$-dimensions where there is conformal symmetry the structure of the 
underlying Green's functions at the fixed point is constrained by conformal 
symmetry as well as crossing symmetry. The ideas derive from very early work on 
scalar field theories in three dimensions such as that of 
Refs. \cite{190,191,192,193,194,195,196} and \cite{197}. There exponent 
estimates were deduced by iteratively solving a general skeleton 
Schwinger-Dyson equation for a theory with only a $3$-point vertex. That vertex
is nested throughout the equation and representative equations, suitably 
regularized, can be solved numerically. Since then the large $N$ critical point 
method \cite{34,35} was extended in 
Refs. \cite{198,199,200,201} and \cite{202} to examine the operator product 
expansion of Green's functions as well as the operator algebra to several 
orders in $1/N$. In other words the consequences of the $d$-dimensional 
conformal symmetry was being used to study the field algebra of the $O(N)$ 
universality class which included (\ref{lagnlsm}). Similar ideas were also 
being developed at the same time \cite{203}. These were later applied to the 
problem of trying to see if the strictly two dimensional $c$-theorem of 
Zamolodchikov could be extended to $d$-dimensions. For example, the large $N$ 
method was applied to the operator product expansion for (\ref{lagnlsm}) and 
(\ref{laggn}) in $2$~$<$~$d$~$<$~$4$ in Refs. \cite{204,205} and \cite{206} 
and expressions obtained for the generalizations of the central charges for the
energy-momentum tensor, $C_T$, and the conserved current associated with the 
internal symmetry group, $C_J$. The large $N$ $d$-dimensional expressions for 
$C_T$ and $C_J$ have recently been studied in the various ultraviolet complete 
theories \cite{159} which extends the range to $2$~$<$~$d$~$<$~$6$. This has
been examined for higher dimensions and agreement found \cite{207}. Overall 
such large $N$ studies have complemented the direct perturbative approach where
the central charges for both conserved quantities can be computed to high loop 
order in, for example, Refs. \cite{208,209} and \cite{210}. In other words at
the Wilson-Fisher fixed point of each critical theory the large $N$ and 
perturbative charges are in agreement. Recent related work using large $N$ 
techniques can be found in Refs. \cite{211,212} and \cite{213} for instance.

{\begin{figure}[ht]
\begin{center}
\includegraphics[width=10.0cm,height=2.9cm]{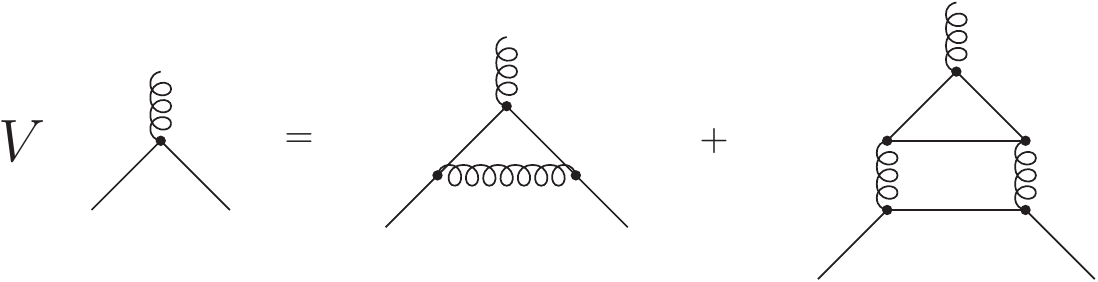}
\end{center}
\caption{Graphs defining the large $N$ conformal bootstrap at $O(1/N)$.}
\label{bootcr}
\end{figure}}

While the early bootstrap progamme of Refs. \cite{195,196} and \cite{197} in 
a strictly three dimensional scalar theory was used as a seed idea for the 
modern numerical bootstrap revolution, it also provided the basis for an 
analytic bootstrap using the large $N$ expansion in {\em arbitrary} dimensions.
This was developed originally in Ref. \cite{36} for the $O(N)$ universality 
class containing (\ref{lagnlsm}) and to avoid confusion will be referred to as 
the large $N$ conformal bootstrap. The beauty of the construction \cite{36} is 
that the critical exponent $\eta$ of the $\phi^i$ field was determined as a 
function of $d$ to $O(1/N^3)$ analytically\footnote{We note that in 
Ref. \cite{36} there was a typographical error in one term of the expression 
for $\eta_3$ which was corrected in Ref. \cite{214} and noted also in 
Ref. \cite{57} which was the six loop $\MSbar$ evaluation of the anomalous 
dimension of $\phi^i$ in $O(N)$ $\phi^4$ theory.}. At criticality an algebraic 
consistency equation can be deduced from the vertex function 
$V$~$=$~$V(\tilde{z},\alpha,\beta_\sigma;\delta,\delta^\prime)$ where 
$\tilde{z}$ is an amplitude, $\alpha$ and $\beta_\sigma$ are the field 
exponents. The quantities $\delta$ and $\delta^\prime$ are regularizing
parameters which regularize divergences associated with vertex subgraphs
\cite{195,196,197} and are effected by shifting the exponents of the fields of
two of the external legs. This is similar to (\ref{chishift}) and both are
indicative of the fact that this approach using the critical point large $N$ 
formalism is also in effect perturbation theory in the vertex anomalous 
dimension. The leading order consistency equation for the vertex function is 
graphically illustrated in Fig. \ref{bootcr} with the higher order graphs given
in Ref. \cite{36}. In Fig. \ref{bootcr} like Fig. \ref{sigcr} there are no 
self-energy corrections on the propagators. At the Wilson-Fisher fixed point 
the asymptotic scaling forms of the propagators are used and include the 
anomalous dimension in the exponent. This quantifies the contribution from the 
virtual propagator corrections so that including decorations on lines would 
overcount. However there is a difference between Fig. \ref{bootcr} and Fig. 
\ref{sigcr} which is in part the essence of the bootstrap. This is that the 
vertices are also dressed which is what in effect is represented by the dot at 
each vertex. Indeed the vertex expansion of Fig. \ref{bootcr} is in terms of 
primitive graphs with no vertex subgraphs within the ordering of the expansion 
by $1/N$. We recall that this means that a $\sigma$ line counts a factor of 
$1/N$ and a closed $\phi^i$ loop counts $N$. 

{\begin{figure}[ht]
\begin{center}
\includegraphics[width=7.8cm,height=2.6cm]{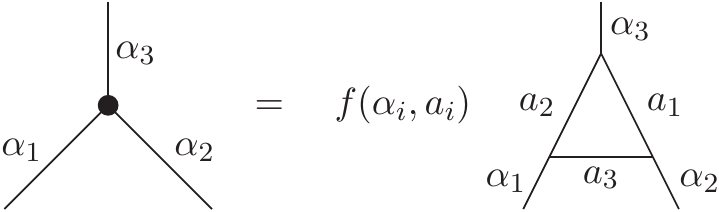}
\end{center}
\caption{Definition of Polyakov's conformal triangle for a scalar vertex.}
\label{cntrcr}
\end{figure}}

The evaluation of the vertex function using the primitive graphs as the basis
for the large $N$ conformal bootstrap formalism \cite{36,193,194,195,196,197} 
relies on the fact that the dot of the vertex is defined to be what is termed a
Polyakov conformal triangle \cite{36,193,194}. This is defined graphically in 
Fig. \ref{cntrcr} in coordinate space representation for general scalar 
propagators with exponents $\alpha_i$. The dotted vertex is replaced by a one 
loop graph where the exponents of the internal lines, $a_i$, are chosen so that
each of the new vertices connecting to an external line are unique in the sense
of Refs. \cite{34,35} and \cite{196}. In other words the sum of the exponents
of the three lines joining at a vertex is equal to the spacetime dimension. For
Fig. \ref{cntrcr} this means that 
\begin{eqnarray}
a_1 ~+~ a_2 ~+~ \alpha_3 &=& 2 \mu \nonumber \\
a_2 ~+~ a_3 ~+~ \alpha_1 &=& 2 \mu \nonumber \\
a_3 ~+~ a_1 ~+~ \alpha_2 &=& 2 \mu ~. 
\label{bootrel}
\end{eqnarray}
These conditions also apply to the situation when the vertex regularizing
parameters decorate the overall primitive graph of the large $N$ bootstrap
expansion. In Appendix A we have discussed at length the derivation and 
properties of unique vertices as well as their power in computing massless 
scalar and fermion Feynman graphs particularly. With the uniqueness property 
the one loop integral of Fig.  \ref{cntrcr} can be related to the original 
vertex with the proportionality, $f(\alpha_i,a_i)$, reflecting the exponent 
dependence. In practical terms for (\ref{lagnlsm}) $\alpha_i$ will involve 
$\alpha_\phi$ and $\beta_\sigma$. For the conformal triangle application to the
bootstrap the key point is that the replacement of a dressed vertex by 
Polyakov's conformal triangle means the complicated $3$-point graphs of Fig. 
\ref{bootcr} can be evaluated using conformal methods \cite{36}. In particular 
if one applies a conformal transformation of the form of (\ref{confmap}) to 
each of the variables of a graph in Fig. \ref{bootcr} then the choice of the 
internal lines of (\ref{bootrel}) for all vertices means that the $3$-point 
vertex graphs immediately collapse to $2$-point integrals. These have 
structural similarities to the graphs of Fig. \ref{sigcr} and similar 
techniques which were used to evaluate those graphs can be applied to the 
graphs of the bootstrap equations. For instance given that there are three 
different possible ways of applying a conformal transformation to a $3$-point 
graph this gives enough information to determine the graphs of Fig. 
\ref{bootcr} to the next order in $1/N$ as discussed in Ref. \cite{36}. This 
is necessary as part of the process to find an expression for $\eta_3$. 

These tools provide the value of the vertex function when the vertex
regularizing parameters $\delta$ and $\delta^\prime$ are nonzero. Once
$V(\tilde{z},\alpha_\phi,\beta_\sigma;\delta,\delta^\prime)$ is available at
a specific order in $1/N$ one has to extract the value for $\eta$ at that
order. Like the earlier $2$-point construction to find $\eta$ the function
contains several variables which aside from $\eta$ are 
$\tilde{z}$~$=$~$f^2 A^2B$ and $\chi_\sigma$. The value of the latter is only
required at one order less than that of $\eta$. Therefore three equations are
needed in order to be able to solve for the value of $\eta$. The formalism
to derive these was presented in Ref. \cite{197}, which introduces the
regularization procedure, and are \cite{36}
\begin{eqnarray}
1 &=& V(\tilde{z},\alpha_\phi,\beta_\sigma;0,0) \nonumber \\
p(\alpha_\phi) &=& s \tilde{z} \left. \frac{\partial ~}{\partial \delta^\prime}
V(\tilde{z},\alpha_\phi,\beta_\sigma;\delta,\delta^\prime) 
\right|_{\delta=\delta^\prime=0} \nonumber \\
\frac{2}{N} p(\beta_\sigma) &=& s \tilde{z} \left. 
\frac{\partial ~}{\partial \delta} 
V(\tilde{z},\alpha_\phi,\beta_\sigma;\delta,\delta^\prime) 
\right|_{\delta=\delta^\prime=0}
\label{cnbooteqn}
\end{eqnarray}
where $s$ is a function that depends on the $\alpha_\phi$ and $\beta_\sigma$. 
In practice its value is not needed since it can be eliminated to produce
\begin{equation}
\frac{N p(\alpha_\phi)}{2p(\beta_\sigma)} ~=~ \left.
\frac{ \left[ 1 + 2 \chi_\sigma \frac{\partial ~}{\partial \delta^\prime}
V(\tilde{z},\alpha_\phi,\beta_\sigma;\delta,\delta^\prime) \right]}
{ \left[ 1 + 2 \chi_\sigma \frac{\partial ~}{\partial \delta}
V(\tilde{z},\alpha_\phi,\beta_\sigma;\delta,\delta^\prime) \right]}
\right|_{\delta=\delta^\prime=0}
\end{equation}
which is sufficient to find $\eta_3$ in $d$-dimensions once the explicit
expansion for $\tilde{z}$ is available from the first equation of 
(\ref{cnbooteqn}). The explicit evaluation of all the contributing graphs is an
exercise separate from the formal construction of the key equations. In light 
of the earlier sections on relating large $N$ exponents in the $d$-dimensional 
universal theory to the perturbative renormalization group functions in various
critical theories, it is worth emphasising not only how much information is 
contained in $\eta_3$ of Ref. \cite{36} and its usefulness for recent six and 
seven loop computations \cite{57,58,59} but how far {\em ahead} of its time was
the result. Moreover it represents the limit that has been achieved so far in 
terms of orders of expansion in $1/N$ in this or any other model. Similar 
results have been determined for (\ref{laggn}) and variations on the core 
Gross-Neveu Lagrangian where there are extra symmetries such as chiral 
symmetry. For several relevant articles for (\ref{laggn}) see, for example, 
Refs. \cite{169} and \cite{172}.

\subsection{Conformal field theory connections}

While our large $N$ critical exponent focus has in part been in relation to the
connection with field theories in integer dimension spacetimes, the fact that 
information is available on exponents for a continuous value of the spacetime 
dimension is of interest due to overlap with other techniques. At this point we
have to stress that we are not suggesting the rigorous existence of field 
theories in noninteger dimensions in general. However as an aside we note that
the class of scalar field theories which has one self-interaction of the form
$\phi^r$ has a critical dimension of $2r/(r-2)$ as noted in Ref. \cite{215}, 
for instance. For $r$~$\geq$~$7$ as well as for $r$~$=$~$5$ the critical 
dimension is rational and the latter case has an application in condensed 
matter physics \cite{216,217}. These situations aside in the universal theory 
context where a core interaction dictates the properties of the underlying 
fixed point one has in principle data on that theory for all $d$. For instance,
it is possible to plot the behaviour of the critical exponents to $O(1/N^2)$ or 
$O(1/N^3)$ in the case of $\eta$ as well as $C_T$ and $C_J$ as a function of 
$d$ as illustrated in Ref. \cite{159}. In the case of critical exponents for 
several universality classes based on the Gross-Neveu-Yukawa structure their 
qualitative behaviour matches closely with studies using the functional 
renormalization group, which we noted earlier. 

A specific example where this can be seen clearly is for the chiral 
Heisenberg-Gross-Neveu model which is believed to be in the universality class 
of a phase transition in graphene \cite{218,219,220,221}. In Ref. \cite{218}
the behaviour of exponents $1/\nu$, $\eta_\phi$ and $\eta_\psi$ are plotted as 
a function of $d$ in Figs. $1$, $2$ and $3$ of that article with the 
corresponding large $N$ plots given in Fig. $7$ of Ref. \cite{221}. Both 
sets of graphs are for the case of $N$~$=$~$4$. Similar analyses have been 
carried out for the original Ising Gross-Neveu model of (\ref{laggn}) but the 
behaviour of the exponents in $d$-dimensions is not the same as for the chiral 
Heisenberg-Gross-Neveu case. However the separate results from the large $N$ 
and functional renormalization group methods are effectively parallel and 
certainly in qualitative agreement. 

One consequence is that there are several analytic ways now of estimating 
exponents in three dimensions which is the spacetime of interest. While for 
many years the original $\epsilon$ expansion of exponents in either two or four
dimensions offered a way of extracting estimates in three dimensions after
resummation, nowadays one has to have a large number of loop orders in order to
be competitive with say Monte Carlo results. For example, the detailed and 
comprehensive study of Ref. \cite{58} to extract exponent estimates in three 
dimensions from six loop $\phi^4$ theory represents the current state of the 
art in terms of $\epsilon$ summation. It has yet to be extended to the seven
loop case of Ref. \cite{59}. Indeed it has reopened the debate concerning the 
asymptotic behaviour of the strictly four dimensional $\phi^4$ $\beta$-function
at large loop order which is discussed in Ref. \cite{222}. For instance, early
ideas in this direction \cite{223,224,225,226,227,228,229,230} were unable to 
benefit from the high loop order data that has been revealed in more recent 
years. In light of the possibility of reliably studying field theory for 
arbitrary $d$ the information contained in the $\epsilon$ expansion can be 
viewed in a new way. Again this has been examined in the case of the Ising 
Gross-Neveu model \cite{231}. The approach is to collect the information 
contained in the $\epsilon$ expansion of the two theories in the universality 
class near two and four dimensions and construct a Pad\'{e} approximant to the 
exponent in $d$-dimensions. For the Ising Gross-Neveu model of (\ref{laggn}) a 
matched approximant has been precisely constructed from the four loop 
renormalization group functions of (\ref{laggn}) and (\ref{laggny}) which have 
been computed over several years \cite{158,160,232,233,234}. Moreover the 
behaviour of the three highlighted exponents in $2$~$<$~$d$~$<$~$4$ 
qualitatively matches that from the other two techniques. Indeed the 
comprehensive analysis of Ref. \cite{231} has demonstrated that the 
combination of the data from the various approaches can give accurate three 
dimensional estimates similar to using Monte Carlo simulations. The latter by 
the very nature of the technique is strictly three dimensional. 

While we have noted the complementary aspects of the large $N$ method with
other techniques such as the perturbative $\epsilon$ expansion and the
functional renormalization group technique the first two are in effect 
different views of the same underlying technique. By contrast the functional
renormalization group deploys fundamental aspects of the renormalization
group to gain structural data on the theory. It is nonperturbative in nature.
However in being able to study field theories with the spacetime dimension not
fixed to be an integer value it is also possible to carry out large $N$ 
studies with the functional renormalization group method. Early work in this 
direction can be found in Refs. \cite{235} and \cite{236}. However in recent 
years several aspects of that work has come into question since results 
apparently contradict results from other methods. An example of where this was 
observed is Ref. \cite{176} where scalar $O(N)$ models were studied. The 
resolution was subsequently provided in Ref. \cite{177} where the assumptions 
in the development of the core functional renormalization group equations in 
the large $N$ limit were re-examined.  

\section{Discussion}	

In this review we have endeavoured to give a balance between background to the
development of the large $N$ expansion in quantum field theories relating to 
particle physics and more recent directions. In the main in the latter case 
this has concerned the large $N$ critical point formalism of 
Refs. \cite{34,35} and \cite{36} although there are other large $N$ 
approaches which we have not discussed at length \cite{237,238}. The main 
feature of the critical point approach is the use of the underlying universal 
field theory living at the $d$-dimensional Wilson-Fisher fixed point. This 
allows one to compute renormalization group invariant data through the 
$d$-dimensional critical exponents. The current viewpoint is that the 
information derived from these complement analyses from other methods. It is 
perhaps in this general context that one ought to view the overall technique. 
In studying problems of interest such as phase transitions, there are a variety
of techniques such as the $\epsilon$ expansion or perturbation theory, 
conformal bootstrap and the functional renormalization group which can equally 
be used as investigation tools. However each in their own way and approximation
computes data on the same quantity. Rather than compete the current movement is
now towards pooling that information in order to improve the overall picture. 
This has been evident in the exponent estimates in the Gross-Neveu analysis of 
Ref. \cite{231} where there is a hope that equivalent reliable precision will 
be obtained for similar and other theories. In this context we have focused 
throughout on two key theories which are the nonlinear sigma model and the 
Gross-Neveu model as well as their four dimensional counterparts which are 
$\phi^4$ theory and the Gross-Neveu-Yukawa models respectively. This is partly 
because not only are their perturbative renormalization group functions known 
to very high loop order but the leading three terms of the large $N$ critical 
exponents are available. With recent activity in critical field theories 
revealing the concept of emergent symmetries where apparently different 
theories can be connected at one fixed point, there is now a need to develop 
the critical point large $N$ method to a new set of models.

One such model is the $CP(N)$ model which is a generalization of 
(\ref{lagnlsm}) where there is a $U(1)$ gauge field. This has been studied
over many years in the large $N$ expansion in different contexts. See 
Refs. \cite{142,144,239,240,241,242,243,244,245} and \cite{246} for several 
instances. At present there is interest in the critical properties of the 
$CP(N)$ model in three dimensions. One example of this is that for a specific 
value of $N$ it has connections with a possible duality 
\cite{247,248,248,250,251} in an extension of (\ref{laggn}) where a QED sector 
is appended. To probe the duality further requires data on the critical 
exponents to high precision for the $CP(N)$ model side of the equivalence. In 
this context there are only a few leading order large $N$ critical point 
studies \cite{142,144} with no $O(1/N^2)$ exponents known in $d$-dimensions. In
principle the necessary formalism is available to carry out the computations to
$O(1/N^2)$. However such calculations will be tedious due to the presence of 
the gauge field. As is discussed in the Appendices the useful calculational 
technique of conformal integration for scalar and scalar-Yukawa type vertices 
has no parallel for a scalar-gauge or fermion-gauge vertex. While this can be 
circumvented by methods reviewed in the Appendices for practical purposes of
determining specific integral values, it is perhaps indicative of the potential
conflict between gauge and conformal symmetry indicated briefly earlier and in 
Ref. \cite{252}. For instance, this limitation of conformal integration in a 
gauge sector would suggest that the development of the Polyakov triangle and 
conformal vertex for a gauge theory is technically difficult if not impossible.
Moreover it is not clear if there are obstructions to applying the modern
manifestation of the conformal bootstrap to gauge theories for related reasons.
That aside in Ref. \cite{147} an attempt was made to develop the basic 
equations introduced in Ref. \cite{197} similar to (\ref{cnbooteqn}) for a 
gauge theory but these were not placed in the large $N$ context. While we have 
noted this in the context of the $CP(N)$ model which is a scalar theory the 
same potential limitations would apply to any extension to QED or QCD. For the 
other major symmetry of field theoretic interest, supersymmetry, the large $N$ 
critical point formalism can be applied without any deep problems. For instance 
exponents in the supersymmetric nonlinear sigma model have been computed 
\cite{253,254}. Indeed early large $N$ work on this specific model 
\cite{255,256,257,258} and the $CP(N)$ extension \cite{259,260} revealed a mine
of very rich properties due to the underlying supersymmetry. In the latter case
the critical point formalism has again only been applied at leading order 
\cite{261,262} in $1/N$. The relation of large $N$ results to the understanding
of lattice computations can also be found in Refs. \cite{83} and \cite{263}.

Finally we note that any review of a topic while there is still development
activity in the area can merely represent a brief snapshot of the subject
before it is superseded by new breakthroughs. Although the large $N$ methods in
general have served the development and understanding of quantum field theories
well over a period of half a century they can only ever be a complementary tool
to our overall understanding of the part of Nature we wish to describe. However
at this particular point exciting new ideas are being examined which we have 
alluded to such as the hope that gravity can be finally unveiled through 
Weinberg's asymptotic safety vision \cite{126} as well as the potential of 
emergent symmetries to give insight into the theory which lies beyond the
Standard Model.  
 
\section*{Acknowledgments}

This work was supported in part by the STFC Consolidated Grant ST/L000431/1 and
a DFG Mercator Fellowship. Also the author has great pleasure in alphabetically 
thanking 
Dr. J. Babington,
Dr. J.M. Bell,
Prof J. Bl\"{u}mlein,
Dr. D.J. Broadhurst,
J. Hackett, 
Dr. J.R. Honkonen,
Prof I. Jack,
Dr. H. Kissler,
Prof D. Kreimer,
Prof S. Moch,
Prof A.J. Niemi,
Prof H. Osborn,
Dr. E. Panzer,
Dr. M.M. Scherer,
Dr. R.M. Simms,
Dr. S. Teber
and 
Dr. J.A.M. Vermaseren
for many valuable discussions, encouragement, interest and support over either 
early, many or recent years on foundational aspects of material in this 
article. The hospitality of LPTMC, Sorbonne University, Paris and the
Mathematical Physics Group at Humboldt University, Berlin, where part of the 
work was carried out, is also gratefully acknowledged. The figures were 
prepared with {\sc Axodraw} \cite{264} and various calculations could not have
been achieved over the years without the symbolic manipulation language 
{\sc Form} \cite{265,266}.

\appendix

{\begin{figure}[hb]
\begin{center}
\includegraphics[width=11.5cm,height=3.0cm]{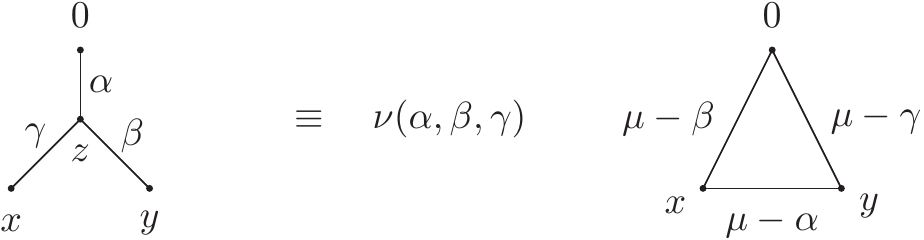}
\end{center}
\caption{Unique vertex integration when 
$\alpha$~$+$~$\beta$~$+$~$\gamma$~$=$~$2\mu$ in coordinate space 
representation.}
\label{uniqcr}
\end{figure}}
 
\section{Computational methods}

As the large $N$ critical point formalism involves (massless) propagators with 
nonunit powers unlike conventional perturbation theory the well-developed and
standard techniques to evaluate Feynman integrals cannot always be applied. By 
this we mean analytic methods ranging from the early introduction of 
integration by parts \cite{267,268,269} and the Gegenbauer polynomial approach 
\cite{270} to the Mellin-Barnes representation of propagators \cite{271,272}. 
The success of several of these have led to computer algebra packages to 
determine the poles and finite parts of three, four and higher loop Feynman 
graphs. These include {\sc Mincer} \cite{273,274} and its four loop successor 
{\sc Forcer} \cite{275,276}, which are efficient tools in the main for 
evaluating massless $2$-point functions in four dimensional theories, as well 
as the packages which were developed to implement the systematic integration by 
parts algorithm of Laporta \cite{277} such as {\sc Reduze} \cite{278,279}, 
{\sc Fire} \cite{280}, {\sc Litered} \cite{281,282}, {\sc Air} \cite{283} and 
{\sc Kira} \cite{284}. While the Laporta method has the benefit of handling 
huge numbers of Feynman integrals to complete the evaluation of a Green's 
function it requires the explicit values of what is termed master integrals for
the final stage. This is where other techniques are relevant such as 
Mellin-Barnes and Schwinger parameter representations of Feynman graphs. 
Packages for the implementation of these are available such as {\sc Mb} 
\cite{285} and {\sc Hyperint} \cite{286}. A recent comprehensive review of 
these techniques has been provided in Ref. \cite{287} while we will focus on 
another method almost exclusively used for the large $N$ critical point 
technique.

{\begin{figure}[ht]
\begin{center}
\includegraphics[width=8.5cm,height=3.0cm]{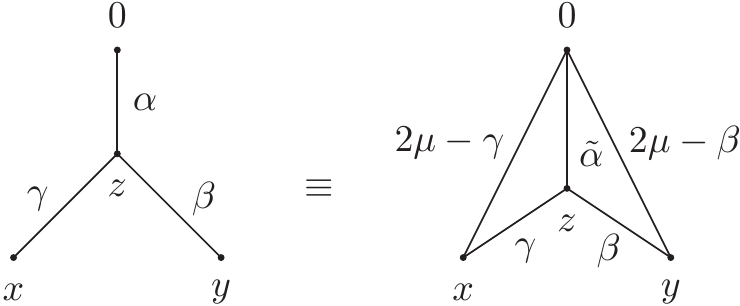}
\end{center}
\caption{Application of conformal transformation to coordinate space vertex.}
\label{cnfrcr}
\end{figure}}

While these and other techniques are more than valuable in pushing back our
conventional perturbative knowledge of the renormalization group functions of 
many theories to very high loop order, there is another method which is central
to the class of Feynman diagrams which arise in the critical point large $N$ 
approach. It is termed conformal integration or uniqueness and was introduced 
in Ref. \cite{196} specifically for three dimensions. Thereafter it was 
extended to $d$-dimensions in Refs. \cite{34} and \cite{35} as the conditions
when the uniqueness method can be applied are naturally satisfied for the large
$N$ calculations at the $d$-dimensional Wilson-Fisher fixed point of the scalar
and fermion theories we have discussed. While this is the main direct 
application of conformal integration, it has been used in certain cases for 
evaluating finite integrals in perturbation theory. For instance, in 
Refs. \cite{288} and \cite{289} two loop finite massless off-shell master 
$3$-point integrals were evaluated in terms of the polylogarithm function. In 
that particular computation the master integral was analytically regularized 
and the uniqueness rule applied after an integration by parts. Moreover there 
is a comprehensive introduction to the application of uniqueness to very high 
loop order integrals in the lectures of Ref. \cite{290}. Details are provided 
there on some of the more technical developments of the method. Other sources 
can be found in Refs. \cite{55,291,292} and \cite{293} and \cite{294}. In 
the case of Ref. \cite{291} it was used to determine the value of the four 
loop massless $2$-point zigzag graph for the final remaining analytic number 
required for the five loop $\beta$-function of $N$~$=$~$1$ $\phi^4$ theory in 
four dimensions \cite{55}. This was subsequently extended to five loops for the
$O(N)$ case \cite{56} as well as six \cite{57} and seven \cite{58} loops. These
were all for the $\MSbar$ scheme. The latter two evaluations used the more 
modern approaches bannered under the heading of applications of algebraic 
geometry. To be more concrete the rule for integrating over what is called a 
unique vertex is given in Fig. \ref{uniqcr}. To summarize when the sum of the 
exponents of three propagators is equal to the spacetime dimension \cite{35} 
then the integration over the vertex coordinate $z$ can be carried 
out\footnote{In this Appendix we use $z$ as a coordinate which is not to be 
confused, for instance, with the combination of amplitudes used in 
(\ref{nlsmsde}).}. This rule is also sometimes referred to as the star-triangle
rule. 

{\begin{figure}[ht]
\begin{center}
\includegraphics[width=12.0cm,height=9.5cm]{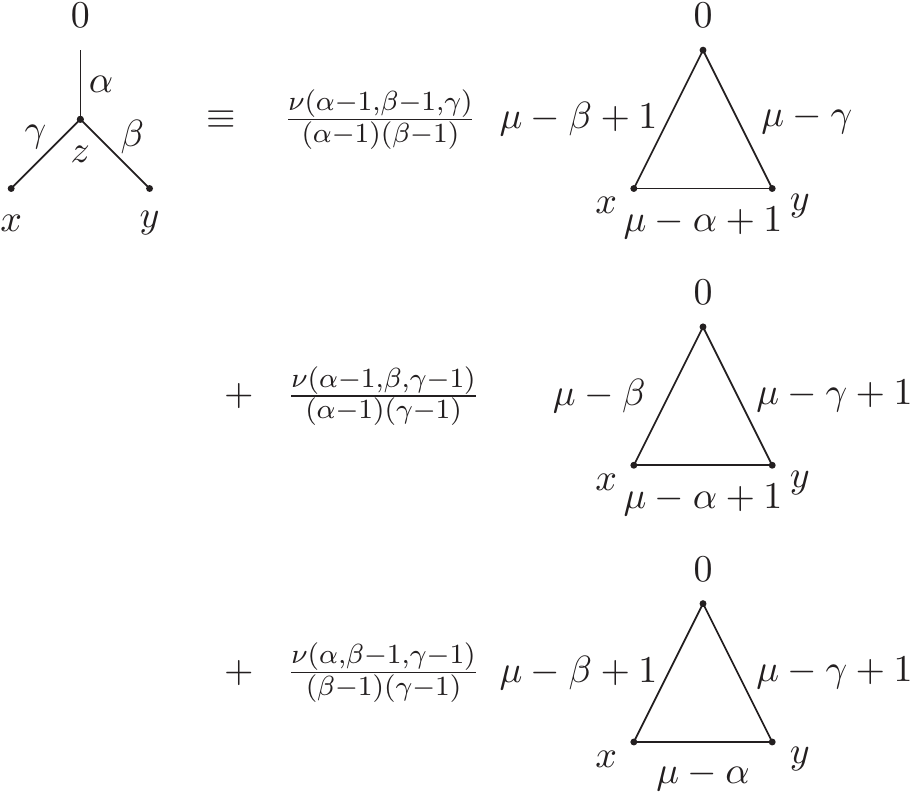}
\end{center}
\caption{Integration rule for one step from uniqueness vertex when 
$\alpha$~$+$~$\beta$~$+$~$\gamma$~$=$~$2\mu$~$+$~$1$ in coordinate space 
representation.}
\label{stepcr}
\end{figure}}

There are various ways of verifying the relation of Fig. \ref{uniqcr}. One is
to carry out the explicit integration over $z$ using Feynman parameters which 
produces a one parameter integral involving a hypergeometric function dependent
on arbitrary $\alpha$, $\beta$ and $\gamma$. The final integration cannot be 
completed in general for the three independent values of the exponents in 
general. However when the uniqueness condition \cite{34,35} 
\begin{equation}
\alpha ~+~ \beta ~+~ \gamma ~=~ 2 \mu
\end{equation}
is satisfied the hypergeometric function collapses to a simple form allowing
the final one parameter integration to proceed. When a vertex satisfies this 
criterion it is called a unique vertex. By contrast when the sum of the 
exponents on the lines comprising a triangle sum to $\mu$ then it is termed a 
unique triangle and can be replaced by a unique vertex. This is evident from 
the right hand side of the equation in Fig. \ref{uniqcr}. A more elegant way of
deriving the result is to apply the conformal transformation of 
(\ref{confmap}). For example, using
\begin{equation}
x_\mu ~\to~ \frac{x_\mu}{x^2}
\end{equation}
and similar mappings for the other coordinates gives
\begin{equation}
(x-y)^2 ~\to~ \frac{(x-y)^2}{x^2 y^2} ~~,~~
d^d z ~\to~ \frac{d^d z}{(z^2)^{2\mu}}
\label{conftr}
\end{equation}
where the Jacobian associated with the measure is indicated in the second
equation. The application of (\ref{conftr}) to the integral on the left side of
the equation in Fig. \ref{uniqcr} is illustrated in Fig. \ref{cnfrcr} where for 
shorthand we have set 
$\tilde{\alpha}$~$=$~$2\mu$~$-$~$\alpha$~$-$~$\beta$~$-$~$\gamma$ and used the 
top external leg as the reference origin for the transformation. Choosing the 
origin at one of the other external points would lead to a different graph
topology under a conformal transformation but the same overall uniqueness
relation. The graph on the right hand side of the equation in Fig. 
\ref{cnfrcr} cannot be completed for arbitrary values of the exponents. However
if the exponent of the $z^2$ propagator is zero or a nonnegative integer then 
the resulting simple integration over the $z$-vertex can be completed. When 
$\tilde{\alpha}$~$=$~$0$ then the result of Fig. \ref{uniqcr} emerges. In the
case when $\tilde{\alpha}$~$=$~$-$~$n$ where $n$ is a positive integer then a 
more involved rule emerges \cite{214,295} which is called $n$ steps from 
uniqueness. The one step case is given in Fig. \ref{stepcr}. A complementary 
rule can be derived from this case by taking its Fourier transform \cite{214}. 
It is illustrated in Fig. \ref{trspcr} and corresponds to a one step from 
unique triangle. Generalizations of both rules are available in 
Refs. \cite{156} and \cite{214} with a mixed rule given in Ref. \cite{290}.

{\begin{figure}[hb]
\begin{center}
\includegraphics[width=12.0cm,height=9.5cm]{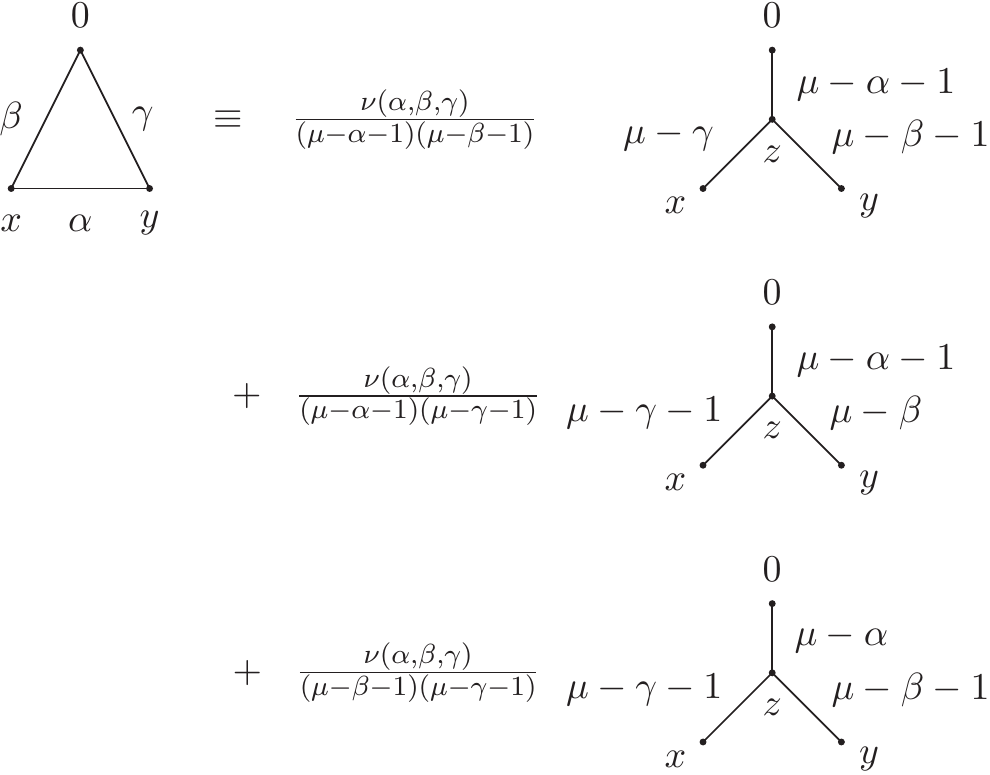}
\end{center}
\caption{Integration rule for one step from uniqueness triangle when 
$\alpha$~$+$~$\beta$~$+$~$\gamma$~$=$~$\mu$~$-$~$1$ in coordinate space 
representation.}
\label{trspcr}
\end{figure}}

One benefit of the rule of Fig. \ref{stepcr} is that one can derive relations
between different Feynman integrals. For $O(1/N^2)$ computations even though 
there are high loop graphs to be determined for the skeleton Schwinger-Dyson 
equations in the detailed evaluation the more intractable ones reduce to two 
loop scalar self-energy graphs with nonunit propagator exponents. To extract 
the required value, if it cannot be carried out in a simple direct way, 
necessitates a type of large $N$ Laporta algorithm where the integral is 
manipulated to what is now termed as master integrals. While the Laporta 
algorithm used for explicit perturbative computations is systematic and 
encoded in symbolic manipulation computer programmes, the situation for the 
large $N$ graphs has yet to be systematized. This is partly because there is no
systematic way of finding a concrete termination point in the algorithm. For 
perturbative integrals with unit exponent propagators integration by parts can 
nullify this power on one propagator which reduces the graph to one of a lower 
topology. Hence, in principle, it is easier to determine. With nonunit 
exponents in the large $N$ critical point approach it is rare that integration
by parts or the approach we will review here will reduce an integral to a lower
topology. Instead what one has to effect is the reduction of a graph to a 
situation where a vertex or a triangle becomes unique whence one can carry out 
an evaluation. An example of this is given in the next section.

{\begin{figure}[ht]
\begin{center}
\includegraphics[width=10.5cm,height=7.5cm]{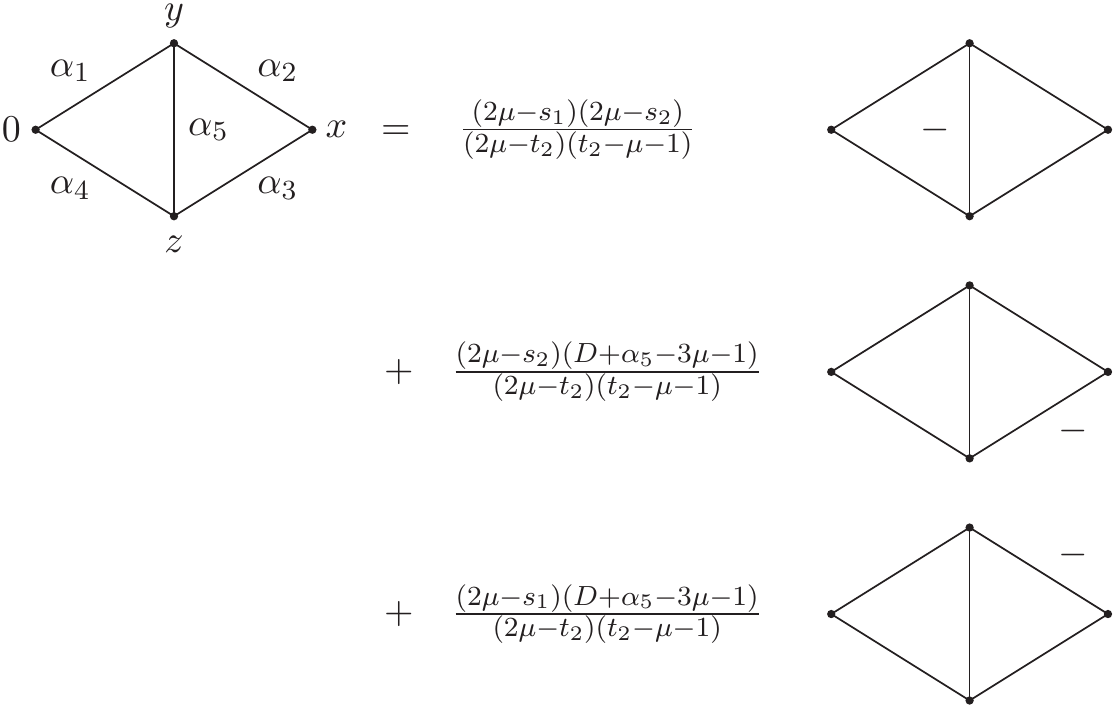}
\end{center}
\caption{Integration rule for two loop self-energy in coordinate space
representation.}
\label{redncr}
\end{figure}}

An example however of the type of relations which be derived using the one step
from uniqueness \cite{295,296} is illustrated in Fig. \ref{redncr} where 
$\alpha_i$ are arbitrary exponents. The distribution of these exponents on the 
propagators in the graphs on the right hand side of the equation are the same 
as the corresponding one on the left except where there is a minus sign. In 
that case the power of the exponent is reduced by unity. The parameters 
appearing in the coefficients of the graphs in Fig. \ref{redncr} correspond to 
those introduced in Ref. \cite{35} and are related to the $\alpha_i$. The full
set are
\begin{eqnarray}
s_1 &=& \alpha_1 + \alpha_2 + \alpha_5 ~~~,~~~
t_1 ~=~ \alpha_1 + \alpha_4 + \alpha_5 ~~~,~~~
D ~=~ \sum_{i=1}^5 \alpha_i \nonumber \\
s_2 &=& \alpha_3 + \alpha_4 + \alpha_5 ~~~,~~~
t_2 ~=~ \alpha_2 + \alpha_3 + \alpha_5 ~.
\end{eqnarray}
As an aside the general properties and structure of the two loop self-energy
graph of the left hand side of the relation in Fig. \ref{redncr} has now been 
well-established. For instance, an early investigation of its full set of
symmetries was provided in Refs. \cite{297} and \cite{298} while the all 
orders $\epsilon$ expansion of the case when 
$\alpha_i$~$=$~$1$~$+$~$a_i \epsilon$ where $d$~$=$~$4$~$-$~$2\epsilon$ was 
given in Ref. \cite{299} having built on earlier high order expansions 
\cite{35,139} for instance. The result of Ref. \cite{299} established that the
series expansion only contains multiple zeta values. The relation of Fig. 
\ref{redncr} is not the only such relation as one can derive similar ones. 
However, one benefit it has is that it reduces the sum of exponents at least at
one vertex or for one triangle. So it is akin to the integration by parts 
reduction that is at the heart of the Laporta algorithm. Moreover the 
associated prefactors are not singular at any integers as is the case in 
similar relations. This is evident in the more complete set of these relations 
that have been listed in the Appendix of Ref. \cite{300}.

{\begin{figure}[hb]
\begin{center}
\includegraphics[width=12.0cm,height=3.0cm]{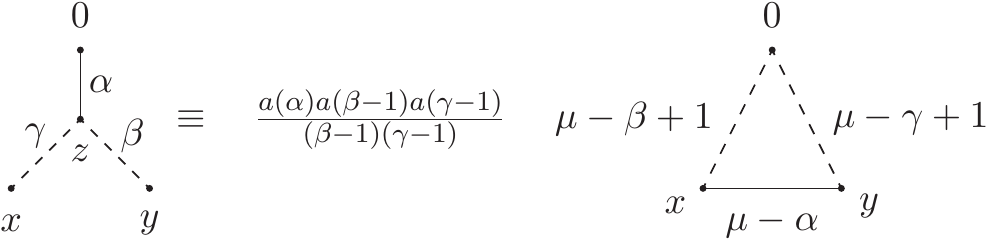}
\end{center}
\caption{Unique vertex integration for a Yukawa type vertex when 
$\alpha$~$+$~$\beta$~$+$~$\gamma$~$=$~$2\mu$~$+$~$1$ in coordinate space 
representation.}
\label{gnuniqcr}
\end{figure}}

While the focus of the conformal integration has been on scalar vertices the
rules can be generalized to other cases. For instance, rules for vertices with 
tensor structure involving the vectors of the three propagator meeting at $z$ 
in Fig. \ref{uniqcr} have been given in 
Refs. \cite{156,293,301,302} and \cite{303}. Another direction is the 
construction of a uniqueness rule for vertices involving fermions which can be 
applied to Yukawa type theories. The basic unique vertex for the Gross-Neveu 
model was provided in Ref. \cite{146} and shown in Fig. \ref{gnuniqcr}. Due to
the presence of fermion propagators the criterion for integration differs from 
that of the purely scalar case in that the sum of the exponents of the 
propagators meeting at a vertex have to be one higher than the spacetime 
dimension. The derivation of this rule can be carried out in either of the two 
ways indicated for the full scalar vertex. In order to use the conformal 
transformation approach the rules of (\ref{conftr}) imply
\begin{equation}
( \xslash - \yslash ) ~ \to ~ -~ \frac{\yslash ( \xslash - \yslash ) \xslash}
{x^2 y^2} ~=~ -~ \frac{\xslash ( \xslash - \yslash ) \yslash} {x^2 y^2} 
\label{conftrfer1}
\end{equation}
where there are two ways of writing the result of the transformation. One is
more useful than the other depending on the structure of the neighbouring
propagator. So for example when one is dealing with a scalar-fermion vertex 
whose Feynman rule does not involve a $\gamma$-matrix then applying a conformal 
transformation to successive strings of numerators which appear in a fermion
propagator we have
\begin{equation}
( \xslash - \zslash ) ( \zslash - \yslash ) ~ \to ~
\frac{\xslash ( \xslash - \zslash ) ( \zslash - \yslash ) \yslash}
{x^2 y^2 z^2} ~.
\label{conftrfer2}
\end{equation}
This implies that transforming the vertex on the left side of that shown in
Fig. \ref{gnuniqcr} the internal propagator structure is preserved. Again we 
have chosen the origin for the conformal transformation to be at the top of the
scalar propagator. Under the transformation it is the exponent of this line 
which provides for the definition of the uniqueness condition similar to that 
shown in Fig. \ref{cnfrcr}. The denominator factors in (\ref{conftrfer2}) lead 
to the appearance of unity in the condition for a scalar-fermion uniqueness 
condition which is 
$\alpha$~$+$~$\beta$~$+$~$\gamma$~$=$~$2\mu$~$+$~$1$. While the rule of Fig. 
\ref{gnuniqcr} is extremely useful for large $N$ critical point computations in
the Gross-Neveu universality classes, it can be generalized to $n$ steps away
from uniqueness similar to the rule of Fig. \ref{stepcr}. The case of 
$n$~$=$~$1$ is given in Fig. \ref{figgnstcr} where
\begin{equation}
\tilde{\nu}(\alpha,\beta,\gamma) ~=~ 
\frac{\nu(\alpha-1,\beta-1,\gamma-1)}{(\alpha-1)(\beta-1)(\gamma-1)} ~.
\end{equation} 

{\begin{figure}[ht]
\begin{center}
\includegraphics[width=12.5cm,height=12.0cm]{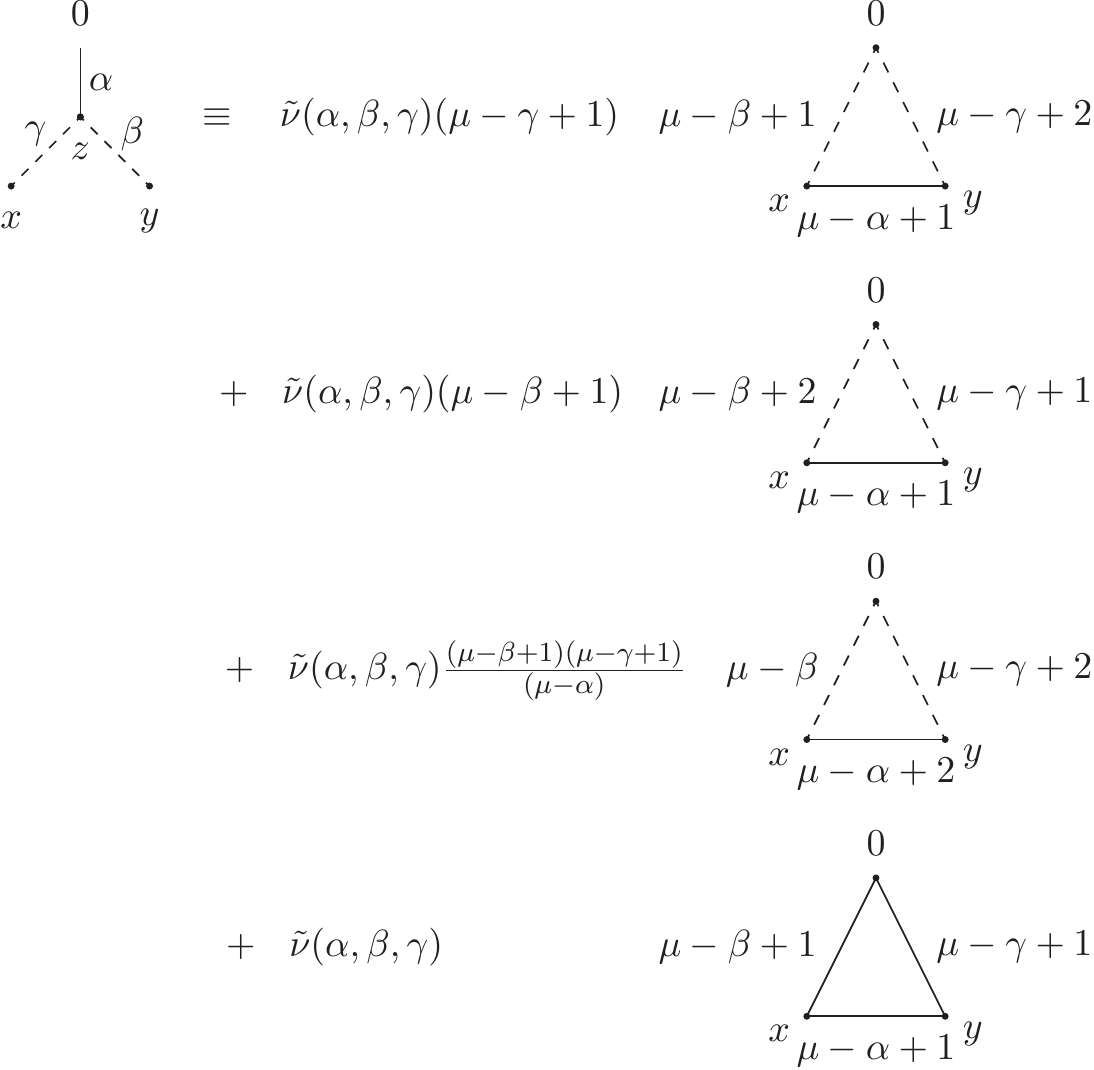}
\end{center}
\caption{Integration rule for one step from uniqueness for a Yukawa type vertex
when $\alpha$~$+$~$\beta$~$+$~$\gamma$~$=$~$2\mu$~$+$~$2$ in coordinate space 
representation.}
\label{figgnstcr}
\end{figure}}

Although the two main uniqueness rules for a pure scalar and Yukawa vertices 
have proved to be of significant benefit to evaluating the underlying Feynman
graphs needed for high order large $N$ computations the extension to gauge
theories is not straightforward. By this we mean the usual gluon-quark vertex 
of QCD rather than the vertex of a scalar gauge theory. This is because a
technical complication arises due in essence to the $\gamma$-matrix structure 
in the vertex. In addition the propagator of the gauge field of 
(\ref{ferphoprop}) in coordinate space is comprised of two terms. Although the 
first is not unlike the scalar propagator the second part is in effect a rank
$2$ tensor. Ignoring the $\gamma$-matrix of the fermion propagator form of 
(\ref{ferphoprop}) then in that case one is dealing with a rank $1$ tensor. It
is this in effect which was responsible for the Yukawa vertex having a
uniqueness condition requiring the exponents at the vertex to sum to
$2\mu$~$+$~$1$ in contrast to the rank $0$ case of $2\mu$. Therefore on these 
grounds one would expect the uniqueness condition for a gauge-fermion vertex 
for a gauge field in a general gauge to be $2\mu$~$+$~$2$. To apply such a rule
to large $\Nf$ gauge theories, however, one would require the exponent sum to 
be $2\mu$~$+$~$1$ on dimensional grounds given in (\ref{expqed}). However it 
turns out that there is a situation where for a certain value of the gauge 
parameter this uniqueness condition could be replaced by one with a smaller 
value such as $2\mu$~$+$~$1$ which would then be of potential use. For it to be
useful for practical purposes though, it should not produce an evaluation where 
the original gauge is not preserved. A detailed study of conformal integration
for the specific case of a gauge-fermion vertex has been discussed at length in 
Refs. \cite{301,302} and \cite{303}. In addition the rule for a general 
vertex with tensor structure is summarized in 
Refs. \cite{156} and \cite{293}. In Refs. \cite{301,302} and \cite{303} the
special case for a unique gauge-fermion vertex for a value of the sum of the 
vertex exponents being lower than $2$ above the spacetime dimension was 
discussed. This is the case when either (\ref{confcond1}) or (\ref{confcond2}) 
is satisfied. In other words the asymptotic form of the gauge field propagator 
is proportional to $\Lambda_{\mu\nu}$. The reason for the simplification can be
seen by applying a conformal transformation to the gauge-fermion vertex in the 
same way as in (\ref{cnfrcr}) when the gauge field propagator involves 
$\Lambda_{\mu\nu}(z)$. After the transformation part of the string of 
$\gamma$-matrices involves
\begin{equation}
\zslash \gamma^\nu \zslash \Lambda_{\mu\nu}(z) ~=~
\zslash \gamma_\mu \zslash ~-~ 2 z_\mu \zslash ~.
\end{equation}
Using the Clifford algebra of the $\gamma$-matrices produces $\gamma_\mu$. For
any other gauge this simplification would not happen. Therefore to complete
the integration the analogous exponent $\bar{\alpha}$, similar to that in 
previous cases, is set to zero which means that the uniqueness condition in 
this case is the same as the scalar-fermion case. However the $z$-integration 
has to be completed and this is not similar to the simpler vertices as it 
involves a rank $2$ tensor. Consequently one cannot write the integral under 
this uniqueness condition as a triangle involving the original propagators of 
the vertex nor indeed as one term. Therefore we have not provided a graphical 
representation of the rule for this particular gauge choice but it can be found
in Ref. \cite{303} as well as in Ref. \cite{150,156}. Another way of viewing 
the lack of a uniqueness rule analogous to those of (\ref{cnfrcr}) and 
(\ref{gnuniqcr}) for a general gauge field propagator is that under a conformal
transformation the gauge which one is carrying out the calculation in is not 
preserved. While this effectively rules out applying direct conformal 
transformations on large $N$ graphs in a gauge theory, the underlying Feynman 
graphs themselves can still be evaluated using indirect conformal methods. By
this we mean the tedious decomposition of the graph into the constituent scalar
Feynman integrals by taking the spinor traces and then using the scalar vertex
uniqueness rule as well as integration by parts on these individual integrals. 

{\begin{figure}[ht]
\begin{center}
\includegraphics[width=11.0cm,height=10.0cm]{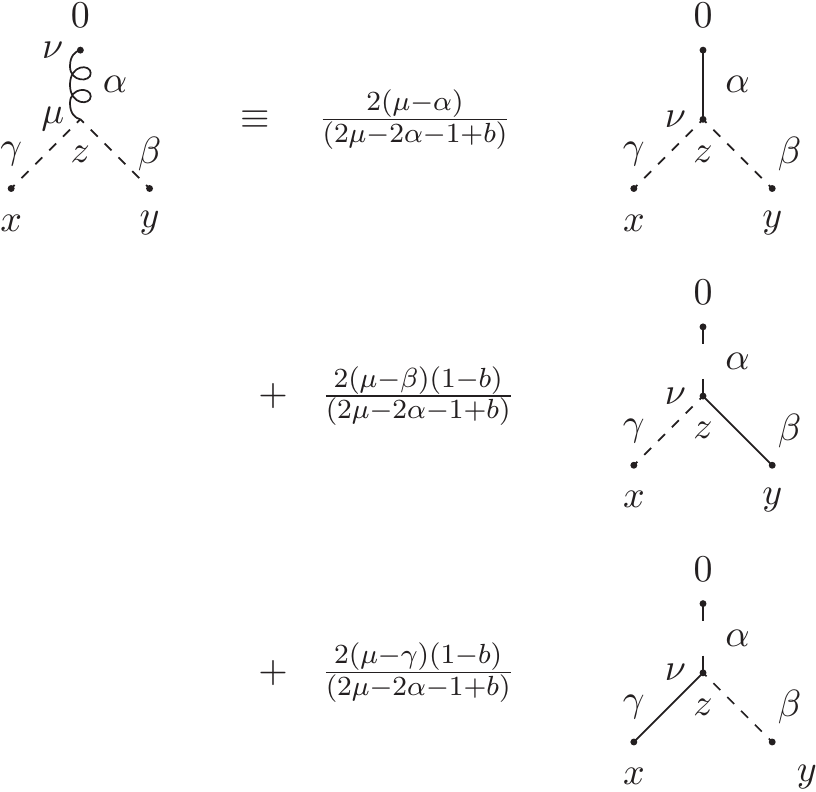}
\end{center}
\caption{Integration by parts rule for gauge-fermion vertex in 
coordinate space representation.}
\label{glipcr}
\end{figure}}

While such an approach is in principle not problematic one can still use 
properties of the gauge-fermion vertex at the start of a large $N$ gauge theory
integral evaluation. Aside from the rule provided in 
Refs. \cite{301,302} and \cite{303} another rule for the gauge vertex can be 
constructed and partially simplifies the complications stemming from the second
term of the gauge propagator in (\ref{ferphoprop}). This is achieved by using 
integration by parts on the second term inside the coordinate space 
representation of the vertex. The rule is illustrated in Fig. \ref{glipcr} 
where the curly line corresponds to the gauge field propagator for nonzero 
gauge parameter $b$ of (\ref{ferphoprop}).  We have indicated the location of 
the Lorentz indices $\mu$ and $\nu$. When it is adjacent to the vertex it 
indicates the $\gamma$-matrix contracted with one of the indices on the gauge 
propagator. When it appears at the outer end of this type of propagator it 
contracts the $\gamma$-matrix of the adjacent gauge-fermion vertex. For the 
final two vertices the Lorentz index $\nu$ is the label of the $z^\nu$ vector 
which is denoted by the long dash line. The other dashed line corresponds to a 
standard fermion propagator. The benefit of these final two graphs is that the 
$\gamma$-matrix structure is simpler whereas the first term is trivial in the 
sense that it is the Feynman gauge term of the left side of the equation. 

{\begin{figure}[hb]
\begin{center}
\includegraphics[width=10.5cm,height=3.0cm]{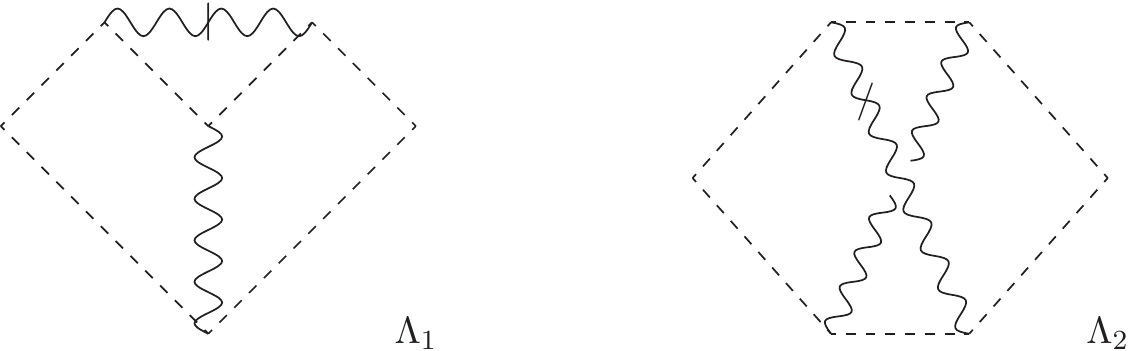}
\end{center}
\caption{Two graphs from the computation of $\nu$ in the Gross-Neveu model.}
\label{int1cr}
\end{figure}}

\section{Examples of integral evaluation}

In this section we give several examples of how the rules of the previous
section are applied in practical situations. In particular we will discuss the
various steps in integrating two graphs which occur in the computation of the
exponent $\nu$ from Ref. \cite{171} which is related to the critical slope of 
the $\beta$-function of (\ref{laggn}) at $O(1/N^2)$. These are illustrated in 
Fig. \ref{int1cr} where the tick line on a $\sigma$ field indicates the 
correction to scaling term of the corresponding asymptotic scaling form of the 
propagator given in (\ref{gaugecor}). For instance for the two dimensional 
$\beta$-function the leading term of the correction exponent is $(\mu-1)$.

{\begin{figure}[ht]
\begin{center}
\includegraphics[width=11.5cm,height=8.0cm]{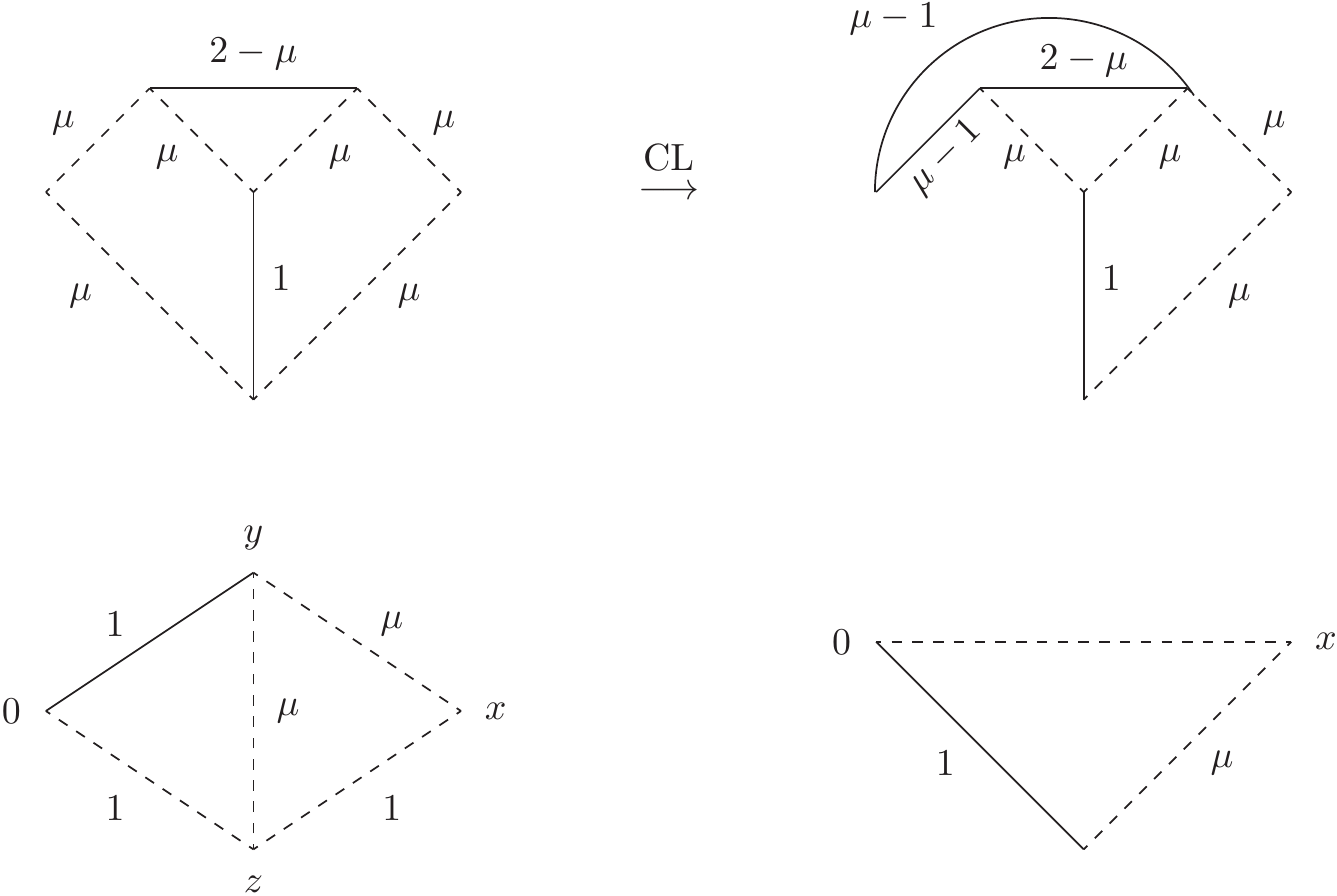}
\end{center}
\caption{Integration steps for the evaluation of $\Lambda_1$.}
\label{gn1cr}
\end{figure}}

First we concentrate on the sequence of steps to evaluate $\Lambda_1$ which are
given in Fig. \ref{gn1cr}. The starting point is the graph of the top left 
where we have included the exponents of the respective lines where there is a 
single trace over the $\gamma$-matrices. In this coordinate space 
representation there are four rather than three integrals to compute since 
there are four internal vertices to integrate over. While this may sound 
daunting the number can be reduced quickly by several of the techniques 
mentioned previously. First if we define the coordinate of the left external 
vertex as the origin then applying the conformal transformations 
(\ref{conftr}), (\ref{conftrfer1}) and (\ref{conftrfer2}) on the graph produces
the graph of the top right of Fig. \ref{gn1cr}. This transformation is denoted 
by CL in Fig. \ref{gn1cr} meaning conformal left \cite{35}. In effecting this 
there is still a trace over the fermion lines even though the graph shows a 
noncontinuous string of fermions. The reason for the break is that in rules
(\ref{conftrfer1}) and (\ref{conftrfer2}) introduce pieces such as 
$(\yslash-\zslash) (\yslash-\zslash)$ which drop out of the trace and alter the
power of the associated propagator. In certain instances a line can disappear 
as the sum of the exponents after all aspects of the transformations have been 
taken into account is zero. This is the case for one such line. While there are
still four vertices to be integrated over in the resulting graph two can be 
carried out immediately. One is the boson fermion chain where the bottom vertex 
is integrated over. The other is the unique triangle which has appeared 
involving the bosonic propagator with exponent $(2-\mu)$. Carrying out this 
integration using the rule of Fig. \ref{uniqcr} in reverse produces an 
integral which in effect is another chain in the coordinate space 
representation of Fig. \ref{chaintcr}. By this we mean that one link in the 
chain is the main integral itself which is some $d$-dependent function times a
propagator with power $1$ and the propagator with power $(2\mu-2)$ stemming 
from the propagator with the correction to scaling. The power $1$ arises from 
the dimensionalities of the propagators as well as that of the integration 
measure. The upshot is that after these two integrations the bottom left graph 
of Fig. \ref{gn1cr} emerges. Here the trace is
$\mbox{Tr}(-\zslash)(\zslash-\yslash)(\yslash-\xslash)(\xslash-\zslash)$. The
final integration is over the $y$-vertex as this is unique from the rule of
Fig. \ref{gnuniqcr}. Applying it and noting that 
$(-\zslash)\zslash$~$=$~$-$~$z^2$ produces the final chain integral of the
graph at the bottom right of Fig. \ref{gn1cr}. While this describes the 
algorithm to evaluate $\Lambda_1$ account has to be kept of the factors
associated with each step. Doing so produces
\begin{equation}
\Lambda_1 ~=~ -~ \frac{2}{(\mu-1)^4 \Gamma^4(\mu)} ~.
\end{equation} 

{\begin{figure}[ht]
\begin{center}
\includegraphics[width=11.5cm,height=7.5cm]{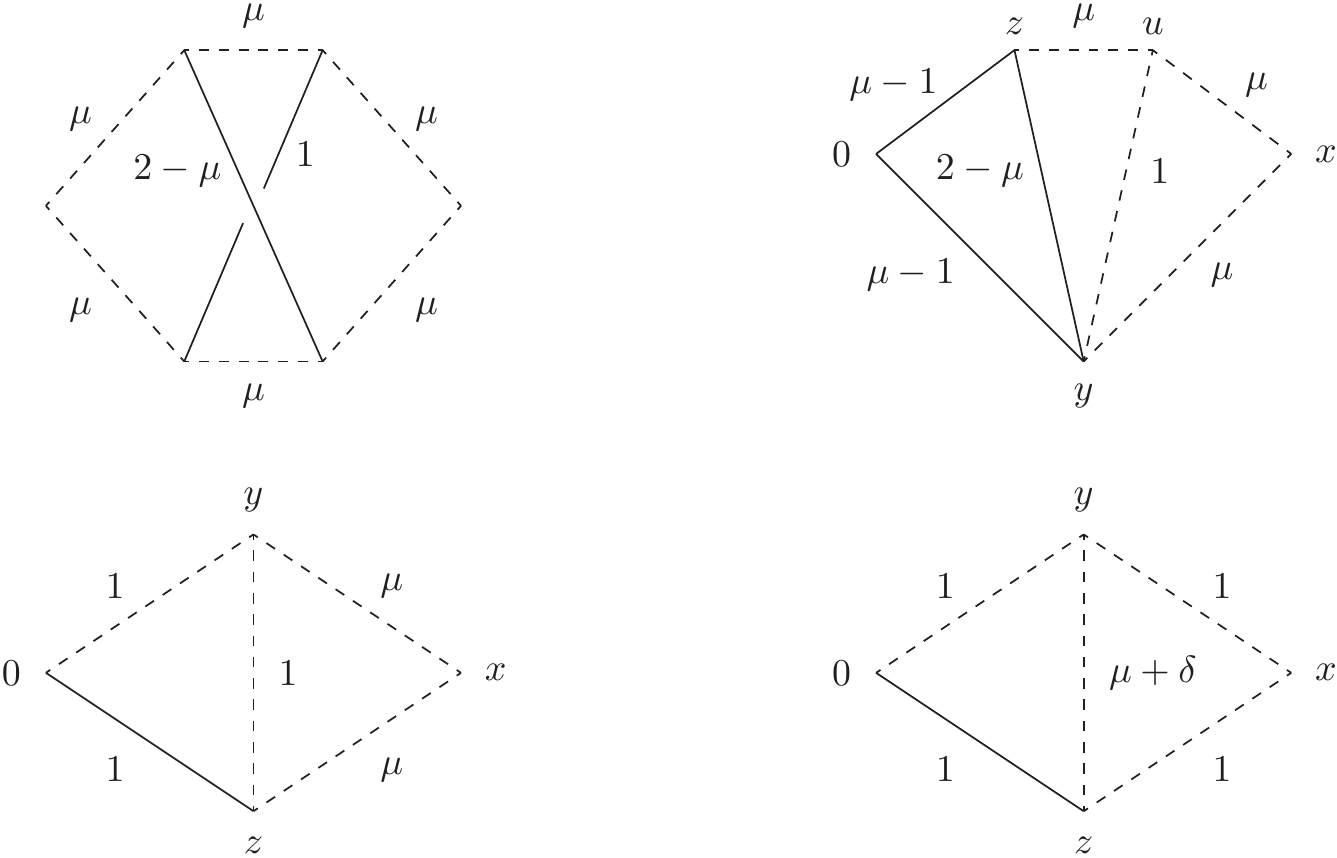}
\end{center}
\caption{Integration steps for the evaluation of $\Lambda_2$.}
\label{gn2cr}
\end{figure}}

The second case we consider is the nonplanar graph of Fig. \ref{int1cr} 
denoted by $\Lambda_2$. It corresponds to the top left graph of Fig.
\ref{gn2cr} where the exponents are included. Again there are four integrations
to be performed which can be reduced to three by first applying the uniqueness 
rule to the lower left vertex and apply the rule of Fig. \ref{gnuniqcr}. 
Applying a conformal left transformation to the resultant graph produces the 
graph at the top right of Fig. \ref{gn2cr} which contains a unique triangle. 
Implementing the unique scalar vertex relation of Fig. \ref{uniqcr} produces 
the graph of the bottom left in Fig. \ref{gn2cr} after several chain 
integrals. The last step is to apply the transformation $\leftarrow$ to the 
right hand external vertex of this graph. While the details of this rule were 
introduced in Ref. \cite{35} for a purely scalar two loop self-energy this 
transformation has a parallel for the scalar-fermion unique vertex of Fig. 
\ref{uniqcr}. The upshot is the graph at the bottom left of Fig. \ref{gn2cr}. 
The steps described so far are not dissimilar to those discussed for 
$\Lambda_1$ and we have given a brief summary for this reason. The key point 
here is that the final evaluation differs in the last step from that for 
$\Lambda_1$. In the graph the trace is
$\mbox{Tr}(-\yslash)(\yslash-\xslash)(\xslash-\zslash)(\zslash-\yslash)$ and we
have noted this since it has to be evaluated at the next step. This creates a
technical problem due to a singularity in several of the resulting scalar
integrals deriving from the exponent of the central propagator. When a scalar 
line has an exponent $\mu$~$+$~$n$ where $n$ is an integer and $n$~$\geq$~$0$ 
then the associated function for this exponent, $a(\mu+n)$, will be divergent. 

{\begin{figure}[ht]
\begin{center}
\includegraphics[width=12.0cm,height=2.2cm]{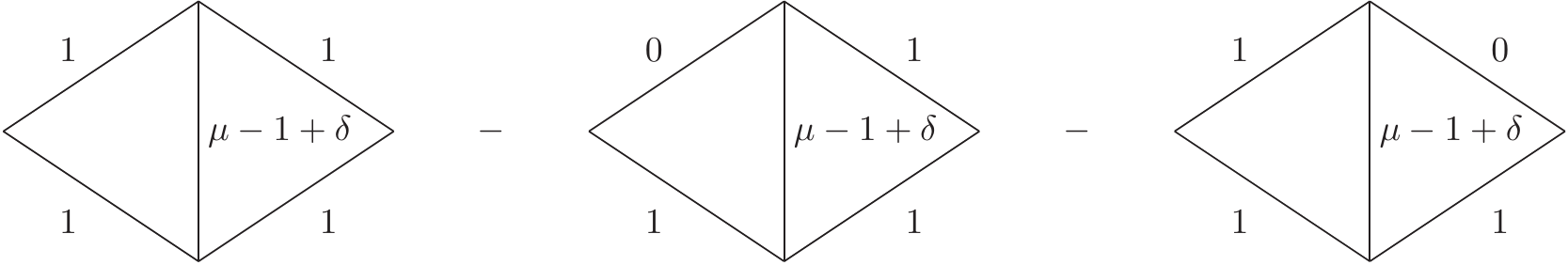}
\end{center}
\caption{Scalar integrals contributing to the evaluation of $\Lambda_2$.}
\label{gn3cr}
\end{figure}}

To circumvent the singularity after taking the trace a temporary regularization
is added to the exponent of this line which is denoted by $\delta$ in Fig. 
\ref{gn2cr}. In this and other cases where a temporary regularization is
introduced to overcome hidden singularities the location where the regulator is
placed is not unique. Moreover an appropriate choice is one where there is no 
obstruction to subsequent integration and the intermediate or temporary 
regularizing parameter must be located on the analogous lines in each of the 
graphs. For instance, it is not difficult to imagine placing it in such a way 
that a nonproblematic unique vertex within a graph ceases to be unique for a 
nonzero $\delta$. Writing
\begin{equation} 
\mbox{Tr}(-\yslash)(\yslash-\xslash)(\xslash-\zslash)(\zslash-\yslash) ~=~
(x-y)^2 \mbox{Tr}(-\yslash)(\yslash-\zslash) ~-~ 
(y-z)^2 \mbox{Tr}(-\yslash)(\yslash-\xslash) 
\label{inttr}
\end{equation}
simplifies the earlier trace. The integral corresponding to the first graph of
(\ref{inttr}) is a set of simple chain integrations where the initial one
produces a divergence. Overall it gives
\begin{equation}
-~ \frac{2(\mu-1)}{(\mu-1+\delta)} \nu(1,\mu-1+\delta,\mu-\delta)
\nu(1+\delta,1,2\mu-2-\delta)
\label{div1}
\end{equation}
where the singularity is in the third argument of the first $\nu$-function.
Taking the trace in the second term of (\ref{inttr}) produces the three
graphs of Fig. \ref{gn3cr}. The first is finite while the final two are
equivalent under reflection. The value for one of these is
\begin{equation}
\nu(1,\mu-1+\delta,\mu-\delta) \nu(1+\delta,1,2\mu-2-\delta) ~.
\label{div2}
\end{equation}
Summing the two contributions from (\ref{div1}) and (\ref{div2}) in the 
context of (\ref{inttr}) gives
\begin{equation}
\frac{2}{(\mu-1)^2} \nu(1,\mu-1,\mu-1) \nu(1,1,2\mu-2) ~+~ O(\delta)
\end{equation}
and hence the temporary regularization $\delta$ can be safely lifted. The value
of the remaining finite graph was given in Ref. \cite{35} so that the sum of 
all contributions to $\Lambda_2$ ultimately gives
\begin{equation}
\Lambda_2 ~=~ \frac{1}{(\mu-1)^2\Gamma^2(\mu)} \left[
3 \left[ \psi^\prime(\mu) - \psi^\prime(1) \right] + \frac{1}{(\mu-1)^2} 
\right] ~.
\end{equation}

\end{document}